\documentclass[aps,pre,reprint,superscriptaddress]{revtex4-1}

\usepackage{subfigure}
\usepackage{amsfonts, amssymb, amsmath}
\usepackage{bm}
\usepackage{graphics}
\usepackage{graphicx}
\usepackage{wasysym}

\usepackage[utf8x]{inputenc}
\usepackage[T1]{fontenc}

\usepackage{color}
\definecolor{darkblue}{rgb}{0,0,0.6}
\definecolor{darkred}{rgb}{0.6,0,0}

\usepackage[colorlinks=true,urlcolor=darkblue,citecolor=darkblue,linkcolor=darkred,hyperfootnotes=false]{hyperref}

\newcommand{\argc}[1]{\left[#1\right]}
\newcommand{\arga}[1]{\left\lbrace #1\right\rbrace }
\newcommand{\argab}[1]{\lbrace #1\rbrace }
\newcommand{\argp}[1]{\left(#1\right)}
\newcommand{\valabs}[1]{\vert #1\vert}
\newcommand{\moy}[1]{\left\langle  #1 \right\rangle }

\newcommand{\clap}[1]{\hbox to 0pt{\hss#1\hss}}
\newcommand{\mathclapinternal}[2]{\clap{$\mathsurround=0pt#1{#2}$}}
\def\mathclap{\mathpalette\mathclapinternal}
\newcommand{\Wbar}{ \mathclap{\phantom{W}\overline{\phantom{I}}} W}

\begin{document}

\title{
Static fluctuations of a thick 1D interface in the 1+1 Directed Polymer formulation: numerical study
}

\author{Elisabeth Agoritsas}
\email[]{Elisabeth.Agoritsas@unige.ch}
\affiliation{DPMC-MaNEP, University of Geneva, 24 Quai Ernest-Ansermet, 1211 Geneva 4, Switzerland}
\author{Vivien Lecomte}
\affiliation{DPMC-MaNEP, University of Geneva, 24 Quai Ernest-Ansermet, 1211 Geneva 4, Switzerland}
\affiliation{Laboratoire Probabilit\'es et Mod\`eles Al\'eatoires (CNRS UMR 7599), Universit\'es Paris VI et Paris VII, B\^atiment Sophie Germain, Avenue de France, 75013 Paris, France}
\author{Thierry Giamarchi}
\affiliation{DPMC-MaNEP, University of Geneva, 24 Quai Ernest-Ansermet, 1211 Geneva 4, Switzerland}

\date{\today}


\begin{abstract}

We study numerically the geometrical and free-energy fluctuations of a static one-dimensional (1D) interface with a short-range elasticity, submitted to a quenched random-bond Gaussian disorder of \textit{finite} correlation length ${\xi>0}$, and at finite temperature~$T$.
Using the exact mapping from the static 1D interface to the 1+1 Directed Polymer (DP) growing in a continuous space, we focus our analysis on the disorder free-energy of the DP endpoint, a quantity which is strictly zero in absence of disorder and whose sample-to-sample fluctuations at a fixed growing `time' $t$ inherit the statistical translation-invariance of the microscopic disorder explored by the DP.
Constructing a new numerical scheme for the integration of the Kardar-Parisi-Zhang (KPZ) evolution equation obeyed by the free-energy,
we address numerically the `time'- and temperature-dependence of the disorder free-energy fluctuations at fixed finite~$\xi$.
We examine on one hand the amplitude $\widetilde{D}_{t}$ and effective correlation length $\tilde{\xi}_t$ of the free-energy fluctuations,
and on the other hand the imprint of the specific microscopic disorder correlator on the large-`time' shape of the free-energy two-point correlator.
We observe numerically the crossover to a low-temperature regime below a finite characteristic temperature ${T_c(\xi)}$, as previously predicted by Gaussian-Variational-Method (GVM) computations and scaling arguments, and extensively investigated analytically in~Ref.~\cite{agoritsas_2012_FHHtri}.
Finally we address numerically the `time'- and temperature-dependence of the roughness $B(t)$, which quantifies the DP endpoint transverse fluctuations, and we show how the amplitude ${\widetilde{D}_{\infty}(T,\xi)}$ controls the different regimes experienced by $B(t)$
-- in agreement with the analytical predictions of a DP `toymodel' approach.

\end{abstract}


\maketitle

\tableofcontents


\section{Introduction}
\label{section-intro}

Consider a material constituted of a large number of elements which interact locally.
Long-range correlations are known to arise in such systems close to the critical point of a second-order phase transition, leading generically to scale-invariant structures~\cite{barabasi_book}.
Another situation where the collective behavior of the system constituents also {induces} correlations at large lengthscales is given by the boundaries of coexisting different phases, which define interfaces.
Examples of such systems range from growth interfaces~\cite{halpin_zhang_1995_PhysRep254,krug_1997_AdvPhys_46_139} of crystals adsorbing dissolved molecules, to domain walls (DWs) in ferromagnetic~\cite{lemerle_1998_PhysRevLett80_849,repain_2004_EurPhysLett68_460,metaxas_2007_PhysRevLett99_217208}
or ferroelectric~\cite{tybell_2002_PhysRevLett89_097601,paruch_2005_PhysRevLett94_197601,pertsev_2011_JApplPhys110_052001}
thin films, interfaces in turbulent liquid crystals~\cite{takeuchi_2010_PhysRevLett104_230601,takeuchi_2011_scientificReports1_34,takeuchi_2012_JStatPhys147_853}, 
fronts of combustion in burning paper~\cite{zhang_1992_PhysicaA189_383,maunuksela_1997_PhysRevLett79_1515}, fractures in paper~\cite{alava_2006_RepProgPhys69_669}, or contact lines in wetting experiments~\cite{alava_2004_AdvPhys53_83,santucci_2011_EurPhysLett94_46005}.
Such experimental interfaces exhibit a self-similarity at large lengthscales, which is characterized by a `roughness exponent'~$\zeta$~\cite{barabasi_book,krug_1997_AdvPhys_46_139}.
They can be studied in the generic framework of \emph{disordered elastic systems} (DES)~\cite{agoritsas_2012_ECRYS2011}, in which a 1D interface is described as an elastic string fluctuating in a two-dimensional disordered energy potential.
The elasticity of the string tends to minimize its distortions while the disorder --~accounting for inhomogeneities in the underlying medium~--, concurring with the thermal noise, induces ample geometrical fluctuations and yields metastability and glassy properties~\cite{mezard_1990_JPhys51_1831}.
Once the dimensionality, the elasticity and the type of disorder are given, the corresponding Hamiltonian of the DES modelling fully determines the universality class and the value of the roughness exponent $\zeta$.
However, an additional physical ingredient that must be included in a realistic DES modelling of experimental systems is the existence of a finite microscopic width of the interface or equivalently a finite disorder correlation length~${\xi>0}$.
The subtle interplay between this finite width and thermal fluctuations at finite temperature~$T$ raises challenging issues from an analytical point of view, with relevant implications for experimental systems~\cite{agoritsas_2012_ECRYS2011}.

\begin{figure}
 \includegraphics[width=0.95 \columnwidth]{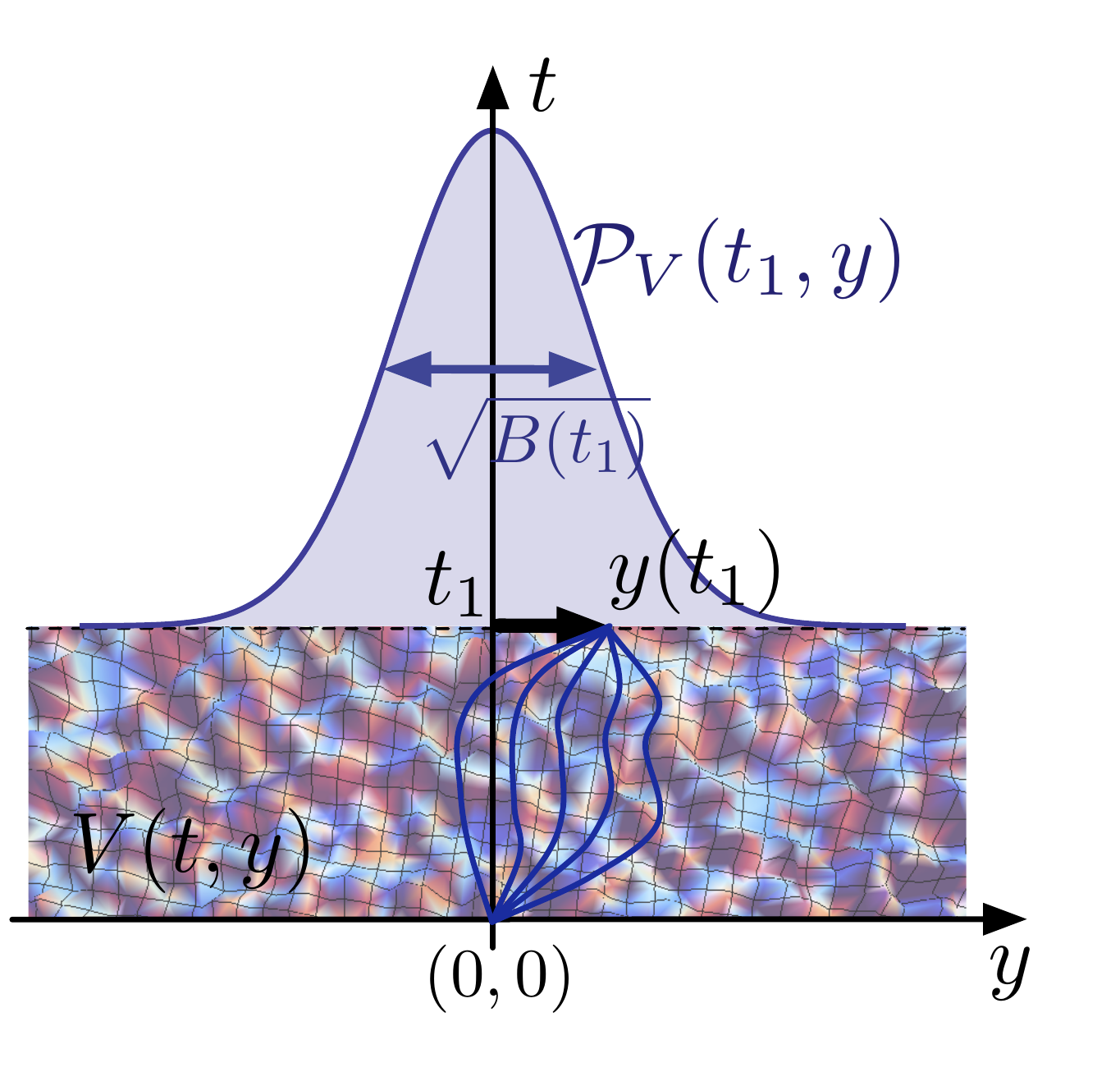}
 \caption{
	The paths represent different realizations of a static 1D interface of length $t_1$ in the 1+1 DP representation: one polymer extremity is attached at a fixed origin ${(0,0)}$ while its endpoint at `time'~$t_1$ is fluctuating.
	In a given disorder realization ${V(t,y)}$, the distribution of the endpoint ${y_1=y(t_1)}$ is denoted by ${\mathcal P_V(t_1,y_1)}$ and its mean variance at `time'~$t_1$ defines the DP `roughness' ${B(t_1)}$.
The DP description of the interface as growing along a `time'-direction allows to use tools of stochastic processes, and amounts, in the language of the 1D interface, to study the effective statistics at a fixed lengthscale~$t_1$, having integrated the fluctuations at shorter lengthscales.
 }
 \label{fig:DP-toymodel-schema}
\end{figure}

For {1D} interfaces, two universality classes of geometrical fluctuations have actually emerged from a theoretical point of view: the Edwards-Wilkinson (EW) class~\cite{edwards_wilkinson_1982_ProcRSocLondA381_17} and the Kardar-Parisi-Zhang (KPZ) class~\cite{kardar_1986_originalKPZ_PhysRevLett56_889}, with respectively ${\zeta_{\text{EW}}=1/2}$ and ${\zeta_{\text{KPZ}}=2/3}$.
A fruitful approach to study the fluctuations of a static 1D interface consists in adopting the 1+1 {Directed Polymer (DP)} description, as illustrated in~Fig.~\ref{fig:DP-toymodel-schema}: in a fixed disorder potential ${V(t,y)}$, a segment of the interface of {length} $t_1$ is described as the path ${y(t)}$ of a DP starting from a fixed origin and growing along a `time'-direction~$t$ up to a final `time'~$t_1$.
With the equivalence between the interface lengthscale and the DP growing `time', a central quantity to study is the free-energy~${F_V(t,y)}$ associated to the trajectories passing through~$y$ at `time'~$t$. {Indeed, a} complete characterization of the fluctuations of the 1D interface then amounts to the determination of the whole statistical distribution of this free-energy as a function of `time'.
This distribution encodes in particular the geometrical fluctuations of the interface as a function of the lengthscale, which are directly measurable experimentally and whose variance defines the \emph{roughness} function~${B(t)}$ (see Fig.~\ref{fig:DP-toymodel-schema}) --~following asymptotically the powerlaw~${B(t)\sim t^{2\zeta}}$.

We focus on the case of a 1D interface with a short-range elasticity and a quenched random-bond Gaussian disorder, which yields a generic continuous model belonging to the KPZ universality class~\cite{kardar_1986_originalKPZ_PhysRevLett56_889,krug_1997_AdvPhys_46_139,praehofer_spohn_2000_PhysRevLett84_4882,corwin_2011_arXiv:1106.1596}.
A large variety of problems actually belong to this class, such as random matrix models~\cite{johansson_2000_CommMathPhys209_437,praehofer_spohn_2000_PhysRevLett84_4882}, the noisy Burgers equation in hydrodynamics~\cite{forster_nelson_stephen_1976_PhysRevLett36_867,forster_nelson_stephen_1977_PhysRevA16_732}, population dynamics in random environments~\cite{comets_yoshida_2011_JTheorProbab24_657}, one-dimensional growth phenomena~\cite{halpin_zhang_1995_PhysRep254}, last-passage percolation~\cite{krug-spohn_1991_Godreche_BegRohu},
dynamics of cold atoms~\cite{kulkarni-lamacraft_2012_arXiv:1201.6363}, or vicious walkers~\cite{spohn_2006_physicaA369_71,rambeau-schehr_2010_EurPhysLett91_60006,forrester-majumdar-schehr_2011_NuclPhysB844-500}.
%
%
There has been a recent increase of interest in this class of problems both in Physics~\cite{spohn_2006_physicaA369_71,kriecherbauer_krug_2010_JPhysA43_403001,sasamoto_spohn_2010_JStatMech2010_P11013} and Mathematics~\cite{amir_arXiv:1003.0443,borodin_corwin_2012_arXiv:1204.1024}:
it has indeed been shown in those references that, for an \emph{uncorrelated disorder} (${\xi=0}$), the complete scaling of the free-energy fluctuations at all `times' has been elucidated, in the sense that the free-energy not only presents the universal roughness exponent ${\zeta_\text{KPZ}=2/3}$~\cite{kardar_1987_NuclPhysB290_582,huse_henley_fisher_1985_PhysRevLett55_2924,johansson_2000_CommMathPhys209_437,balazs_arXiv:0909.4816}, but, once properly rescaled, follows a universal distribution at asymptotically large `time'.
The case of a disordered potential \emph{correlated at short lengthscales} (${\xi>0}$) however challenges possible universality features and proves more difficult to tackle, thus less results are available.
Such correlations are nevertheless particularly relevant to understand experimental results: the $\xi=0$ limit is indeed only an ideal limit since disordered materials always present correlations at a short scale ${\xi>0}$; besides, one can show that thick interfaces (\emph{e.g.} ferromagnetic DWs) are equivalent to point-like interfaces with finite $\xi$, in the DES description~\cite{agoritsas_2010_PhysRevB_82_184207}.
In~Refs.~\cite{agoritsas_2010_PhysRevB_82_184207,agoritsas_2012_ECRYS2011} we have shown, with scaling arguments and Gaussian-Variational-Method (GVM) computations, that a characteristic temperature ${T_c(\xi)}$ separates two regimes for the roughness.
{At} high temperatures well above $T_c$, the microscopic correlation length $\xi$ plays no role and the disorder can as well be assumed to be uncorrelated,
whereas at low temperatures below ${T_c}$ it plays a role \textit{at all lengthscales}, even macroscopically.
The central quantity that controls this temperature crossover turns out to be the asymptotic free-energy amplitude~${\widetilde{D}_{\infty}}$, which also rules the roughness amplitude and characteristic crossover lengthscales.
Actually, the existence of such a temperature-dependent parameter was already hinted numerically in~Ref.~\citep{agoritsas-2012-FHHpenta}, for both a continuous and a discrete DP, as a fitting parameter in order to test a scaling relation between the free-energy global scaling properties and the DP geometrical fluctuations. 
In~Ref.~\cite{agoritsas_2012_FHHtri} we have investigated analytically the `time'- and temperature dependence of the free-energy and geometrical fluctuations at finite~$\xi$, providing a proper justification to a previous DP `toymodel' for the free-energy fluctuations and thus of its corresponding GVM roughness predictions~\cite{agoritsas_2010_PhysRevB_82_184207}; we have in particular obtained an analytical prediction for the full temperature-induced crossover of ${\widetilde{D}_{\infty}(T,\xi)}$.
However, a numerical counterpart to this investigation was needed in order to complete and test these analytical results, and is thus the object of the present study.

In this paper, we propose a novel numerical scheme to integrate the KPZ equation \emph{with colored noise} which governs the evolution of the free-energy ${F_V(t,y)}$, with the so-called `sharp wedge' initial conditions.
Usual procedures are either afflicted with discretization issues~\cite{krug-spohn_1991_Godreche_BegRohu}, and/or by the predominance at short `times' of thermal fluctuations, which obfuscate the genuine effects of disorder.
The numerical evaluation {that} we have designed works directly in the continuum and uses a symmetry warranting to focus on the sole disorder contribution to the free-energy.
It allows us to sample this `disorder free-energy' ${\bar{F}_V(t,y)}$ on a wide range of `times' and temperatures, at fixed $\xi$.
We perform an extensive test of the assumptions made in the analytical approach put forward in our companion paper~\cite{agoritsas_2012_FHHtri}, assessing primarily the validity of our DP `toymodel' with respect to the full free-energy fluctuations and the qualitative agreement of its GVM predictions for the roughness function.
We observe in particular the predicted  monotonous crossover to the low-temperature regime for the amplitude~${\widetilde{D}_{\infty}(T,\xi)}$.
Our numerical approach also probes successfully the KPZ-specific non-linearities together with hallmarks of the non-Gaussianity of the free-energy distribution.
We emphasize that our proposed procedure is of general purpose --~given the richness of the KPZ universality class~--, even if we apply it
here to tackle specifically issues of {the finite width of a 1D interface at finite temperature}.


%
The plan of the paper is as follows.
In~Sec.~\ref{section-DPformulation-numerics} we detail the 1+1 Directed-Polymer model, making the link with the static 1D interface and presenting the open questions at ${\xi>0}$ that we examine.
In~Sec.~\ref{section-numerical-recipe} we discuss the specific difficulties of simulating the KPZ equation and we expose the numerical procedure that we have used in this paper, before moving to the three parts of our numerical results.
Firstly, we study in Sec.~\ref{section-timeevol-Rty-fits} the `time'-evolution of the disorder free-energy  {fluctuations at fixed temperature, via the} two-point correlator of interest denoted~${\bar{R}(t,y)}$. {In order to test our DP toymodel, we measure its} amplitude~$\widetilde{D}_t$ and correlation length~$\tilde{\xi}_t$ assuming different shapes of the correlator.
Secondly, we focus in Sec.~\ref{section-temp-dep-asympt-Rbar-sat} on the large-`time' saturation of this effective disorder correlator and its temperature-dependence, measuring in particular the full temperature-induced crossover of the amplitude~${\widetilde{D}_{\infty}(T,\xi)}$.
Thirdly and last, we investigate in Sec.~\ref{section-numerical-roughness} the temperature-dependence of the roughness function $B(t)$ and its effective scale-dependent exponent $\zeta(t)$. 
We present our conclusions in Sec.~\ref{section-conclusion-numerics}.
{Appendices gather part of the technical details and thorough numerical analyses.}


\section{1+1 Directed-Polymer formulation of the static 1D interface}
\label{section-DPformulation-numerics}
In this section, we define the continuous 1+1 DP formulation of {the} 1D interface model that we study, together with our observables of interest at a fixed lengthscale or DP growing `time', as first set in~Fig.~\ref{fig:DP-toymodel-schema}.
{After recalling its known analytical properties, we present the DP toymodel discussed in~Ref.~\cite{agoritsas_2012_FHHtri} and ultimately aimed at grasping the temperature-dependence of the 1D interface geometrical fluctuations.}

\subsection{Model}
\label{subsec:model_observables}

From the point of view of the interface, the energy associated to a segment of length (or `time') $t_1$ is the sum of a short-range quadratic elastic cost and of the total potential energy accumulated along the line in a fixed disorder landscape~$V(t,y)$.

Let us first precise the definition of the \textit{unnormalized} Boltzmann weight ${W_V(t_1,y_1)}$ of trajectories starting at the fixed origin~$(0,0)$ and ending at~$(t_1,y_1)$:
\begin{eqnarray}
 W_V (t_1,y_1)
 = \int_{y(0)=0}^{y(t_1)=y_1} \mathcal{D}y(t) \, e^{- \mathcal{H} \argc{y,V;t_1}/T}
 \label{eq-def-unnorm-Boltzmann-Wv}
\end{eqnarray}
where the DES energy of the line is the sum of the integrated elastic and disorder contributions:
\begin{equation}
 \mathcal{H} \argc{y,V;t_1}
 = \int_0^{t_1} dt \, \argc{\frac{c}{2} \argp{\partial_t y(t)}^2 + V(t,y(t))}
 \label{eq-partial-Hamiltonian-DP}
\end{equation}
with $c$ the elastic constant.
{The} Boltzmann constant is conveniently set to ${k_\text{B}=1}$ such that the temperature has the units of an energy.

Importantly, the disorder potential $V(t,y)$ is assumed to have a Gaussian distribution of zero mean and of \emph{finite transverse correlations} described by a `colored noise' fully characterized by its two-point correlator:
\begin{equation}
  \overline{V(t,y)V(t',y')} = D \cdot \delta(t-t') \cdot R_{\xi}(y-y')
 \label{eq-def-moydis}
\end{equation}
where the statistical average with respect to disorder is denoted by an overline,
$D$ is the disorder strength and $\xi$ its typical correlation length.
We study the case of a \textit{short-range} `random-bond' disorder potential, \textit{i.e.} characterized by a correlation function ${R_{\xi}(y)}$ decaying fast enough at large~$y$ (faster than any power law).
The disorder strength $D$ is defined by fixing the normalization of ${R_{\xi}(y)}$ with ${\int_{\mathbb{R}} dy\:R_{\xi}(y)=1}$. 
The parameter $\xi$ is assumed to fully control the scaling of the correlator by the relation ${R_{\xi}(y)=\frac 1\xi R_{1}(y/\xi)}$ and to fix the characteristic temperature~${T_c(\xi)=(\xi c D)^{1/3}}$~\cite{agoritsas_2010_PhysRevB_82_184207,agoritsas_2012_ECRYS2011}.

The `partition function' $Z_V(t,y)$ of the DP trajectories writes
\begin{equation}
 Z_V(t,y)=\frac{W_V (t,y)}{\Wbar_{V\equiv 0}(t)}
 \label{eq-defZVfromWV}
\end{equation}
where ${\Wbar_{V\equiv 0}(t) = \int_{\mathbb R} dy\: W_{V\equiv 0} (t,y)} $ fixes the normalization of the path-integral~\eqref{eq-def-unnorm-Boltzmann-Wv}.
We refer the reader to~Ref.\cite{bertini-cancrini_1995_JStatPhys78_1377} or~\cite{amir_arXiv:1003.0443} for mathematical constructions in different contexts and to~Ref.~\cite{agoritsas_2012_FHHtri} for a `time'-discrete \textit{à la} Feynman approach.
The fixed origin translates into the initial condition~$Z_V(0,y)=\delta(y)$ and the `time'-evolution is given by a stochastic heat equation \cite{huse_henley_fisher_1985_PhysRevLett55_2924,bertini-cancrini_1995_JStatPhys78_1377,bertini-giacomin_1997_CommMathPhys183_571,alberts-quastel_2012_arXiv:1202.4403v1}:
\begin{equation}
 \partial_t Z_V(t,y)
 = \argc{\frac{T}{2c} \partial_y^2 - \frac{1}{T} V(t,y) } Z_V(t,y)
 \label{eq-FeynmanKac-Wvnorm}
\end{equation}
that can be seen as a Langevin equation with multiplicative spatiotemporal `noise' ${V(t,y)}$, and may be understood as a `Feynman-Kac' formula \cite{feynman_1948_RevModPhys20_367,kac_1949_TransAmerMathSoc65_1,kardar_2007_StatPhysFields,agoritsas_2012_FHHtri}.
In absence of disorder, ${Z_{V \equiv 0}(t,y)}$ is a normalized Gaussian of variance ${B_{\text{th}}(t)=\frac{Tt}{c}}$.
Note that the normalization~${\Wbar_{V\equiv 0}(t)}$ in~\eqref{eq-defZVfromWV} is essential to obtain the stochastic heat equation~\eqref{eq-FeynmanKac-Wvnorm} and is usually hidden in the definition of the path-integral~\eqref{eq-def-unnorm-Boltzmann-Wv}.

With the disorder-dependent normalization at fixed `time' $t$:
\begin{equation}
 \Wbar_V (t)
 \equiv \int_{-\infty}^{\infty} dy \cdot  W_V (t,y)
 \label{eq-WV-normalization}
\end{equation}
we can define the probability distribution function {(PDF)} of the DP endpoint, respectively at fixed disorder $V$ and after the disorder average:
\begin{equation}
 \mathcal{P}_V(t,y)
 \equiv \frac{W_V (t,y)}{\Wbar_V (t)}
 \, , \;
 \mathcal{P}(t,y)
 =\overline{\mathcal{P}_V(t,y)}
 \label{eq-def-PDF-yt}
\end{equation}
We emphasize that, due to different normalizations, the probability distribution ${\mathcal{P}_V(t,y)}$ is \emph{not} the same as the DP partition function ${Z_V(t,y)}$~\eqref{eq-defZVfromWV}:
only the latter evolves with the stochastic heat equation~\eqref{eq-FeynmanKac-Wvnorm}
but they coincide in absence of disorder: ${\mathcal{P}_{V \equiv 0}(t,y) = Z_{V \equiv 0}(t,y)}$.

The corresponding free-energy, defined by ${F_V(t,y) \equiv -T\log Z_V(t,y)}$, obeys the KPZ equation~\cite{huse_henley_fisher_1985_PhysRevLett55_2924,kardar_1986_originalKPZ_PhysRevLett56_889}:
\begin{equation}
 \partial_t F_V  (t,y)
 =	\frac{T}{2c} \partial_y^2 F_V (t,y)
	- \frac{1}{2c} \argc{\partial_y F_V (t,y)}^2
	+ V(t,y)
 \label{eq-FeynmanKac-FV}
\end{equation}
whose specificities arise from the non-linear term ${\argc{\partial_y F_V (t,y)}^2}$;
in absence of this non-linearity~\eqref{eq-FeynmanKac-FV} becomes the EW equation~\cite{edwards_wilkinson_1982_ProcRSocLondA381_17}.
The pure thermal contribution $F_{V\equiv 0}(t,y)$ can be explicited from the (diffusive) solution of the problem without disorder:
\begin{eqnarray}
 Z_{V\equiv 0}(t,y)
 &=& e^{-y^2/(2B_{\text{th}}(t))}/\sqrt{2 \pi B_{\text{th}}(t)}
 \label{eq-thermal-ZV} \\
 F_{V \equiv 0} (t,y)
 &=& F_{\text{th}} (t,y) + \frac T2 \log \frac{2\pi Tt}{c}
 \label{eq-FzeroV} \\
 F_{\text{th}} (t,y)
 &=& \frac{cy^2}{2t}
 \, , \quad
 B_{\text{th}}(t) = \frac{Tt}{c}
 \label{eq-FB-thermal}
\end{eqnarray}
and once removed from the total free-energy, it allows us to focus on the sole disorder contribution, which presents helpful statistical properties.
Indeed, the \emph{disorder free-energy} defined by
\begin{align}
  \bar{F}_V(t,y) \equiv F_V(t,y)- F_{V\equiv 0}(t,y)
\label{eq-genericdecompositionFvFbarV}
\end{align}
obeys the Statistical Tilt Symmetry (STS), which states that, \textit{at fixed `time' $t$}, the distribution of ${\bar{F}_V (t,y)}$ is invariant by translation along the transverse direction~$y$:
\begin{equation}
 \bar{\mathcal{P}} \argc{\bar{F}_V (t,y+Y)}
 = \bar{\mathcal{P}} \argc{\bar{F}_V (t,y)}
 \label{eq-STS-PFbarV}
\end{equation}
We refer the reader to Ref.~\cite{mezard_1990_JPhys51_1831,fisher_huse_1991_PhysRevB43_10728,hwa_1994_PhysRevB49_3136,ledoussal_2003_PhysicaA317_140,amir_arXiv:1003.0443} for previous discussions of the STS, and to~Ref.~\cite{agoritsas_2012_FHHtri} for a derivation at finite~$\xi$.
Moreover, this definition of the disorder free-energy implies that ${\bar{F}_V (t,y)}$ evolves with a `tilted' KPZ equation at~${t >0}$:
\begin{equation}
 \begin{split}
 \partial_t \bar{F}_V  (t,y)
 =&	\frac{T}{2c} \partial_y^2 \bar{F}_V (t,y)
 	- \frac{1}{2c} \argc{\partial_y \bar{F}_V (t,y)}^2 \\
 &	- \frac{y}{t} \partial_y \bar{F}_V (t,y) + V(t,y)
 \end{split}
 \label{eq-FeynmanKac-FbarV}
\end{equation}
Note that neglecting the non-linearity in this evolution does not yield back the EW equation, since the linear tilt ${\frac{y}{t} \partial_y \bar{F}_V (t,y)}$ stems itself from the KPZ non-linearity in~\eqref{eq-FeynmanKac-FV}.
The initial condition ${Z_V(0,y)=\delta(y)}$, which is difficult to express for the free-energy ${F_V(t,y)}$ (this is the `sharp-wedge' initial condition~\cite{sasamoto_spohn_2010_NuclPhysB834_523}), translates simply for the disorder free-energy into ${ \bar{F}_{V} (0,y) \equiv 0 }$, since the Dirac $\delta$-function is absorbed by the thermal ${Z_{V\equiv 0}(0,y)=\delta(y)}$.

\subsection{Disorder free-energy fluctuations and DP toymodel}
\label{subsec:disorder-free-energy-DPtoymodel}

We may assume that the scaling of the distribution ${\bar{\mathcal{P}} \argc{\bar{F},t}}$ is in large part controlled by the translation-invariant two-point disorder correlators defined by:
\begin{eqnarray}
 \bar{C} (t,y_2-y_1)
 \equiv & \overline{\argc{\bar{F}_V(t,y_1)-\bar{F}_V (t,y_2)}^2} &
 \label{eq-def-corr-FbarFbar2} \\
 \bar{R} (t,y_2-y_1)
 \equiv & \overline{ \partial_y \bar{F}_V(t,y_1) \partial_y \bar{F}_V(t,y_2)} &
 \label{eq-def-corr-etaeta}
\end{eqnarray}
where the functions ${ \bar{R}(t,y)}$ and ${\bar{C}(t,y)}$ are even functions of $y$ and are related through ${\partial_y^2 \bar{C}(t,y) = 2 \bar{R}(t,y)}$~\cite{agoritsas_2012_FHHtri}.
{Note that although the study of~${\bar{C}(t,y)}$ is a natural choice as it characterizes the second moment of~$\bar{F}_V$, we have argued in~Ref.~\cite{agoritsas_2012_FHHtri} that the study of~${\bar{R}(t,y)}$ yields a clearer physical picture. In particular, its value at~${y=0}$ characterizes the fluctuations of the KPZ nonlinearity in the evolution equation~\eqref{eq-FeynmanKac-FbarV} as~${\bar{R}(t,0)=\overline{\argc{\partial_y \bar{F}_V(t,y)}^2}}$.}

For an \emph{uncorrelated disorder}, \emph{i.e.} for~${R_\xi(y)=\delta(y)}$ {in~\eqref{eq-def-moydis}} or equivalently ${\xi=0}$, it has been shown that the features shared in the KPZ universality class {consist not only of} the value of roughness scaling exponent ${\zeta_{\text{KPZ}}=2/3}$ but also {of} the asymptotic \textit{distribution} of the free-energy.
More precisely, the disorder free-energy $\bar{F}_V  (t,y)$ scales in distribution at large `times' according to:
\begin{align}
\hspace{-3mm}  \bar F_V(t,y)& \stackrel{d}{=}  \Big(\frac{\widetilde D_\infty^2}{c}t\Big)^{\frac 13} 
  \mathcal A_2\big(y/\sqrt{B_{\text{RM}}(t)}\big)
\;\;\ {\text{as}}\;\ {t \to \infty}
\label{eq-scalingFbarV}
\\
B_{\text{RM}}(t) & \sim \Big(\frac{\widetilde D_\infty}{c^2}\Big)^{\frac 23} \:t^{\frac 43}
\, , \quad \widetilde{D}_{\infty} = \frac{cD}{T}
\label{eq-scalingFbarV_BRM}
\end{align}
where
$\mathcal A_2(\bar y)$ is the so-called Airy$_2$ process~\cite{praehofer-spohn_2002_JStatPhys108_1071}, independently of the system DES parameters $\arga{c,D,T}$, and $B_{\text{RM}}(t)$ is the roughness in the asymptotic `random manifold' (RM) regime.
As announced in the introduction,
the parameter $\widetilde D_\infty$ is central in our analysis, being common to the free-energy scaling in distribution~\eqref{eq-scalingFbarV} and to the scaling of the large-`time' roughness amplitude~\eqref{eq-scalingFbarV_BRM}.
The scaling relations~\eqref{eq-scalingFbarV}-\eqref{eq-scalingFbarV_BRM} express the fact that {in addition to the universality of the exponents, the amplitudes and the free-energy distribution are also universal} in the $\xi=0$ KPZ class.

{At ${\xi=0}$ the mere relation ${\widetilde D_\infty=\frac{cD}{T}}$ holds exactly. However, the crucial point for the short-range correlated case ${\xi>0}$ is that it must be generalized to}
\begin{equation}
 \widetilde{D}_{\infty}(T,\xi) \equiv f(T,\xi) \cdot \frac{cD}{T}
 \label{eq-Dtilde-infty-finterp}
\end{equation}
where ${f(T,\xi)}$ is an interpolating parameter such that {the correct uncorrelated limit ${\widetilde D_\infty(T,0)=\frac{cD}T}$ is recovered, with} ${f(T,0) \equiv 1}$.
In~Ref.~\cite{agoritsas_2012_FHHtri} we have predicted a monotonous behavior of $\widetilde D_\infty$ as a function of temperature~$T$, described by ${f(T,\xi)}$ being the solution of the equation ${f^\gamma \propto \big[\frac{T}{T_c(\xi)}\big]^\gamma (1-f)}$.
{Note that the exponent~$\gamma$ takes different values~${\gamma>0}$} depending on the approximation scheme considered, {but always predicts a saturation of the amplitude at zero temperature: ${\widetilde D_\infty(0,\xi)=\frac{cD}{T_c}}$.}

In~Ref.~\cite{agoritsas-2012-FHHpenta}, we have shown numerically with other co-authors that the scaling relation~\eqref{eq-scalingFbarV} {can be extended to the correlated case ${\xi>0}$ with a modified Airy$_2$ process (still to be characterized) and by replacing the asymptotic ${B_{\text{RM}}(t)}$ in~\eqref{eq-scalingFbarV_BRM} by the actual roughness~${B(t)}$.
The characterization of this modified Airy$_2$ process will however} require the understanding of a scaling relation valid on the \emph{full transverse range}~$y$
{, and this generalized scaling relation has actually been tested by the numerical collapse of the full correlator~${\bar{C}(t,y)}$ \eqref{eq-def-corr-FbarFbar2} at different `times' and temperatures.}
In addition, in the companion paper~\cite{agoritsas_2012_FHHtri}, we have centered our analytical study {on the \textit{local} scaling properties of the correlator~${\bar{R}(t,y)}$ \eqref{eq-def-corr-etaeta}, which actually characterizes the fluctuations of the KPZ non-linearity in~\eqref{eq-FeynmanKac-FbarV}.}
We have showed that this \emph{a priori} perplexing observable captures in fact the entire crossover from the low- to the high-temperature regime, through its \textit{small}-$y$ behavior.
In a DP toymodel approach, we have thus advocated that, at least for ${\valabs{y} \lesssim \xi}$ and sufficiently large `times', the correlator ${\bar{R} (t,y)}$ takes the form
\begin{equation}
 \bar{R}(t,y)
 \approx \widetilde{D}_t \cdot \mathcal{R}_{\tilde{\xi}_t}(y)
 \, , \;
 \int_{\mathbb{R}} dy \cdot \mathcal{R}_{\tilde{\xi}_t}(y) \equiv 1
 \label{eq-toymodel-def-Rbar-functional}
\end{equation}
where the normalized function ${\mathcal R_{\tilde{\xi}_t} (y)}$ is assumed to take a stabilized form closely related to the disorder correlator ${R_\xi(y)}$ defined by~\eqref{eq-def-moydis}, in the sense that these correlator share the same scaling in ${\tilde{\xi}_t \propto \xi}$.
All the possible `time'-dependence is then hidden in two effective parameters $\widetilde{D}_t$ and $\tilde{\xi}_t$, and at asymptotically large-`time' a steady-state is reached with
${\bar{R}(\infty,y) \equiv \widetilde{D}_{\infty} \cdot \mathcal{R}_{\tilde{\xi}_{\infty}}(y) }$
and in particular ${\bar{R}(\infty,0)\sim \widetilde{D}_{\infty}/\tilde{\xi}_{\infty}}$.
{This behavior suggests the definition of a \emph{saturation `time'}~$t_{\text{sat}}$ above which
\begin{equation}
 \bar{R}(t,\valabs{y} \lesssim \xi)
 \approx \widetilde{D}_{\infty} \cdot \mathcal{R}_{\tilde{\xi}_{\infty}}(y) 
 \equiv \bar{R}_{\text{sat}}(y)
  \label{eq-toymodel-def-Rbar-functional-saturation}
\end{equation} 
The crossover of~${\widetilde{D}_{\infty}(T,\xi)}$ announced in~\eqref{eq-Dtilde-infty-finterp} stems from the subtle interplay of the KPZ nonlinearity feedback with the disorder~${V(t,y)}$ for short `times'~${t \leq t_{\text{sat}}}$.}

\begin{figure}
 \includegraphics[width=\columnwidth]{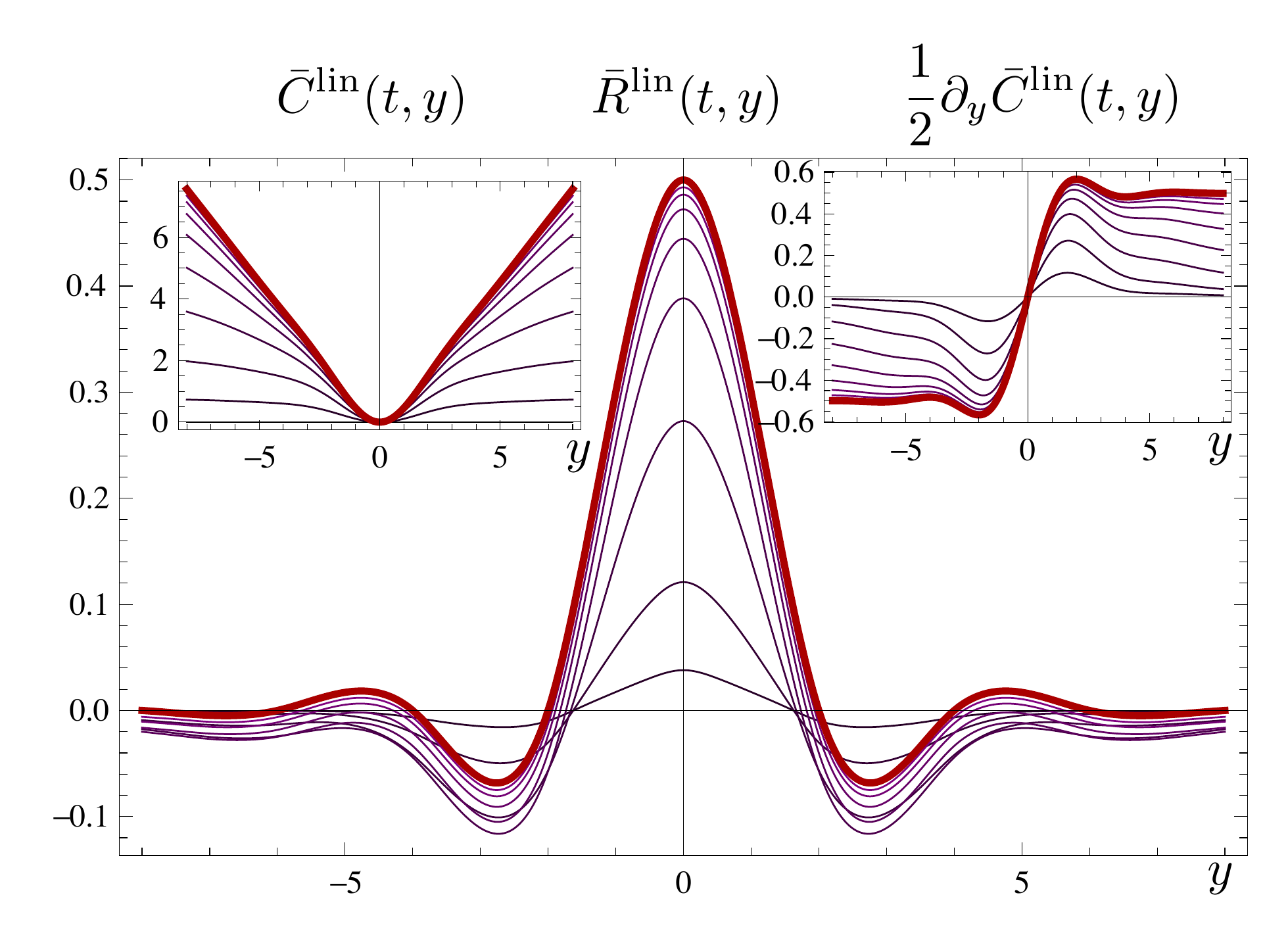}
 \caption{
 (Color online) The finite-`time' correlator $\bar{R}^\text{lin}(t,y)$ (thin purple lines) for the `CubicS' disorder correlator ${R_\xi(y)=R_{\xi}^{\text{CubicS}} (y)}$ defined in \eqref{eq-def-RCubicS-1}-\eqref{eq-def-RCubicS-2}, plotted as a function of $y$ for different `times' and compared to its infinite-`time' limit $R_\xi(y)$ (thick red line). The central peak develops with increasing `times' from the flat initial condition ${\bar{R}(0,y)\equiv 0}$.
 Parameters are ${\xi=2}$, ${c=1}$, ${D=1}$, and ${T=1}$.
 \textit{Left inset}:~same behavior for ${\bar{C}^{\text{lin}}(t,y)}$.
 \textit{Right inset}:~same behavior for ${\frac12 \bar{C}^{\text{lin}}(t,y)=\int_0^y dy' \, \bar{R}^{\text{lin}}(t,y')}$.
 }
 \label{fig:finitetimescaling-Cbarlin-Rbarlin_cubicS}
\end{figure}

The Ansatz~\eqref{eq-toymodel-def-Rbar-functional} has been inspired by the infinite-`time' limit of the free-energy fluctuations, respectively for the uncorrelated case~\cite{huse_henley_fisher_1985_PhysRevLett55_2924} and for the linearized version of the tilted KPZ equation~\eqref{eq-FeynmanKac-FbarV} {presented in}~\cite{agoritsas_2012_FHHtri} (whose solution is denoted thereafter by a subscript~`lin').
{In both cases the fluctuations are} Gaussian and thus fully characterized by their two-point correlators:
\begin{eqnarray}
 \bar R (\infty,y)
 & \stackrel{(\xi=0)}{=} & \frac {cD}T R_{\xi=0}(y) 
 \, , \;
 R_{\xi=0}(y)=\delta(y)
 \label{eq-Rbarinfzeroxi} \\
 \bar{R}^{\text{lin}}(\infty,y)
 &=& \widetilde{D}_{\infty} \cdot R_{\xi}(y)
 \, , \;
 \widetilde{D}_{\infty}  = \frac{cD}{T}
 \label{eq-infty-time-Rxi-lin}
\end{eqnarray}
As discussed in~\cite{agoritsas_2012_FHHtri}, the linearized case --~which is \textit{not} equivalent to a standard EW evolution for ${\partial_t {F}_V (t,y)}$~\cite{edwards_wilkinson_1982_ProcRSocLondA381_17}~-- has been solved exactly at all `times', predicting Gaussian fluctuations and yielding the following decomposition:
\begin{equation}
\bar{R}^{\text{lin}}(t,y)
 =\frac{cD}{T} \argc{ R_{\xi}(y) - b^{\text{lin}}(t,y)}
 \label{eq-decomposition-Rbarlin-1}
\end{equation}
with~${\lim_{t \to \infty} b^{\text{lin}}(t,y) = 0}$.
This behavior is depicted in Fig.~\ref{fig:finitetimescaling-Cbarlin-Rbarlin_cubicS}, for the specific disorder correlator that will be used in our numerical simulations ${R_{\xi}^{\text{CubicS}} (y)}$, defined in \eqref{eq-def-RCubicS-1}-\eqref{eq-def-RCubicS-2}.
Note that the disorder free-energy fluctuations are non-Gaussian in the full non-linearized case, except in the very specific limit of infinite `time' and uncorrelated disorder ${\xi=0}$, requiring thus the complete set of $n$-point correlator to characterize them. 
Nevertheless, the solution \eqref{eq-decomposition-Rbarlin-1} suggests a similar decomposition for the exact two-point correlator ${\bar{R}(t,y)}$, separating its asymptotic amplitude ${\widetilde{D}_{\infty}}$ and shape ${\mathcal{R}_{\xi}(y)}$ from a finite-`time' contribution ${b(t,y)}$, under the exact constraint that ${\int_{\mathbb{R}} dy \, \bar{R}(t,y) =0}$ at~${t<\infty}$~\cite{agoritsas_2012_FHHtri}~:
\begin{eqnarray}
 &\bar{R}(t,y)
 = \widetilde{D}_{\infty} \cdot \argc{ \mathcal{R}_{\xi}(y) - b(t,y)}
 & \label{eq-infty-time-Rxi-bumps-1} \\
 &  \int_{\mathbb{R}} dy \, \mathcal{R}_{\xi}(y) \equiv 1
 \, \Rightarrow \,
 \int_{\mathbb{R}} dy \, b(t,y) = 1 \, \forall t
 & \label{eq-infty-time-Rxi-bumps-2} \\
 & \lim_{t \to \infty} b(t,y) =0 &
 \label{eq-infty-time-Rxi-bumps-3}
\end{eqnarray}
The scaling of~${b(t,y)}$ corresponds to the `wings' of the alternative correlator $\bar{C}(t,y)$~\eqref{eq-def-corr-FbarFbar2}, which rescale with the roughness ${B(t)}$ as studied extensively in~Ref.~\cite{agoritsas-2012-FHHpenta}.

It is striking that the ${\xi=0}$ result~\eqref{eq-Rbarinfzeroxi} holds in the same form in the linearized and non-linearized evolutions of the disorder free-energy, with the imprint of the microscopic disorder correlator at large `times' being such that~${\mathcal{R}_{\tilde{\xi}_{\infty}}(y)=R_{\xi=0}(y)}$. This implies that any modification of both the amplitude~$\widetilde{D}_{\infty}$ and the shape~${\mathcal{R}(y)}$, along with the non-Gaussianity of the free-energy fluctuations, must stem from the KPZ nonlinearity in~\eqref{eq-FeynmanKac-FbarV}.

From an analytical point of view, the validity of the Ansatz~\eqref{eq-toymodel-def-Rbar-functional} has still to be asserted.
{The exact contribution ${b(t,y)}$ and the asymptotic temperature-dependent shape~${\mathcal{R}(y)}$ remain unknown for the time being, and hence the present numerical study is a first step in the characterization of the modified Airy$_2$ process at~${\xi>0}$ in~\eqref{eq-scalingFbarV}.}
In this numerical paper we examine specifically the validity of the DP toymodel centered on~\eqref{eq-toymodel-def-Rbar-functional}, by measuring the `time'-evolution of the correlator~${\bar{R}(t,y)}$ and characterizing its small-$y$ features with respect to the microscopic disorder correlator {${R_{\xi}(y)}$} (see~Sec.~\ref{section-timeevol-Rty-fits} and~\ref{section-temp-dep-asympt-Rbar-sat}).
{We discuss moreover the existence of the saturation `time' $t_{\text{sat}}$ and the consequent characterization of ${\bar{R}_{\text{sat}}(y)}$ with $\arga{ \widetilde{D}_{\infty}, \tilde{\xi}_{\infty}, \mathcal{R} }$ following the definition~\eqref{eq-toymodel-def-Rbar-functional-saturation}.}

\subsection{Geometrical fluctuations and roughness}
\label{subsec:model_geometrical-fluctuations}

The geometrical fluctuations of the polymer endpoint at `time'~$t$, or equivalently of the interface at a given lengthscale~$t$, are of high relevance because they are directly accessible, both numerically and experimentally.
{Their PDF~${\mathcal{P}(t,y)}$ \eqref{eq-def-PDF-yt} can be characterized mainly by its variance, namely} the roughness function ${B(t)}$ and by an effective roughness exponent ${\zeta(t)}$, respectively defined as
\begin{align}
 B(t)
 & \ \equiv\  \overline{\moy{y(t)^2}}
 \ =\  \int_\mathbb R dy\, y^2\, \mathcal{P}(t,y)
 \label{eq-def-roughness-zeta}
\\
\
 \zeta (t)
 & \ \equiv \ \frac12 \frac{\partial \log B(t)}{\partial \log t}
 \label{eq-def-roughness-zeta-zeta}
\end{align}
where the brackets $\moy{\mathcal{O}}$ denotes the statistical average over thermal fluctuations for an observable~$\mathcal{O}$.
It is known that the roughness is characterized at small~$t$ by the EW exponent $\zeta_{\text{EW}}=1/2$ and at asymptotically large~$t$ by the KPZ exponent $\zeta_{\text{RM}}=\zeta_{\text{KPZ}}=2/3$ {in the so-called random-manifold (RM) regime}.
{Beyond these exponents, the amplitudes of the corresponding powerlaws} are in fact of interest and their scaling{s} write
{\begin{equation}
  B(t) =
  \begin{cases}
   \: B_{\text{th}}(t)= \frac{T}{c} t^{2 \zeta_{\text{EW}}} &\text{for}\ t\to 0
   \\
    B_{\text{RM}}(t) \sim \big(\frac{\widetilde D_\infty}{c^2\vphantom{\text{Î}}}\big)^{\frac 23} \:t^{2\zeta_{\text{KPZ}}} &\text{for}\ t\to \infty
  \end{cases}
  \label{eq-roughnessregimes_smalllarget}
\end{equation}
as already defined respectively in~\eqref{eq-FB-thermal} and in~\eqref{eq-scalingFbarV_BRM}.}
The large-`time' result is known to hold exactly with ${\widetilde D_\infty=\frac{cD}T}$ for an uncorrelated disorder (${\xi=0}$).
In a GVM approach based on our DP toymodel~\cite{agoritsas_2010_PhysRevB_82_184207,agoritsas_2012_ECRYS2011,agoritsas-2012-FHHpenta} we have moreover showed that it should also hold for a correlated disorder (${\xi>0}$) for all temperatures, {with ${\widetilde{D}_\infty(T,\xi)}$ characterizing the amplitude of the free-energy fluctuations, as announced in~\eqref{eq-scalingFbarV}-\eqref{eq-scalingFbarV_BRM}-\eqref{eq-Dtilde-infty-finterp}}.

{An important crossover lengthscale is the `Larkin length'~$L_c$~\cite{Larkin_model_1970-SovPhysJETP31_784} which marks the beginning of the RM regime and is closely related to~${\widetilde{D}_{\infty}(T,\xi)}$~\cite{agoritsas_2012_FHHtri}. Note that by consistency the Larkin length should be larger than the saturation `time' for the free-energy fluctuations (defined by~\eqref{eq-toymodel-def-Rbar-functional-saturation}) so we expect~${t_{\text{sat}}\leq L_c}$.}

One aim of this paper is to provide a numerical check of these GVM predictions (see~Sec.~\ref{section-numerical-roughness}), especially in order to probe intermediate regimes and lengthscales between the two opposite asymptotic regimes of the roughness~\eqref{eq-roughnessregimes_smalllarget}.
{This numerical study of the roughness at~${\xi>0}$ provides moreover an indirect test of the validity of our DP toymodel, on which these GVM predictions are based.}


\section{Numerical recipe}
\label{section-numerical-recipe}

In this section we discuss the possible numerical approaches to these problems and we expose our numerical procedure.
All the numerical parameters are gathered in~Appendix~\ref{A-numericalrecipe}.

In the next three sections, we present the different numerical results obtained, which are respectively:
in~Sec.~\ref{section-timeevol-Rty-fits}
the `time'-evolution of the correlator {at fixed temperature ${\bar{R}(t,y)}$~\eqref{eq-def-corr-etaeta} and its parameters $\lbrace \tilde{\xi}_t,\widetilde{D}_t \rbrace$~\eqref{eq-toymodel-def-Rbar-functional}};
in~Sec.~\ref{section-temp-dep-asympt-Rbar-sat}
the temperature-dependence of the asymptotic effective correlator ${\bar{R}_{\text{sat}}(y)}$ {and $\arga{ \widetilde{D}_{\infty}, \tilde{\xi}_{\infty}, \mathcal{R} }$ according to~\eqref{eq-toymodel-def-Rbar-functional-saturation}};
and in~Sec.~\ref{section-numerical-roughness}
the temperature-dependence of the roughness $B(t)$~{\eqref{eq-def-roughness-zeta}} and its logarithmic slope~${\zeta(t)}$~{\eqref{eq-def-roughness-zeta-zeta}.
In addition,
Appendix~\ref{A-time-evolution-Rbar-et-al} gathers the details of a quantitative test of the DP toymodel by comparing three fitting functions for~${\mathcal{R}(y)}$
and
Appendix~\ref{A-section-fluctuation-MeanMeanFbar} presents a self-consistency check of our numerical procedure by analyzing the `time'- and temperature-dependence of the mean value~${\overline{\bar{F}_V(t,y)}=-\frac{1}{2c} \int_0^t dt'\, \bar{R}(t',0)}$.}

\subsection{Possible numerical approaches}

The Feynman-Kac evolution equation \eqref{eq-FeynmanKac-FbarV} for the disorder free-energy ${\partial_t \bar{F}_V(t,y)}$ provides the starting point for a numerical study of the geometrical and free-energy fluctuations of the 1+1 DP, and consequently of the static 1D interface, directly in their continuum formulation.
This approach uses an exact property of the model -- the STS \eqref{eq-STS-PFbarV} -- to focus on the effects of the disorder, dissociated from the pure thermal ones.

Among the numerical procedures previously used to tackle this problem, we can mention
firstly the DP under the solid-on-solid (SOS) constraint~\cite{kardar_1985_PhysRevLett55_2923,bustingorry_2010_PhysRevB82_140201,agoritsas-2012-FHHpenta} where the polymer lives on a discretized lattice;
secondly the semi-continuous 1D interface, discretized along its internal dimension but with each point living in a \textit{continuous} 1D splined random potential~\cite{kolton_2005_PhysRevLett94_047002};
and thirdly the continuous DP that we present thereafter.
Those approaches are of course complementary, especially for the investigation of the large- \textit{versus} small-lengthscales and high- \textit{versus} low-temperature properties, if a suitable translation from the specific numerical parameters to the physical ones $\arga{c,D,T,\xi}$ is provided~\cite{agoritsas-2012-FHHpenta,alberts-quastel_2012_arXiv:1202.4403v1}.
The SOS model numerical approach consists in solving the discrete equivalent of the stochastic heat equation~\eqref{eq-FeynmanKac-Wvnorm} onto a partition function defined on a {lattice. This} approach however proves difficult to control numerically in the low-temperature regime, since the DP endpoint is exponentially concentrated in favorable regions of the potential.
An alternative approach thus consists in focusing on a discrete analog of the KPZ equation~\eqref{eq-FeynmanKac-FV}.
In fact, the discretization of the non-linear term proves to be a highly non-trivial problem even without disorder, as Kruskal and Zabusky~\cite{zabusky_kruskal_1965_PhysRevLett15_240} realized a long time ago in the related problem of simulating soliton solutions of the Korteweg-de Vries equation. A well-behaved lattice discretization {which leads to the correct continuum limit is exposed in~\cite{krug-spohn_1991_Godreche_BegRohu}}; we refer the reader to~\cite{quastel_lecture-notes-arizona_2012} for a detailed discussion.

Our procedure is actually based on the continuous analogue of the transfer-matrix method of a DP on a lattice with the SOS constraint, in the sense that it is performed after integration over thermal fluctuations since it follows the evolution of the partition function ${Z_V(t,y)}$ with the lengthscale (namely the Feynman-Kac equation \eqref{eq-FeynmanKac-FbarV}), for many individual disorder configurations.
The continuous-limit formulation has two advantages:
first the discretization issue in numerics is pushed back to a problem of numerical integration of partial differential equations, so even `small' lengthscales can be studied without discretization artifacts;
secondly in the continuous limit the numerical parameters are directly the physical ones of the analytical model.
We refer the reader to Ref.~\cite{bloemker_numerics-Burgers_preprint} for mathematical results on the convergence of discretization schemes of {the} KPZ equation to the solution of the continuous equation, in the situation where the disorder is spatially correlated.


\subsection{Detailed procedure}
\label{section-numerical-recipe-detailedprocedure}

Computing ${\bar{F}_V(t,y)}$ for individual disorder configurations ${V(t,y)}$ up to a maximal `time' $t_m$, we can measure directly the geometrical and free-energy fluctuations stemming at thermodynamic equilibrium from thermal fluctuations exploring a given random potential.
Then averaging these fluctuations over many disorder configurations we have access to the quantities of interest defined in Sec.~\ref{section-DPformulation-numerics}.

\paragraph{Finite box and microscopic discretized grid.}

The Feynman-Kac evolution equation \eqref{eq-FeynmanKac-FbarV} is defined in a continuous limit with $t>0$ and  ${y \in \mathbb{R}}$.
In numerics we work necessarily on a microscopic discretized grid in both $(t,y)$ variables and in a finite box ${(t,y) \in \argc{0,t_m}\times\argc{-y_m,y_m}}$ without periodic boundary conditions.
However, in experimental realizations of 1D interfaces we also have some microscopic cut-off in lengthscales, ultimately the crystal parameter, and a macroscopic cut-off due to scarcer statistics at larger lengthscales.

\begin{figure}
 \includegraphics[width=0.9\columnwidth]{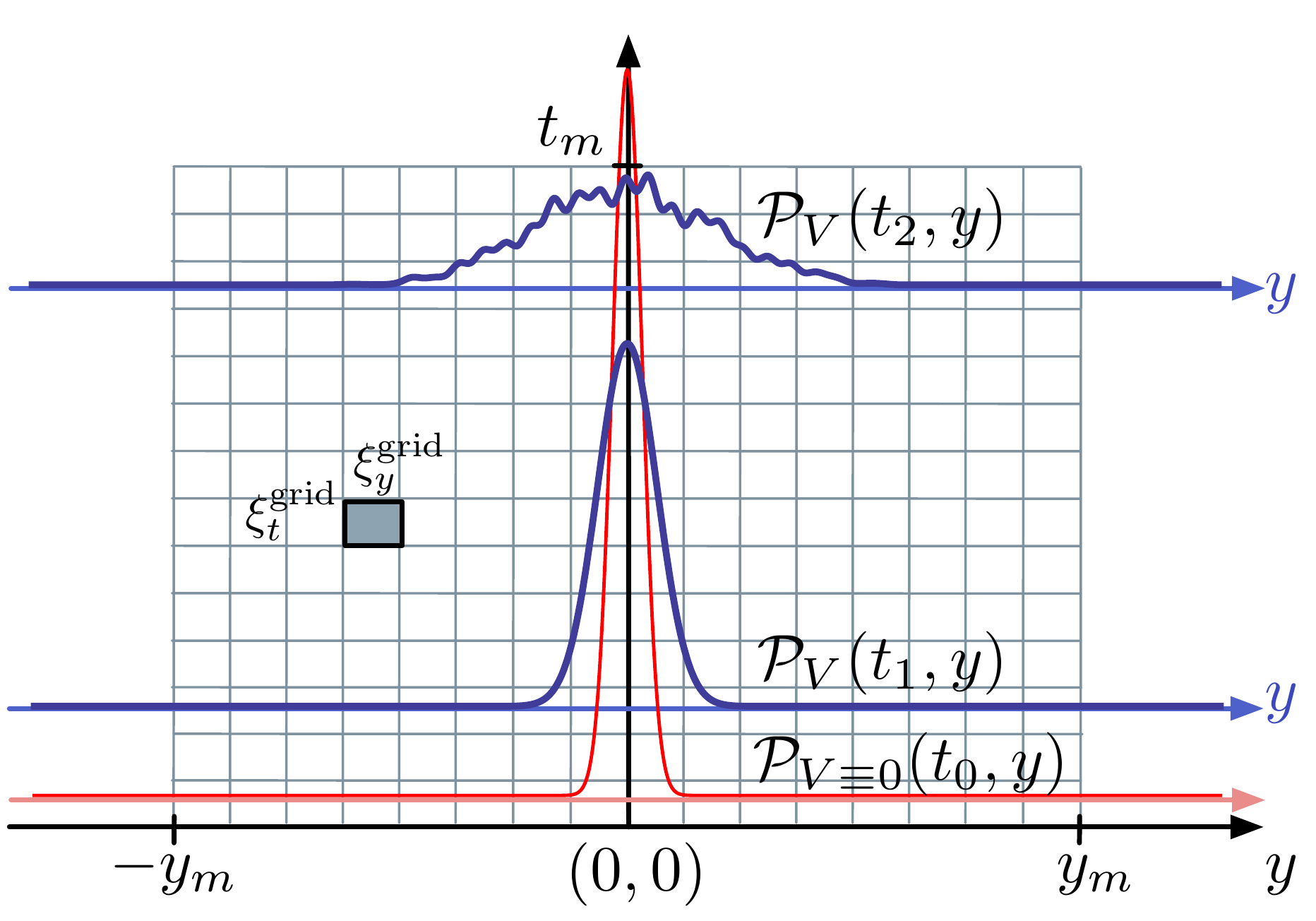}
 \caption{
 (Color online)
 Sketch of the numerical procedure for the 1+1 continuous DP in the finite box $\argc{0,t_m}\times \argc{-y_m,y_m}$ with the initial condition at ${t_0=0.1}$ taken purely thermal ${\mathcal{P}(t_0,y)=Z_{V \equiv 0}(t_0,y)}$~\eqref{eq-thermal-ZV} (red curve).
 There are three levels of discretization grids by decreasing scale:
	the grid for the random potential ${\xi^{\text{grid}}_t=1}$ and ${\xi^{\text{grid}}_y=2}$ (the gray box on the left side);
	the linear grids for the recording of data ${\Delta t^{\text{lin}}}=t_m/80$ and ${\Delta y^{\text{lin}}=2 y_m/100}$,
		and the logarithmic grid ${t^{\text{log}}_j= t_0 \cdot (t_m/t_0)^{j/80} }$ with ${j = 0 \dots 80}$;
	the microscopic grid for the numerical integration adapted in both ${(t,y)}$-directions by \texttt{Mathematica}.
 At a given configuration of disorder $V(t,y)$ as illustrated in~Fig.~\ref{fig:RCubicS-graph-compil}, the PDF ${\mathcal{P}_V(t,y)}$ is computed at increasing `time' following \eqref{eq-FeynmanKac-FbarV}, with the boundary condition {${\partial_y \bar{F}_V(t,\pm y_m)=0}$} and the \textit{ad hoc} normalization \eqref{eq-numerical-normalization}, as illustrated at ${t_0<t_1<t_2}$ (blue curves).
See Fig.~\ref{fig:DP-toymodel-schema} for the interpretation in terms of DP trajectories.
 }
 \label{fig:graph-numrecipe-total} 
\end{figure}

\paragraph{Generation of an individual disorder configuration.}

In order to generate individual disorder configurations for a given set of DES parameters,
we first define a two-dimensional grid of spacing ${\arga{\xi^{\text{grid}}_t,\xi^{\text{grid}}_y} = \arga{t_m/L_t , y_m/L_y}}$ (cf. Fig.~\ref{fig:graph-numrecipe-total}).
On each point of the grid we pick a set of random numbers according to a normal distribution of variance ${D^{\text{grid}}}$ (the `strength of disorder') and the smooth random potential ${V(t,y)}$ is obtained by interpolating between the grid points with a 2D cubic spline as illustrated in Fig.~\ref{fig:RCubicS-graph-compil} (\textit{top}).
As detailed in Appendix~\ref{A-effectiveCorrelatorSpline}, its corresponding two-point correlator at fixed `time' is given by the translation-invariant:
\begin{eqnarray}
 & \overline{V(t,y)V(t,0)}
 = D^{\text{eff}} \cdot R_{\xi_y}^{\text{CubicS}} (y)
 &\\
 & D^{\text{eff}}
 = D \cdot R_{\xi_t}^{\text{CubicS}} (0)
 = D^{\text{grid}} \xi_y^{\text{grid}}
 &
 \end{eqnarray}
with the function ${R_{\xi}^{\text{CubicS}}}$ made explicit in \eqref{eq-def-RCubicS-1}-\eqref{eq-def-RCubicS-2}.
The values of those parameters are fixed for all our numerical computations to
\begin{equation}
\xi^{\text{grid}}_t=1
\; , \; \xi^{\text{grid}}_y=2
\; , \; D^{\text{grid}}=4
\; \Rightarrow \; D=D^{\text{eff}}=8
\label{eq-fixed-parameters-microscopic disorder}
\end{equation}
Note that the physical parameter used in analytical arguments is $D$ \eqref{eq-def-moydis} and not $D^{\text{eff}}$, but they coincide with this particular choice ${\xi_t^{\text{grid}}}$.

\paragraph{Numerical integration at fixed disorder.}

We have chosen to follow the evolution of the disorder free-energy ${\bar{F}_V(t,y)}$ \eqref{eq-FeynmanKac-FbarV} because among its counterparts
{\textit{(i)}~${\partial_y F_V(t,y)}$} is too noisy,
{\textit{(ii)}}~${F_V(t,y)}$ includes ${F_{\text{th}}(t,y)=\frac{cy^2}{2t}}$~\eqref{eq-FB-thermal} that hides the disorder-induced fluctuations of $\bar{F}_V(t,y)$ except at large `times',
and {\textit{(iii)}} the exponential in ${W_V(t,y) \propto e^{-F_V(t,y)/T}}$ reduces the numerical resolution.
Moreover, by subtracting the exact ${F_{\text{th}}(t,y)}$ for ${y \in \mathbb{R}}$ directly in the evolution equation, we get rid of the finite-box artifacts that would have arisen  {already in absence of disorder due to} the pure elastic contribution.
The numerical integration of the differential equation \eqref{eq-FeynmanKac-FbarV} was performed using a numerical algorithm included in \texttt{Mathematica}
\footnote{\texttt{Mathematica} tutorial for the numerical integration of differential equations: \href{http://reference.wolfram.com/mathematica/tutorial/NDSolvePDE.html}{http://reference.wolfram.com/mathematica/tutorial /NDSolvePDE.html}.}, which adapts the numerical discretization in $y$ at each `time'-step in order to minimize the numerical errors.
As emphasized in~\cite{gallego_2007_PhysRevE76_051121}, pseudo-spectral spatial derivatives are more stable than finite differences for simulating the KPZ equation. We enforced this choice through the \texttt{"DifferenceOrder"->"Pseudospectral"} option to the \texttt{NDSolve} integrator.
The main limitation was that the larger the maximum `time' $t_m$, the longer the computation time.
Moreover, the lower the temperature the more the numerical solution of the Feynman-Kac {equation~\eqref{eq-FeynmanKac-FbarV}} is sensitive to the spatial variations of the random potential, thus dictating a smaller grid discretization in order to minimize the numerical error and increasing considerably the computation time.
The number of disorder configurations that have been considered for a given set of DES parameters is thus a compromise between the convergence of the disorder average and a reasonable computation time (as summarized in Appendix~\ref{A-numericalrecipe}).

\paragraph{Initial condition.}

If, as discussed in Sec.~\ref{subsec:model_observables}, the disorder free-energy ${\bar F_V(0,y)}$ is uniformly 0 at initial `time' ${t=0}$, the equation evolution~\eqref{eq-FeynmanKac-FbarV} for ${\bar F_V(t,y)}$ is however singular at ${t=0}$. We thus slightly shift the initial condition to `time' ${t_0=0.1}$, assuming that this procedure is stable as $t_0\to 0$. This {choice} corresponds to taking ${\bar{F}_V(t_0,y) \equiv 0}$, or, in terms of the partition function, ${Z_{V }(t_0,y)\approx Z_{V \equiv 0}(t_0,y)}$ as defined by~\eqref{eq-thermal-ZV}.
For the uncorrelated case ${\xi=0}$, this asymptotics ${Z_{V }(t_0,y)\stackrel{t_0\to 0}\sim Z_{V \equiv 0}(t_0,y)}$ has indeed been shown to be correct~\cite{gueudre_2012_PhysRevE86_041151}, and we assume that this results also holds in the less singular case ${\xi>0}$.
We assume that at large `times' the DP has completely forgotten this initial condition, {and we have checked numerically that this is the case for the particular choice for~${t_0}$. However, at short `times' above $t_0$, the DP}  behavior should carry some artifact for the disorder-induced quantities $\bar{F}_V$ and {${\partial_y \bar{F}_V}$}.

\paragraph{Boundary conditions.}

We impose at each `time' that {${\partial_y \bar{F}_V(t,\pm y_m)=0}$}.
This is equivalent to the statement that we have ${W_{V}(t, y) \approx W_{V \equiv 0}(t,y)}$ {for ${\valabs{y} \geq y_m}$} and thus the normalized DP endpoint probability {can be approximated as} ${\mathcal{P}_V(t,y) \approx Z_{V \equiv 0}(t,y) \approx 0}$~\eqref{eq-thermal-ZV}.
This boundary condition can be physically correct only for `times' such that ${\sqrt{B(t)} < y_m}$, else the DP `senses' inevitably the boundaries of the finite box.
This choice of boundary condition implies for the normalization ${ \Wbar_V (t)}$ \eqref{eq-WV-normalization} that the contribution for {displacement} $y$ outside {the box} $\argc{-y_m,y_m}$ is exactly known analytically:
\begin{equation}
\begin{split}
  & \frac12 \Wbar_V^{\text{num}} (t) 
   = \int_0^{y_m} dy \cdot e^{- \argp{F_{\text{th}}(t,y)+\bar{F}_V(t,y)}/T} + \text{cte}(t,y_m)
  \\
 &\text{cte}(t,y_m)
  = \int_{y_m}^{\infty} dy \cdot e^{- F_{\text{th}}(t,y)/T}
 \; , \;
 F_{\text{th}}(t,y)=\frac{cy^2}{2t}
 \label{eq-numerical-normalization}
\end{split}
\end{equation}
{These definitions yield back the normalized distribution} $\mathcal{P}_V(t,y)$ as defined by \eqref{eq-def-PDF-yt} (cf. Fig.~\ref{fig:graph-numrecipe-total}).

\paragraph{Recorded data.}

We have actually recorded, on a microscopic grid linear in $y$, the following quantities defined throughout {Sec.~\eqref{section-DPformulation-numerics}}:
on one hand the disorder free-energy ${\bar{F}_V(t,y)}$ and {its derivative ${\partial_y \bar{F}_V (t,y)}$},
with their mean values $\overline{\bar{F}_V(t,0)}$ and {$\overline{\partial_y \bar{F}_V(t,0)}$}
and their respective two-point correlators ${\bar{C}(t,y)}$ and ${\bar{R}(t,y)}$ {\eqref{eq-def-corr-etaeta}-\eqref{eq-def-corr-FbarFbar2}};
and on the other hand
the PDFs ${\mathcal{P}_V (t,y)}$ and ${\mathcal{P} (t,y)}$,
their {first} corresponding moments ${\overline{\moy{y(t)^k}}}$ and in particular the roughness ${B(t)\equiv \overline{\moy{y(t)^2}}}$
with the roughness exponent $\zeta(t)$.
We have recorded those quantities using a microscopic grid \textit{linear} in $t$, and in parallel for the roughness-related quantities we have used a microscopic grid \textit{logarithmic} in $t$ in anticipation of powerlaws determination.
Note that the even parity of the correlators has been explicitly used to increase their statistics, so the measured ${\bar{C}(t,y)}$ and ${\bar{R}(t,y)}$ are symmetric by construction:
{\begin{eqnarray}
 & \bar{R}(t,y) = \frac12 \overline{\argc{\partial_y \bar{F} (t,y) \cdot \partial_y \bar{F} (t,0) + \partial_y \bar{F} (t,-y) \cdot \partial_y \bar{F}(t,0)}} &
 \nonumber \\
 & \bar{C}(t,y) = \frac12 \overline{\argc{\bar{F}(t,y)-\bar{F}(t,0)}^2+\argc{\bar{F}(t,-y)-\bar{F}(t,0)}^2} & \nonumber
\end{eqnarray}}

\paragraph{Fitting functions for ${\bar{R}(t,y)}$.}

The correlator ${\bar{R}(t,y)}$ is not known exactly at finite `time', and we have thus only postulated its generic form ${\bar{R}(t,y) \approx \widetilde{D}_t \cdot \mathcal{R}_{\tilde{\xi}_t}(y)}$ in
\eqref{eq-toymodel-def-Rbar-functional}.
Following the DP toymodel assumption \eqref{eq-toymodel-def-Rbar-functional}, we have systematically extracted the typical width $\tilde{\xi}_t$ and amplitude $\widetilde{D}_t$ for three different function $\mathcal{R}_{\tilde{\xi}}(y)$ with the chosen normalization ${\int_{\mathbb{R}} dy \, \mathcal{R}_{\tilde{\xi}}(y) \equiv 1}$.
Firstly $\mathcal{R}_{\tilde{\xi}}(y)$ is taken to be a Gaussian function, whose single feature is given by its variance $2 \tilde{\xi}^2$:
\begin{equation}
 \mathcal{R}^{\text{Gauss}}_{\tilde{\xi}} (y)
 = \frac{e^{-y^2/(4\tilde{\xi}^2)}}{\sqrt{4 \pi} \tilde{\xi}}
 \label{eq-def-RGauss}
\end{equation}
Secondly we add phenomenologically two negative oscillations of the observed correlator, by using a cardinal sinus of period $\tilde{\xi}/\pi$ with a Gaussian envelope function:
\begin{equation}
 \mathcal{R}^{\text{SincG}}_{\tilde{\xi}} (y)
 = e^{-y^2/(4\tilde{\xi}^2)}\cdot \frac{\sin \argc{\pi y/\tilde{\xi}}}{\pi y \, {\operatorname{Erf}}(\pi)}
 \label{eq-def-RSincG}
\end{equation}
And thirdly we consider the same function as the exact two-point correlator of the microscopic random potential $\mathcal{R}^{\text{CubicS}}_{\tilde{\xi}} (y)=R^{\text{CubicS}}_{\tilde{\xi}} (y)$ defined in \eqref{eq-def-RCubicS-1}-\eqref{eq-def-RCubicS-2}.
${\lbrace \tilde{\xi}_t,\widetilde{D}_t \rbrace}$ are reliable quantities if they do not depend on the choice of the fitting  function, except for a numerical constant {depending solely on the choice between} $\mathcal{R}^{\text{Gauss}}$, $\mathcal{R}^{\text{SincG}}$ and $\mathcal{R}^{\text{CubicS}}$.
In Appendix~\ref{A-effectiveCorrelatorSpline}, we have compared the fit of the correlator $\overline{V(t,y)V(0,0)}$ with respect to $\mathcal{R}^{\text{Gauss}}$, $\mathcal{R}^{\text{SincG}}$ and $\mathcal{R}^{\text{CubicS}}$, and determined the numerical constants for passing from one to the others, as a consistency check of this procedure on this well-controlled correlator.


\section{`Time'-evolution of $\bar{R}(t,y)$ at fixed temperature}
\label{section-timeevol-Rty-fits}

In this section we study in detail the `time'-evolution of the correlator ${\bar{R}(t,y)}$~\eqref{eq-def-corr-etaeta} {at fixed temperature,
from the point of view of our DP toymodel~\eqref{eq-toymodel-def-Rbar-functional}.
We thus focus on the evolution of its shape around~${y=0}$, characterized by the effective parameters~$\arga{\widetilde{D}_t,\tilde{\xi}_t}$.
Note that its behavior at large transverse displacements~$y$ has already been investigated from a different perspective in~Ref.~\cite{agoritsas-2012-FHHpenta} on the equivalent correlator~${\bar{C}(t,y)}$~\eqref{eq-def-corr-FbarFbar2}.}

\begin{figure}
 \subfigure{\includegraphics[width=\columnwidth]{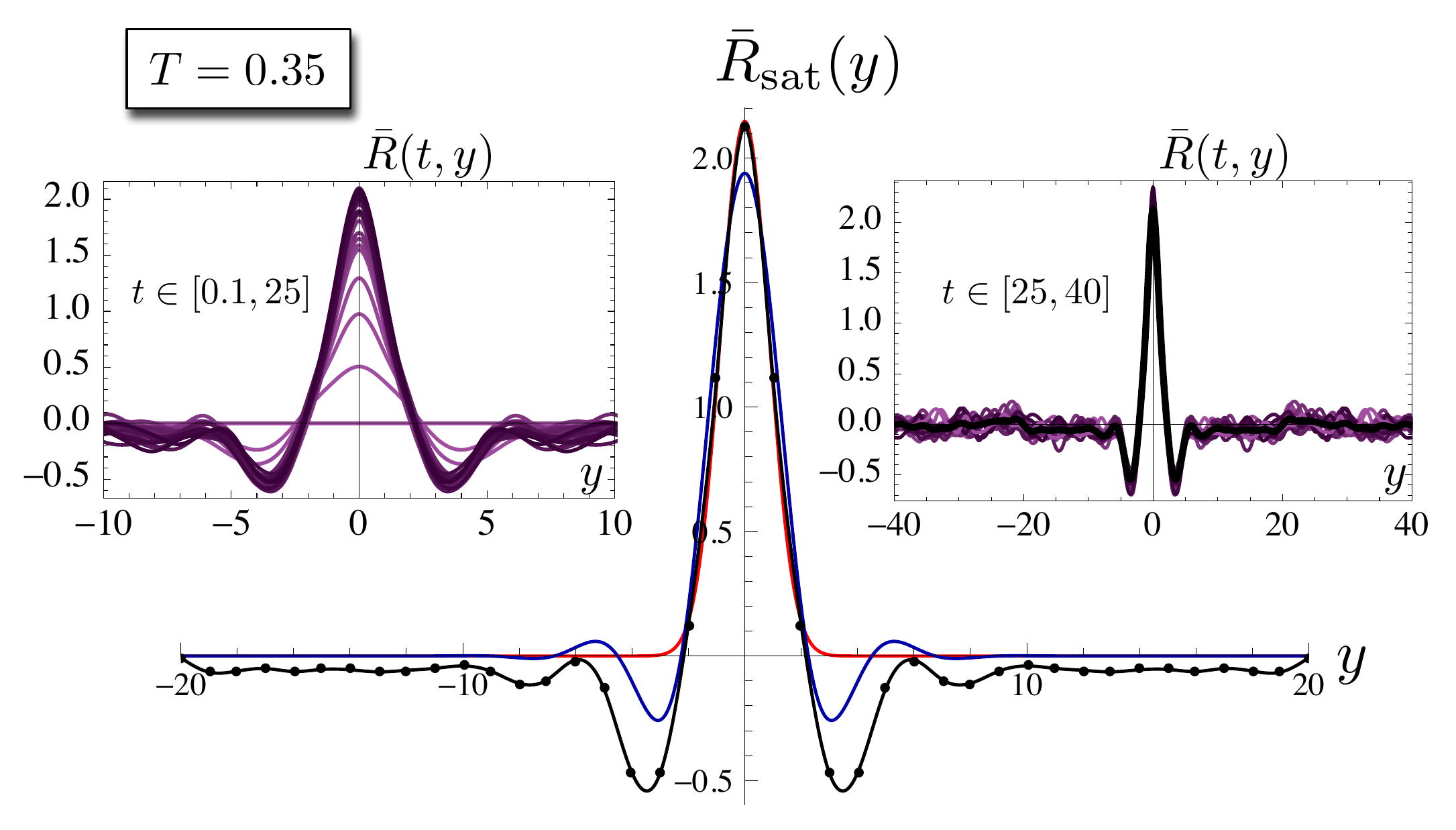}  \label{fig:lowT-T0-35-Rbar-compil}}
 \subfigure{\includegraphics[width=\columnwidth]{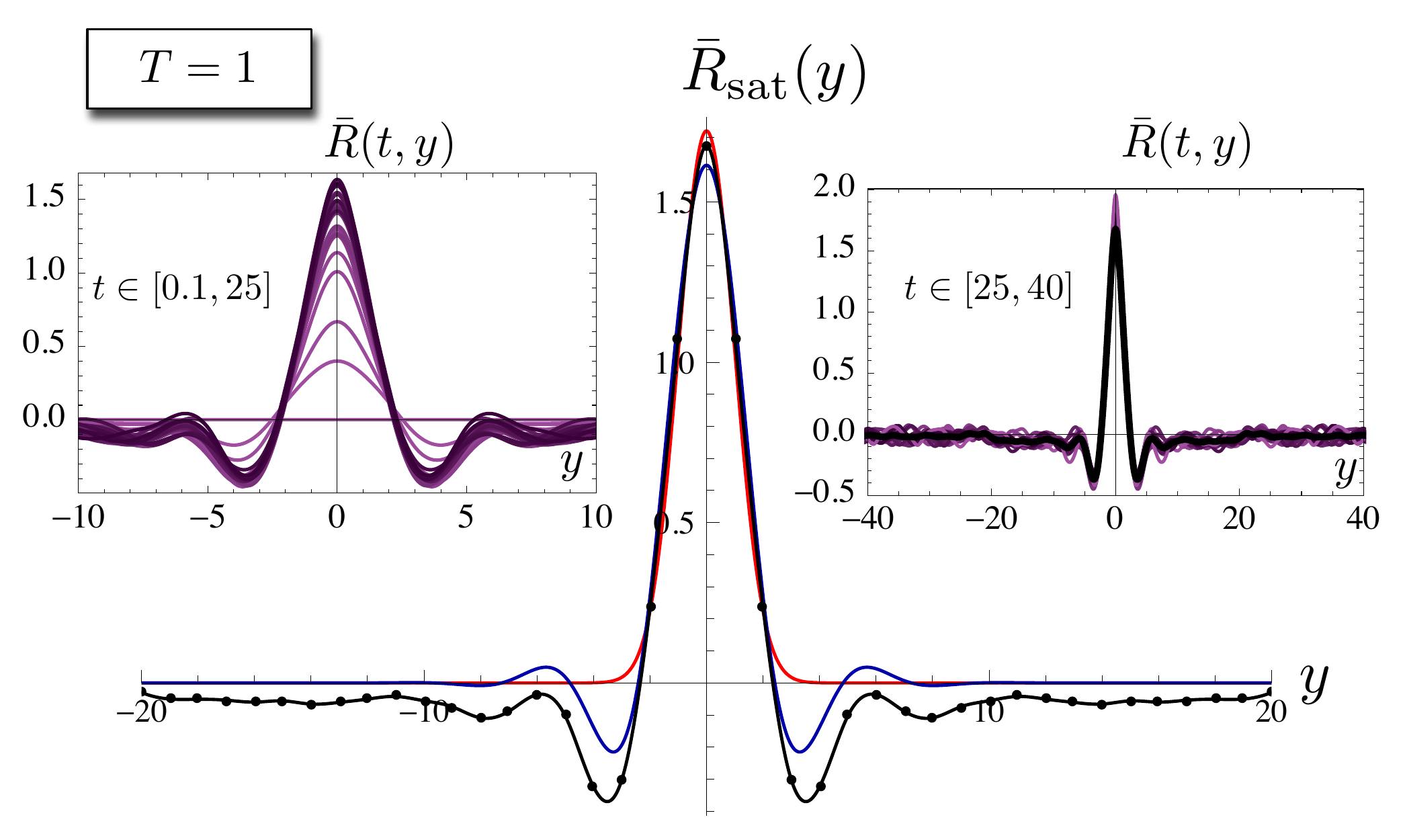}  \label{fig:aroundTc-T1-Rbar-compil}}
 \subfigure{\includegraphics[width=\columnwidth]{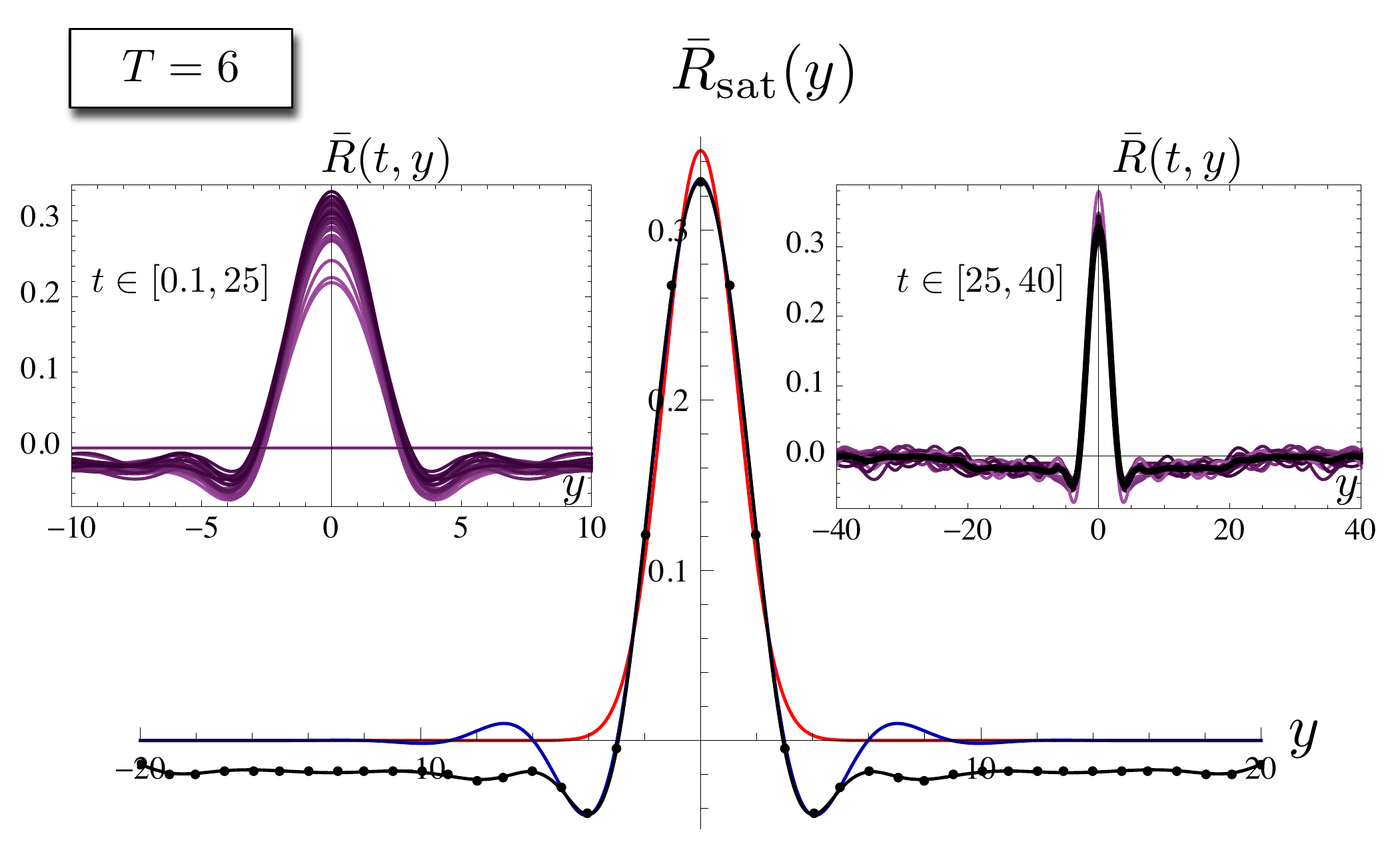}  \label{fig:highT-T6-Rbar-compil}}
 \caption{
 (Color online)
 Effective disorder correlator ${\bar{R}(t,y)}$ measured numerically at fixed temperature ${T\in \arga{0.35,1,6}}$.
 The different `times' ${t \in \argc{0.1,40}}$ are separated in two subsets $\argc{0,t_{\text{min}}}$ (\textit{left}) and $\argc{t_{\text{min}},40}$ (\textit{right}) with ${t_{\text{min}}=25}$, with a `time'-step ${\Delta t =1}$.
 \textit{Left side}~{(${y\in \argc{-10,10}}$)}:~Initial development of the central peak, saturation and accumulation of the curves in the vicinity of ${y=0}$. 
 \textit{Right side}~(${y \in \argc{-40,40}}$):~Large `times' correlators and their average over ${t>t_{\text{min}}}$ (superimposed black curve).
 \textit{Center}:~Saturation correlator ${\bar{R}_{\text{sat}}(y)}$ (black dots) with its Gaussian \eqref{eq-def-RGauss} (red) and `SincG' \eqref{eq-def-RSincG} (blue) fitting functions, `CubicS' collapsing exactly on `SincG' -- cf.~Tab.~\ref{tab:Dtilde-xitilde-sat-3T} for the explicit values of the fitting parameters.
 \label{fig:Rbar-compil-compil}
 } 
\end{figure}

In Fig.~\ref{fig:Rbar-compil-compil} {we have plotted the correlators at three characteristic temperatures ${T \in \arga{0.35,1,6}}$, illustrating respectively the low-, intermediate- and high-temperature regimes of the DP fluctuations -- cf.~Appendix~\ref{A-numericalrecipe} for the complete set of corresponding numerical parameters.}
We can follow the evolution of the correlator, starting by construction from the thermal condition ${\bar{R}(t_0,y) \equiv 0}$ at initial `time' ${t_0=0.1}$.
At small `times' the central peak first increases but quickly saturates, and all the curves start accumulating in the vicinity of ${y=0}$ {(Fig.~\ref{fig:Rbar-compil-compil}, \textit{left side}). This behavior suggests qualitatively the existence of a saturation `time'~$t_{\text{sat}}$}.
Assuming thus that at sufficiently large `times' the correlator has reached its presumably stationary form at small $y$, the saturation correlator ${\bar{R}_{\text{sat}}(y)}$ can be obtained by averaging the correlator over {`times' larger than an arbitrary threshold ${t_{\text{min}}>t_{\text{sat}}}$ (Fig.~\ref{fig:Rbar-compil-compil}, \textit{right side:} superimposed black curve)}.
{This correlator can then be fitted according to~\eqref{eq-toymodel-def-Rbar-functional-saturation} with respect to the three fitting functions $\mathcal{R}^{\text{Gauss}}$, $\mathcal{R}^{\text{SincG}}$ and $\mathcal{R}^{\text{CubicS}}$ defined at the end of~Sec.~\ref{section-numerical-recipe-detailedprocedure} (Fig.~\ref{fig:Rbar-compil-compil}, \textit{center})}.
Having checked that those results are stable for different values ${t_{\text{min}}>10}$, we have chosen arbitrarily ${t_{\text{min}}=25}$ for all temperatures, in order to be safely above the saturation `time' {${t_{\text{sat}} \lesssim 10}$}.

{
So we can distinguish two `time' regimes for the free-energy fluctuations, just by considering~${\bar{R}(t,y)}$:
\textit{(i)}~a short-`time' evolution at~${t < t_{\text{sat}}}$, \textit{a priori} marked by the artificial initial condition;
and \textit{(ii)}~a saturation regime at~${t \geq t_{\text{sat}}}$, when the correlator  around~${y=0}$ has reached a steady-state and is described by the stable function ${\bar{R}_{\text{sat}}(y) = \widetilde{D}_{\infty} \cdot \mathcal{R}_{\tilde{\xi}_{\infty}}(y)}$, as assumed in our DP toymodel~\eqref{eq-toymodel-def-Rbar-functional-saturation}.
The linearized case~Fig.~\ref{fig:finitetimescaling-Cbarlin-Rbarlin_cubicS} and the three temperatures in~Fig.~\ref{fig:Rbar-compil-compil} exhibit qualitatively the same two regimes. However, juxtaposing the high-, intermediate- and low-temperature cases allows to point out the slight temperature-crossover in the \emph{shape} of the asymptotic~${\bar{R}_{\text{sat}}(y)}$, that will be discussed in the next section.
}

{An important point regarding the determination of ${\bar{R}_{\text{sat}}(y)}$ is that  the large-$y$ behavior of~${\bar{R}(t,y)}$ is treated as noise in the averaging procedure.
From the point of view of our DP toymodel~\eqref{eq-toymodel-def-Rbar-functional}, this is equivalent in neglecting the negative excursions~${b(t,y)}$ in~\eqref{eq-infty-time-Rxi-bumps-1} or in the linearized case~\eqref{eq-decomposition-Rbarlin-1} which are known to  move to larger~$y$ with increasing `time'~\cite{agoritsas-2012-FHHpenta}.
It results in a slight displacement below the abscissa axis of the large-$y$ asymptote of ${\bar{R}_{\text{sat}}(y)}$.
We have to cope with this artifact, present in all three temperatures in~Fig.~\ref{fig:Rbar-compil-compil}, since we do not know analytically ${b(t,y)}$ for a correlated disorder~(${\xi>0}$) and thus we cannot remove its contribution before averaging over large `times'.
}

{
In order to characterize quantitatively the two `time' regimes of the free-energy fluctuations, we have measured the evolution of the two fitting parameters~$\arga{\widetilde{D}_t,\tilde{\xi}_t}$, by the bold application of the approximation ${\bar{R}(t,y) \approx \widetilde{D}_t \cdot \mathcal{R}_{\tilde{\xi}_t}(y)}$ of our DP toymodel~\eqref{eq-toymodel-def-Rbar-functional} at all `times'.
We refer the reader to~Appendix~\ref{A-time-evolution-Rbar-et-al} for the detailed quantitative comparison of the three fitting procedures with respectively $\mathcal{R}^{\text{Gauss}}$, $\mathcal{R}^{\text{SincG}}$ and $\mathcal{R}^{\text{CubicS}}$, for the three characteristic temperatures of~Fig.~\ref{fig:Rbar-compil-compil}.
This quantitative comparison shows that:
\textit{(i)}~the approximation of our DP toymodel~\eqref{eq-toymodel-def-Rbar-functional} can be extended even at `times' shorter than~$t_{\text{sat}}$, as the fitting parameters can be obtained with reasonable uncertainties for the low- \textit{versus} high-temperature fits (see~Fig.~\ref{fig:Dtilde-3T}-\ref{fig:xitilde-3T} in~Appendix~\ref{A-time-evolution-Rbar-et-al});
\textit{(ii)}~the three fitting functions for~$\mathcal{R}(y)$ yield consistent values for~$\arga{\widetilde{D}_t,\tilde{\xi}_t}$ and their discrepancies allow to characterize the temperature-crossover of the correlator's shape (see Fig.~\ref{fig:peakMeas-3T} in~Appendix~\ref{A-time-evolution-Rbar-et-al}).
}

\begin{figure}
   \includegraphics[width=\columnwidth]{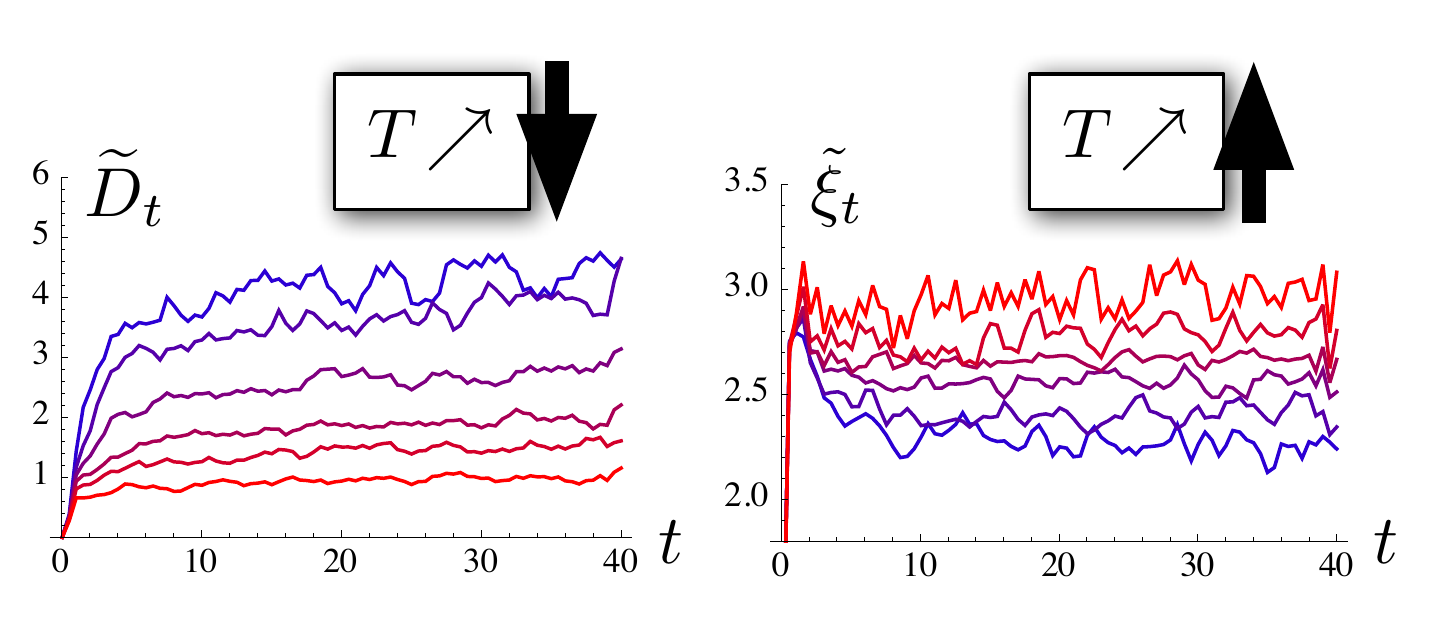} \label{fig:Dtilde-xitilde-t-T}
 \caption{
 (Color online)
 Temperature-dependent {amplitude} $\widetilde{D}_t$ and {typical spread} $\tilde{\xi}_t$ for `SincG' {for ${T \in \arga{0.35,1,2,3,4,6}}$ as listed in~Appendix~\ref{A-numericalrecipe}. See~Fig.~\ref{A-time-evolution-Rbar-et-al-part1} for the error bars of the minimum and maximum temperatures.}
 For increasing temperatures (blue to red), $\widetilde{D}_t$ decreases whereas $\tilde{\xi}_t$ slightly increases (${\xi^{\text{grid}}_y=2}$ for reference), resulting in an overall damping of the correlator ${\bar{R}_{\text{sat}}(y)}$ {(see~Fig.~\ref{fig:MeanMeanFbar-Peak-wrt-T})}.
 \label{fig:Dtilde-xitilde-t-T-Rbarsat}
 }
\end{figure}

{
As pictured in~Fig.~\ref{fig:Dtilde-xitilde-t-T-Rbarsat} for increasing temperatures (\emph{blue to red curves}), we can clearly identify the two regimes of short-`time' \textit{versus} saturation directly on these fitting parameters. 
The saturation is reached faster for the typical spread~$\tilde{\xi}_t$ than for the amplitude~$\widetilde{D}_t$, with~${t_{\text{sat}}}$ slightly decreasing when the temperature increases. On the temperature-range we explore numerically, we have~${t_{\text{sat}} \lesssim 10}$, as already deduced from~Fig.~\ref{fig:Rbar-compil-compil}.
The temperature-dependence of the asymptotic correlator~${\bar{R}_{\text{sat}}(y)}$ can thus be safely obtained from averaging over `times'~${t>t_{\text{min}}=25}$, and will be discussed in the next section.
}

{
As a concluding remark, let's emphasize that the temperature-crossover of $\arga{\widetilde{D}_{\infty}, \tilde{\xi}_{\infty}, \mathcal{R}}$ is conditioned by the short-`time' regime, and thus requests a deeper understanding of the KPZ nonlinearity feedback at `times'~${t < t_{\text{sat}}}$.
Nevertheless, in our numerical measurements, these short `times' are \textit{a priori} altered by the artificial initial condition, whose signature has thus to be investigated otherwise, via the roughness in~Sec.~\ref{section-numerical-roughness} or the study of~${\partial_t \overline{\bar{F}_V(t,y)}}$ in~Appendix~\ref{A-section-fluctuation-MeanMeanFbar}.
}

\section{Temperature-dependence of ${\bar{R}_{\text{sat}}(y)}$}
\label{section-temp-dep-asympt-Rbar-sat}

{In the previous section, we have discussed the `time'-evolution of the correlator~${\bar{R}(t,y)}$ \eqref{eq-def-corr-FbarFbar2}
and pointed out the existence of a characteristic `time'~$t_{\text{sat}}$ that separates a short-`time' regime from a saturation regime.
Now we focus on the temperature-dependence of the large-`time' free-energy fluctuations, in order to observe numerically the crossover from the high- to the low-temperature regime, as predicted analytically when the microscopic disorder is correlated with~${\xi>0}$~\cite{agoritsas_2010_PhysRevB_82_184207,agoritsas_2012_FHHtri} (see~Sec.~\ref{subsec:disorder-free-energy-DPtoymodel}).
}

{
From the point of view of our DP toymodel~\eqref{eq-toymodel-def-Rbar-functional-saturation}, we  examine the saturation correlator~${\bar{R}_{\text{sat}}(y)}$,
discussing on one hand the crossover in the shape $\mathcal{R}(y)$ of its central peak and on the other hand the temperature-dependence of its fitting parameters, \textit{i.e.} the amplitude~$\widetilde{D}_{\infty}$ and typical spread~$\tilde{\xi}_{\infty}$ of the correlator.
}

{In Fig.~\ref{fig:Rbar-compil-compil} we have plotted~${\bar{R}_{\text{sat}}(y)}$ and its fits with respect to the three functions~$\mathcal{R}(y)$ at low-, intermediate- and high-temperature (\textit{center}). We} observe graphically that at high-$T$ the exact microscopic disorder correlator $\mathcal{R}^{\text{CubicS}}(y)$ or alternatively its phenomenological counterpart $\mathcal{R}^{\text{SincG}}(y)$ both correctly encompass the features of the whole peak, including its maximum and its negative anti-correlations.
At low-$T$, {on the contrary,} $\mathcal{R}^{\text{Gauss}}(y)$ appears to be more suited to capture the central peak and its maximum, although completely skipping {its} anti-correlations.
{These anti-correlations} are obviously inherited from the microscopic disorder correlator~${R_{\xi}(y)=\mathcal{R}^{\text{CubicS}}_\xi(y)}$, {plotted in~Fig.~\ref{fig:RCubicS-graph-compil} in~Appendix~\ref{A-effectiveCorrelatorSpline}. They are however} progressively altered when the temperature is lowered.
{As for the negative shift of the large-$y$ asymptote, it is an expected artifact of the averaging procedure (see~Sec.~\ref{section-timeevol-Rty-fits}), and as such it is not encompassed by any of the fitting functions~$\mathcal{R}(y)$.}

{This behavior suggests that at high-$T$ the imprint of the microscopic disorder correlator is recovered in the asymptotic shape~$\mathcal{R}(y)$, as it is exactly the case at infinite `time' for an uncorrelated disorder~\eqref{eq-Rbarinfzeroxi} and for the linearized case with~${\xi>0}$~\eqref{eq-infty-time-Rxi-lin}.
At low-$T$ the shape is analytically not known, but we speculate that it could be a convolution of the disorder correlator~${R_{\xi}(y)}$ and a universal kernel, yielding a modified (non-universal) Airy kernel in~\eqref{eq-scalingFbarV}.
Note that this effect would actually have gone unnoticed if we had focused directly on the parameters $\arga{\widetilde{D}_t,\tilde{\xi}_t}$ or if we had considered the correlator~${\bar{C}(t,y)}$~\eqref{eq-def-corr-FbarFbar2} instead of ${\bar{R}(t,y)=\frac12 \partial_y^2 \bar{C}(t,y)}$.
}

{
We have furthermore characterized quantitatively this temperature-crossover in~Appendix~\ref{A-time-evolution-Rbar-et-al},
on one hand by computing the geometrical ratios ${\frac{\widetilde{D}_{\infty}^{\text{SincG}}}{\widetilde{D}_{\infty}^{\text{Gauss}}}}$ and ${\frac{\tilde{\xi}_{\infty}^{\text{SincG}}}{\tilde{\xi}_{\infty}^{\text{Gauss}}}}$ (see~Appendix~\ref{A-time-evolution-Rbar-et-al-part1}),
and on the other hand by comparing the maximum of the peak ${\bar{R}(t,0)}$ to the values deduced from the different fits (see~Appendix~\ref{A-time-evolution-Rbar-et-al-part2}).
Both analyses support quantitatively the high-$T$ scenario of~${\mathcal{R}(y) \approx \mathcal{R}^{\text{CubicS}}(y)}$.
}

\begin{figure}
 \includegraphics[width=\columnwidth]{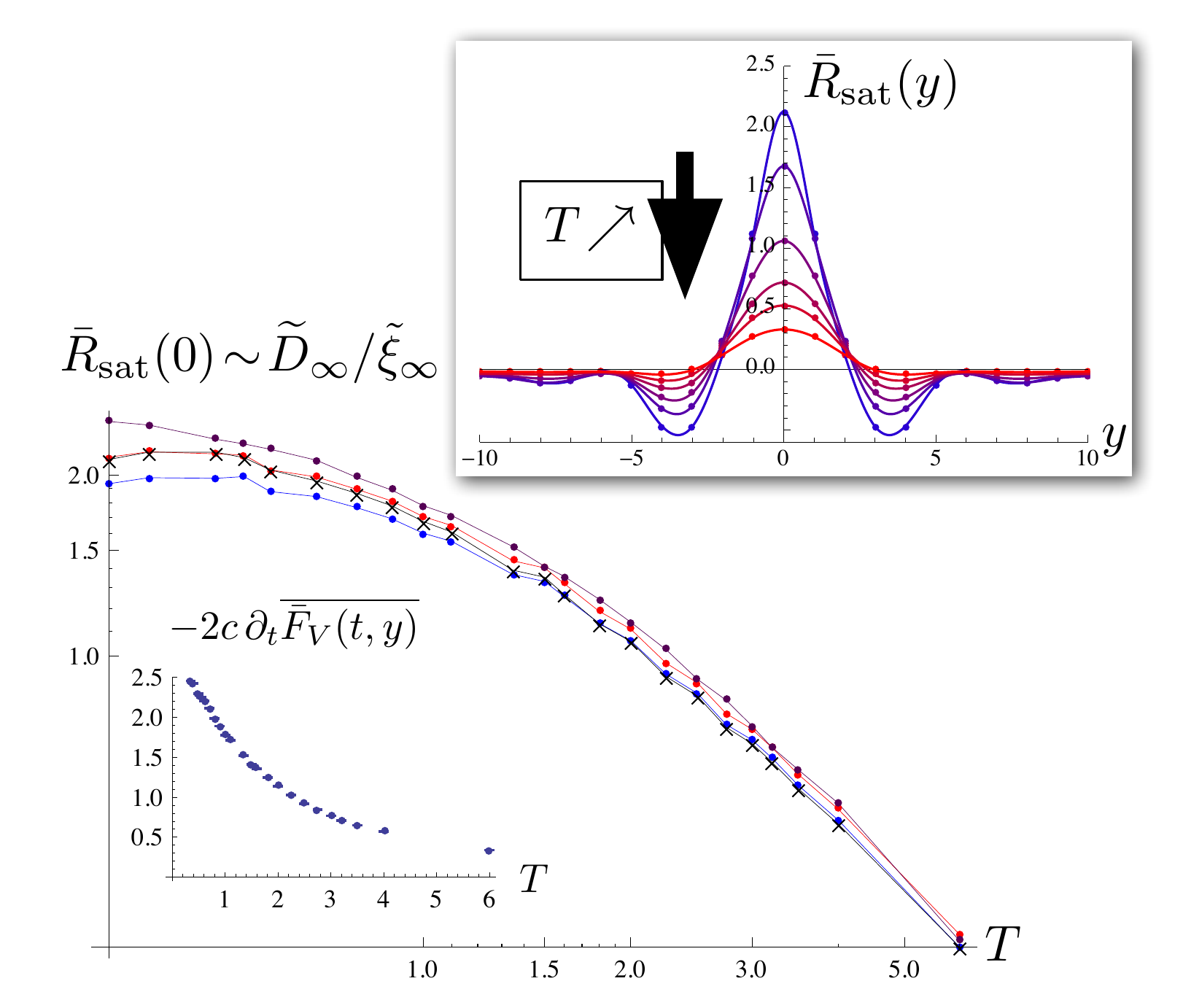}
 \caption{
 (Color online)
 Temperature dependence of the maximum of ${\bar{R}_{\text{sat}}(y)}$ over `times' ${t \in \argc{25,40}}$, measured by three different ways:
 first measured numerically directly $\bar{R}_{\text{sat}}(0)$ (black crosses),
 secondly deduced by the `Gauss' and `SincG' fits as in Fig.~\ref{fig:peakMeas-3T} (respectively red and blue dots),
 and thirdly measured indirectly from the linear slope of ${-{2c} \, \overline{\bar{F}_V(t,y)}}$ in Fig.~\ref{fig:MeanMeanFbar-3T-compil} (purple dots) and systematically slightly overestimated.
 {\textit{Upper inset}:~Saturation correlator~${\bar{R}_{\text{sat}}(y)}$ for ${T \in \arga{0.35,1,2,3,4,6}}$ and ${t_{\text{min}}=25}$ as listed in~Appendix~\ref{A-numericalrecipe}}.
 \textit{Lower inset}:~The excellent linear behavior of ${\overline{\bar{F}_V(t,y)}}$ yields vanishing error bars for the slope on the range ${t \in \argc{25,40}}$.
 \label{fig:MeanMeanFbar-Peak-wrt-T}
 }
\end{figure}

{
We now consider the temperature dependence of the amplitude~${\widetilde{D}_{\infty}}$, which is the central quantity controlling the temperature-crossover at~${\xi>0}$ in our analytical predictions~\cite{agoritsas_2012_FHHtri}, both for the free-energy and the geometrical fluctuations (see respectively~\eqref{eq-scalingFbarV}-\eqref{eq-scalingFbarV_BRM} and~\eqref{eq-roughnessregimes_smalllarget})}.
In~Fig.~\ref{fig:Dtilde-xitilde-t-T-Rbarsat} we observe graphically that
the amplitude $\widetilde{D}_{\infty}$ decreases strongly with $T$ 
whereas the typical spread $\tilde{\xi}_{\infty}$ increases (up to $30\%$) when~$T$~increases.
These behaviors result in an overall damping of the effective disorder correlator ${\bar{R}_{\text{sat}}(y)}$ when the thermal fluctuations are {enhanced, as plotted in the inset of~Fig.~\ref{fig:MeanMeanFbar-Peak-wrt-T}.}
{Comparing quantitatively the values obtained for the three fitting functions~${\mathcal{R}(y)}$ 
in~Appendix~\ref{A-time-evolution-Rbar-et-al-part3},
we find that ${\widetilde{D}_{\infty}(T)}$ exhibits a monotonous crossover from~$~1/T$ at high-$T$ to a saturation~${1/T_c(\xi)}$ at low-$T$. This behavior is qualitatively consistent with the GVM prediction presented after~\eqref{eq-Dtilde-infty-finterp}, but a quantitative test is hindered by numerical constants inherent on one hand to our numerical procedure and on the other hand to the GVM approximation.
The corresponding~${\tilde{\xi}_{\infty}(T)}$ displays in parallel a slight temperature dependence that either corrects the minimal assumption~${\tilde{\xi}_{\infty}\propto \xi}$ of our DP toymodel, or can be attributed to the mismatch between~${\mathcal{R}(y)}$ and our fitting functions. 
}

{We summarize all these temperature dependencies in~Fig.~\ref{fig:Dtilde-xitilde-t-T-Rbarsat}, plotting ${\bar{R}_{\text{sat}}(0) \sim \widetilde{D}_{\infty}/\tilde{\xi}_{\infty}}$ as a function of~$T$. The main result of our numerical study is the observation of the temperature-crossover of the amplitude~${\widetilde{D}_{\infty}}$ and hence of ~${\bar{R}_{\text{sat}}(0)}$, and their saturation below a characteristic temperature~$T_c$.
We actually compare four distinct determinations of this quantity:
\textit{(i)}~a direct measurement on~${\bar{R}_{\text{sat}}(y)}$,
\textit{(ii)}~the values obtained from~${\mathcal{R}^{\text{Gauss}}(y)}$,
\textit{(iii)}~the values obtained from~${\mathcal{R}^{\text{CubicS}}(y)}$ and ${\mathcal{R}^{\text{SincG}}(y)}$ which collapse exactly,
\textit{(iv)}~and an alternative measurement via ${\partial_t \overline{\bar{F}_V(t,y)}}$ which is presented in~Appendix~\ref{A-section-fluctuation-MeanMeanFbar}.
By their good quantitative agreement, this comparison provides a consistency check of our numerical procedure and an additional highlight of the temperature-crossover of the correlator shape~${\mathcal{R}(y)}$.
}
The crossover temperature cannot be sharply determined from~${\bar{R}_{\text{sat}}(0)}$,
nevertheless the definition ${T_c(\xi)=(\xi c D)^{1/3}}$ obtained by GVM~\cite{agoritsas_2010_PhysRevB_82_184207} and by scaling arguments~\cite{agoritsas_2012_FHHtri} predicts with ${\xi=2}$, ${c=1}$ and ${D=8}$ that ${T_c \approx 2.5}$.
Without any corrective numerical constant this value is actually compatible with the crossover of the amplitude~${\widetilde{D}_{\infty}(T,\xi)}$ in~Fig.~\ref{fig:Dtilde-inf-allT-compil} and also of the function $\mathcal{R}(y)$ in~Fig.~\ref{fig:peakMeas-3T} (\textit{bottom left}).

x


\section{Temperature-dependence of the roughness function $B(t)$}
\label{section-numerical-roughness}

{
Up to now we have analyzed in detail the disorder free-energy fluctuations from the point of view of our DP toymodel~\eqref{eq-toymodel-def-Rbar-functional}-\eqref{eq-toymodel-def-Rbar-functional-saturation}, \textit{i.e.} characterizing specifically the two-point correlator ${\bar{R}(t,y)}$~\eqref{eq-def-corr-etaeta}.
Keeping in mind the the translation of the `time' $t$ for the DP endpoint {into} the 1D interface lengthscale, we focus now on the implications of~$t_{\text{sat}}$ and ${\widetilde{D}_{\infty}}$ in the geometrical fluctuations.
We examine specifically their variance as a function of `time', namely the roughness ${B(t)= \overline{\moy{y(t)^2}}}$ as presented in~Sec.~\ref{subsec:model_geometrical-fluctuations}.}

As recalled in~Sec.~\ref{section-DPformulation-numerics} after the definition~\eqref{eq-def-roughness-zeta}, the roughness is expected to display at least two asymptotic powerlaw regimes~\eqref{eq-roughnessregimes_smalllarget}, crossing from a pure thermal behavior ${B(t)\gtrsim B_{\text{th}}(t)=\frac{Tt}{c}}$ at short `times' to the random-manifold ${B_{\text{RM}}(t) \sim [\widetilde{D}_{\infty}/c^2]^{2/3}\, t^{4/3}}$ at large `times'.
Above $T_c$ there is a single crossover Larkin `time' ${t_*(T)=\frac{T^5}{cD^2}}$ corresponding to the intersection of the two asymptotic regimes of~\eqref{eq-roughnessregimes_smalllarget}. This length diverges with $T$, and essentially prevents us from observing the {complete} crossover to the RM regime in our numerical approach in a reasonable computational time for ${T>2}$.
Below $T_c$ however we expect from Ref.~\cite{agoritsas_2010_PhysRevB_82_184207}
the appearance of an intermediate `Larkin-modified' roughness regime ending at the generalized Larkin `time' ${L_c(T,\xi)}$ 
  \begin{align}
    L_c (T,\xi) = 4 \pi \cdot \frac{T^5}{cD^2} \cdot f(T,\xi)^{-5}
    \label{eq-def-Larkin-GVM-DPtoymodel} 
  \end{align}
which marks the beginning of the RM regime, {cf.~Sec.~\ref{subsec:model_geometrical-fluctuations}}.
{It depends on the interpolating parameter ${f(T,\xi)=\widetilde{D}_{\infty}/(\frac{cD}{T})}$ defined in~\eqref{eq-Dtilde-infty-finterp} and which characterizes the crossover between the high- and low-temperature regimes~\cite{agoritsas_2012_FHHtri}}.
{In order to understand how the intermediate `Larkin-modified' regime appears, we focus on the pure disorder component of the roughness:
\begin{align} 
 B_{\text{dis}}(t)=B(t)-\frac{Tt}{c} =  \overline{\moy{y(t)^2}}^c
 \label{eq-def-Bsis-STS}
\end{align}
(this equality is a consequence of the STS~\cite{fisher_huse_1991_PhysRevB43_10728,hwa_1994_PhysRevB49_3136,
ledoussal_2003_PhysicaA317_140,agoritsas_2012_FHHtri}).
We will argue that the two-`time' regimes of~${\bar{R}(t,y)}$, separated by~$t_{\text{sat}}$, are actually transposed to~${B_{\text{dis}}(t)}$: the competition between~${B_{\text{th}}(t)}$ and~${B_{\text{dis}}(t)}$ then yields two or three roughness regimes, respectively at high-$T$ and at low-$T$. In particular, the Larkin `time' is reached when ${B(t) \approx B_{\text{dis}}(t) \approx B_{\text{RM}}(t)}$, implying that~${t_{\text{sat}}\leq L_c}$.
}

\begin{figure}[h]
 \subfigure{\includegraphics[width=0.95\columnwidth]{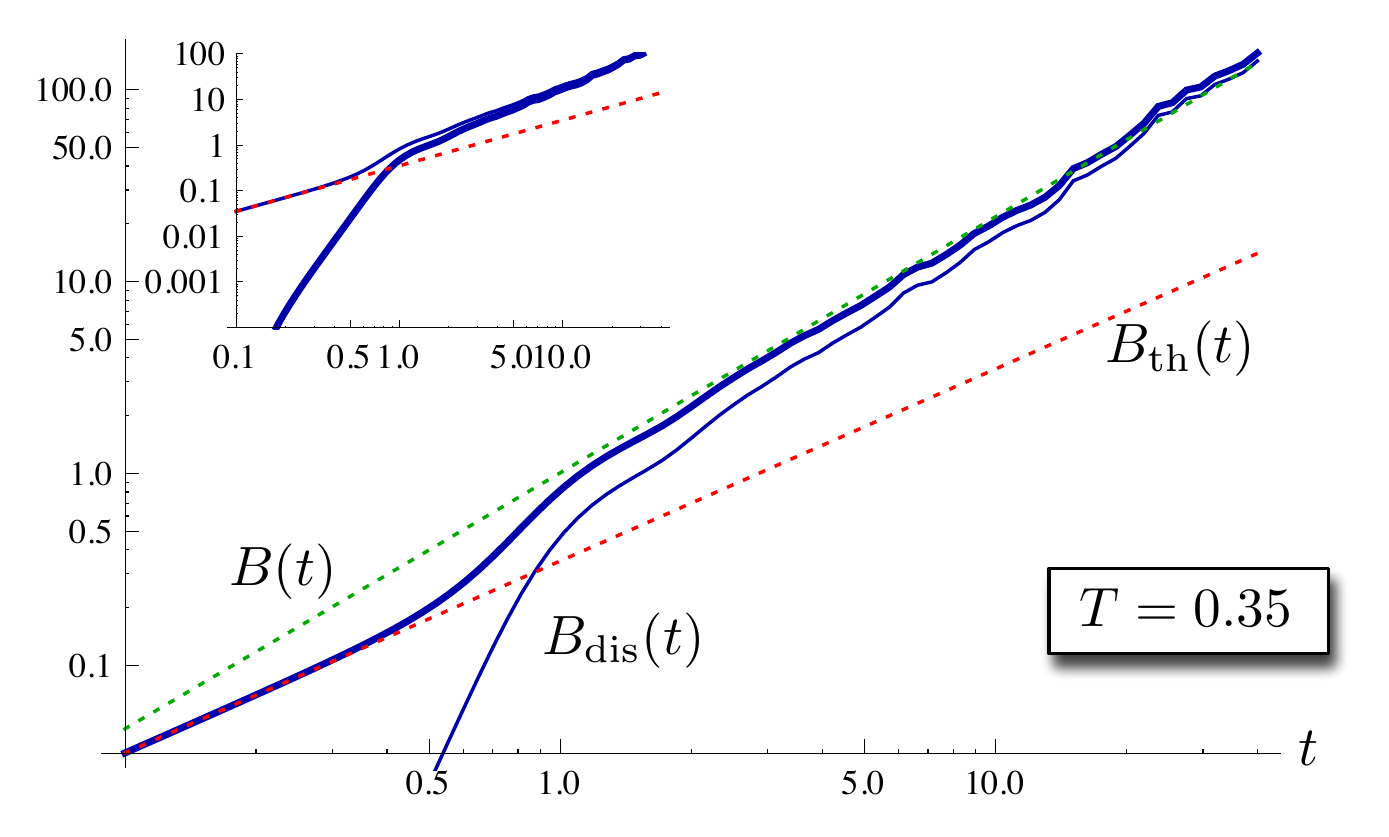}  \label{fig:roughness-T035-dotted}}
 \subfigure{\includegraphics[width=0.95\columnwidth]{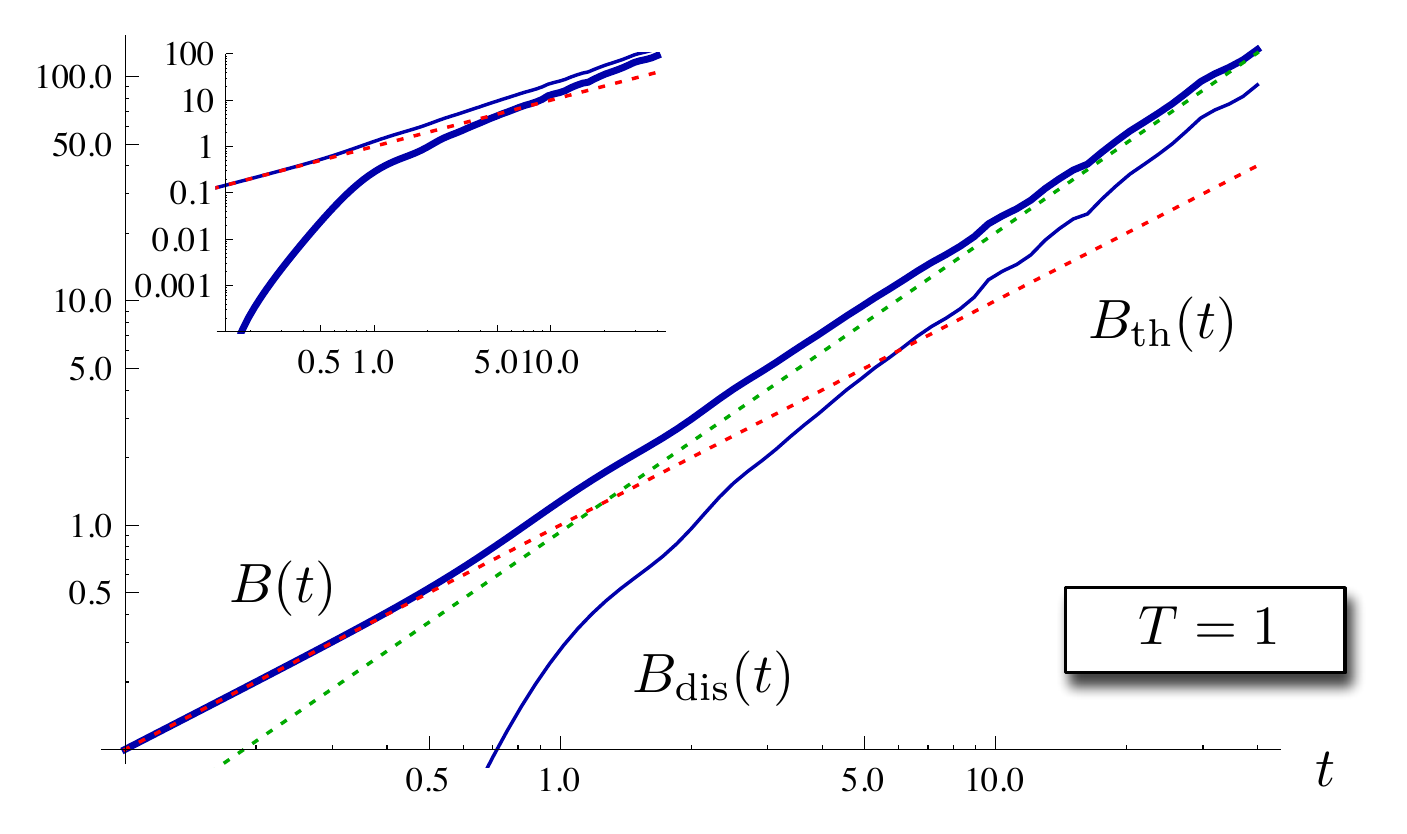}  \label{fig:roughness-T1-dotted}}
 \subfigure{\includegraphics[width=0.95\columnwidth]{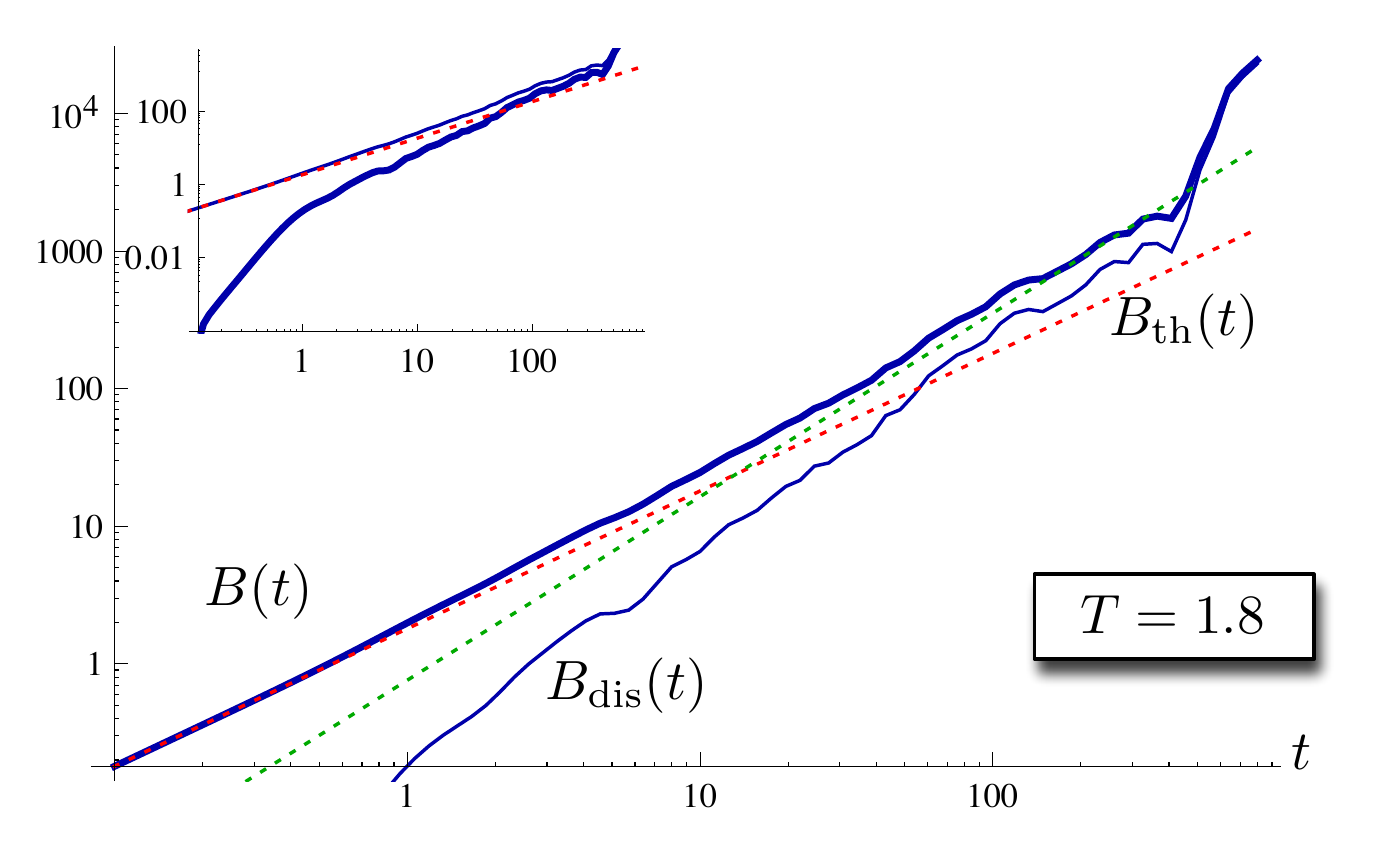}  \label{fig:roughness-T18-dotted}}
 \caption{
 (Color online)
 Roughness ${B(t)=\overline{\moy{y(t)^2}}}$ and its disorder component ${B_{\text{dis}}(t)=B(t)-\frac{Tt}{c} = \overline{\moy{y(t)^2}}^c}$ measured numerically at fixed temperature ${T \in \arga{0.35,1,1.8}}$.
 \textit{Main}:~Focus on the crossover of ${B(t)}$ (thick) from the pure thermal behavior ${B_{\text{th}}(t)=\frac{Tt}{c}}$ (dotted red) at short `times' to the asymptotic ${B_{\text{RM}}(t) \sim \widetilde{D}_{\infty}(T)^{2/3} t^{4/3}}$  (dotted green) at asymptotically large `times'; ${B_{\text{dis}}(t)}$ is the thinner lower curve.
 \textit{Inset}:~Focus on the short-`times' crossover of ${B_{\text{dis}}(t)}$ (thick), in parallel to ${B(t)}$~(thin).
 \label{fig:roughness-3T-dotted}
 } 
\end{figure}

{Although} the GVM framework {yields} a prediction 
for the full temperature-crossover of the DP toymodel (see {\textit{e.g.}} Ref.~\cite{agoritsas_2010_PhysRevB_82_184207} and Appendix~A of~\cite{agoritsas_2012_FHHtri}),
a complete \textit{quantitative} test of this prediction is hindered by the numerical constants that are \textit{a priori} present in the determination of {$\widetilde{D}_t$, $\tilde{\xi}_t$}, $T_c$, $L_c$ and the amplitude of the asymptotic roughness itself, due to previously discussed numerical artifacts and also to the GVM approximation.
Nevertheless, we observe \textit{qualitatively} these different roughness regimes, as plotted at fixed temperature ${T \in \arga{0.35,1,1.8}}$ in~Fig.~\ref{fig:roughness-3T-dotted}, indicating both the total roughness ${B(t)=\overline{\moy{y(t)^2}}}$ and its pure disorder component
In all three graphs we can follow the crossover in `time' from the thermal asymptote to the RM asymptote which stems from the increasing ${B_{\text{dis}}(t)}$ added to ${B_{\text{th}}(t)}$. Note that the RM asymptote has been constructed consistently with~\eqref{eq-roughnessregimes_smalllarget} with a numerical correction fixed once and for all, from the dataset ${T=0.4}$ averaged over ${t \in \argc{25,40}}$, assumed to be already in the RM regime:
\begin{equation}
\frac{B_{\text{RM}}(t)}{t^{4/3}}
 \approx
 \text{corr}_{(T=0.4)} \cdot \frac{3}{2^{2/3}\pi^{1/3}} \argc{\frac{{\widetilde{D}^{\text{SincG}}_{\infty}(T,\xi)}}{c^2}}^{2/3}
 \label{eq-num-roughness-asympt-corrT04}
\end{equation}
with~$\text{corr}_{(T=0.4)}=0.292 \pm 0.008$. %
This RM asymptote is graphically consistent with all the available datasets in the range ${T \in \argc{0.35,1.8}}$, which plays in favor of a numerical corrective factor common to all temperatures and absorbing the discrepancy in ${B_{\text{RM}}(t)}$ stemming from $\widetilde{D}_{\infty}$ and the GVM.
The low-$T$ regime is illustrated by ${T=0.35}$ where the intermediate `Larkin-modified' regime is clearly present, whereas it has disappeared as such already at ${T=1}$.

{We are limited, in our numerical procedure, in the large `times' that we can explore in a reasonable computation time.
Larger `times' have however} been explored at ${T=1.8}$ which is believed to be close to $T_c$ (cf.~Sec.\ref{section-temp-dep-asympt-Rbar-sat}), and is consistent with lower temperatures except for a sudden increase at ${t>400}$ which can clearly be attributed to the finite size of the box ${y \in \argc{-y_m,y_m}}$. {Confining effects due to the finite box $[-y_m,y_m]$ do not appear until ${\sqrt {B(t)} \sim y_m}$, and this condition translates for this dataset into ${B(t)<160^2=2.5 \cdot 10^4}$} and yields an adequate upper bound in~Fig.~\ref{fig:roughness-3T-dotted} (\textit{bottom}).
The corresponding upper bound for most of the datasets is ${B(t)<40^2=160}$ consistently with the range of roughness displayed in all the other graphs.

\begin{figure}
 \includegraphics[width=\columnwidth]{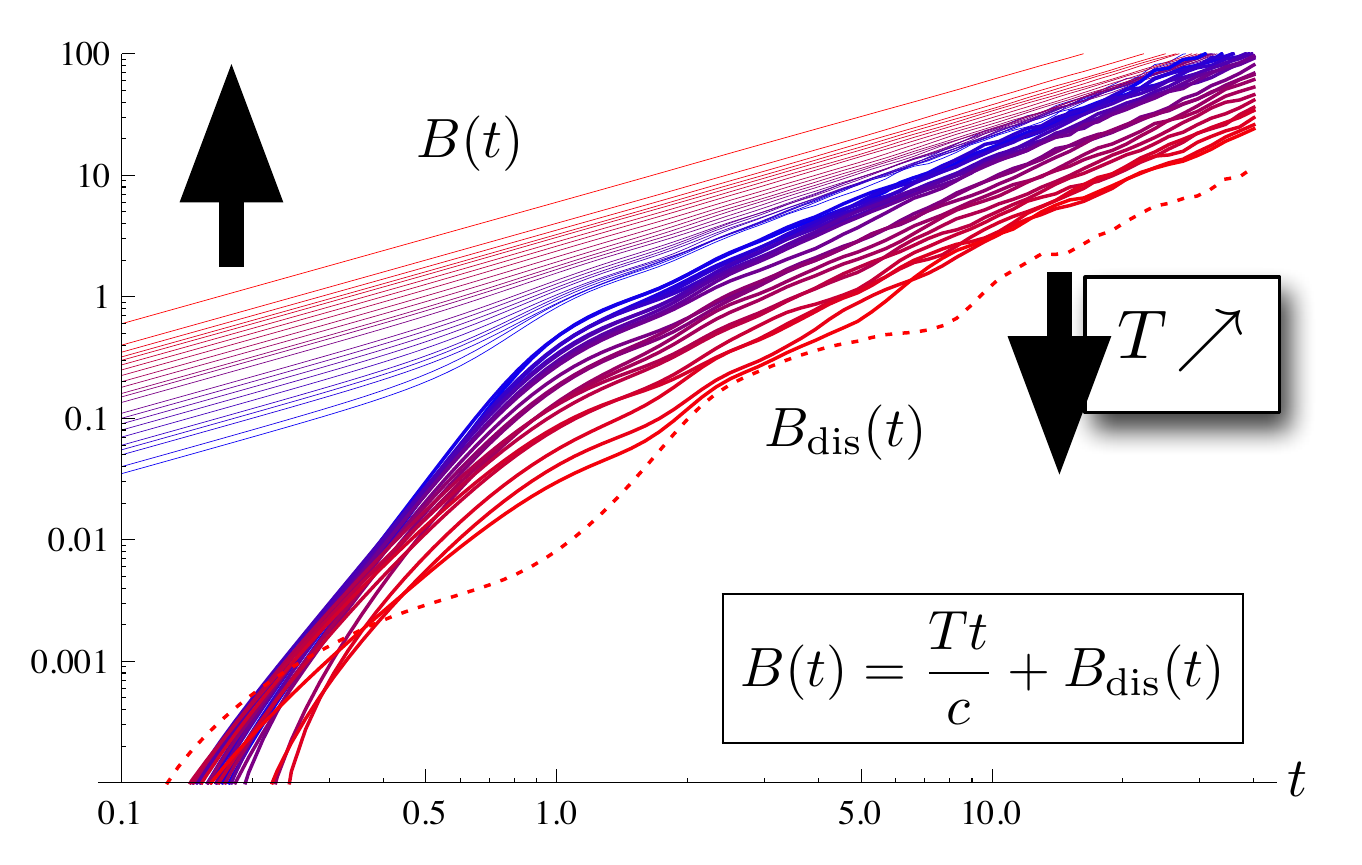}
 \caption{
 (Color online)
 Temperature-dependent roughness ${B(t)= \overline{\moy{y(t)^2}}}$ and its disorder component ${B_{\text{dis}}(t)}$ (in logarithmic scale, respectively top and bottom curves),
 for ${T \in \argc{0.35,6}}$ as listed in~Appendix~\ref{A-numericalrecipe} (blue to red for increasing temperatures).
 ${B_{\text{dis}}(t)}$ for ${T=6}$ is indicated in dotted red.
 \label{fig:roughness-Bdis-allT}
 }
\end{figure}

Gathering in~Fig.~\ref{fig:roughness-Bdis-allT} the roughness ${B(t)}$ and ${B_{\text{dis}}(t)}$ over the range ${T \in \argc{0.35,6}}$,
we observe as expected with increasing temperature that the disorder roughness ${B_{\text{dis}}(t)}$ is progressively damped by thermal fluctuations. {The intermediate `Larkin-modified' regime thus} shrinks with increasing temperature.
The beginning of the RM regime is consequently pushed to larger `times', as it could also be deduced by the condition that ${B_{\text{dis}}(t)}$ becomes comparable to ${B_{\text{th}}(t)}$ close to the Larkin `time' ${t=L_c}$.
From the point of view of the 1D interface, the temperature-dependence of ${B_{\text{dis}}(t)}$ can be physically understood with the following picture:
at small lengthscales the thermal fluctuations make the interface rougher within the local valleys of the disordered free-energy landscape and ${B(t) \gtrsim B_{\text{th}}(t)}$ increases with $T$ as expected;
at large lengthscales on the contrary thermal fluctuations  allow the interface to explore more effectively the free-energy landscape by overcoming some free-energy barriers in order to minimize its elastic energy and thus have ${B_{\text{RM}}(t)}$ decreasing with $T$.
From the point of view of the DP, these behaviors are encoded in the evolution of the translation-invariant distribution ${\bar{\mathcal{P}}\argc{\bar{F},t}}$ \eqref{eq-STS-PFbarV}, the integration of the microscopic disorder $V(t',y')$ explored by the elastic DP over `times' $\argc{0,t}$ defining the effective disorder~${\bar{F}_V(t,y)}$, or equivalently in the moments ${\overline{\moy{y(t)^k}}}$ of the DP endpoint.
The saturation at low-$T$ of the amplitude $\widetilde{D}_{\infty}$ in~Fig.~\ref{fig:Dtilde-inf-allT-compil} and of the peak ${\bar{R}(t,0)}$ in~Fig.~\ref{fig:MeanMeanFbar-Peak-wrt-T} naturally translates into a saturation of the asymptotic amplitude of the roughness \eqref{eq-num-roughness-asympt-corrT04}.

In order to focus on the disorder contribution, we have computed numerically directly ${\bar{F}_V(t,y)}$, but regarding the roughness we had first to construct the total free-energy ${F=F_{\text{th}}+\bar{F}}$, then compute the total roughness $B(t)$ and eventually deduce its disorder component ${B_{\text{dis}}=B-B_{\text{th}}}$.
This quantity is thus more subject to noise at higher $T$ (see \textit{i.e.} ${T=6}$ in~Fig.~\ref{fig:roughness-Bdis-allT}), but especially at low-$T$ it clearly displays two regimes in `time'.
To address the question of a possible powerlaw at short-`times', the logarithmic slope
${\zeta_{(\text{dis})}(t)=\frac12 \frac{\partial B_{(\text{dis})}(t)}{\partial \log t}}$ is plotted in~Fig.~\ref{fig:zeta-logderiv}.
While $\zeta(t)$ crosses over as expected from ${\zeta_{\text{th}}=\zeta_{\text{EW}}=\frac12}$ to ${\zeta_{\text{RM}}=\zeta_{\text{KPZ}}=\frac23}$ \eqref{eq-roughnessregimes_smalllarget} but excludes the definition of an intermediate-`times' powerlaw for ${B(t)}$, ${\zeta_{\text{dis}}(t)}$ on the contrary displays a plateau at low-$T$ which disappears already at~${T=1}$, a tendency confirmed at ${T=1.8}$.
{According to the GVM prediction~\cite{agoritsas_2010_PhysRevB_82_184207} ${B_{\text{dis}}(t)}$ should start in ${\sim \frac{\widetilde{D}}{c^2} t^2 / \tilde{\xi}}$, in which case the value of the plateau would have $1$ as a lower band and only a `time'-dependence of $\widetilde{D}_t$ and/or $\tilde{\xi}_t$ could correct $\zeta_{\text{dis}}$ to match the GVM prediction. {This results is} anyway not to be trusted \textit{a priori} at `times' shorter than $L_c$~\cite{agoritsas_2012_FHHtri}.
The value of this plateau exhibits moreover a slight temperature-dependence, which cannot be accounted for by our DP toymodel but might also simply be an artifact of the thermal condition imposed at ${t_0=0.1}$.
Note finally that all these effects on ${B_{\text{dis}}(t)}$ take place before {${t_{\text{sat}} \lesssim 10 < L_c}$} as estimated on the saturation of the free-energy correlator ${\bar{R}(t,y)}$ in the previous sections.

\begin{figure}
 \includegraphics[width=\columnwidth]{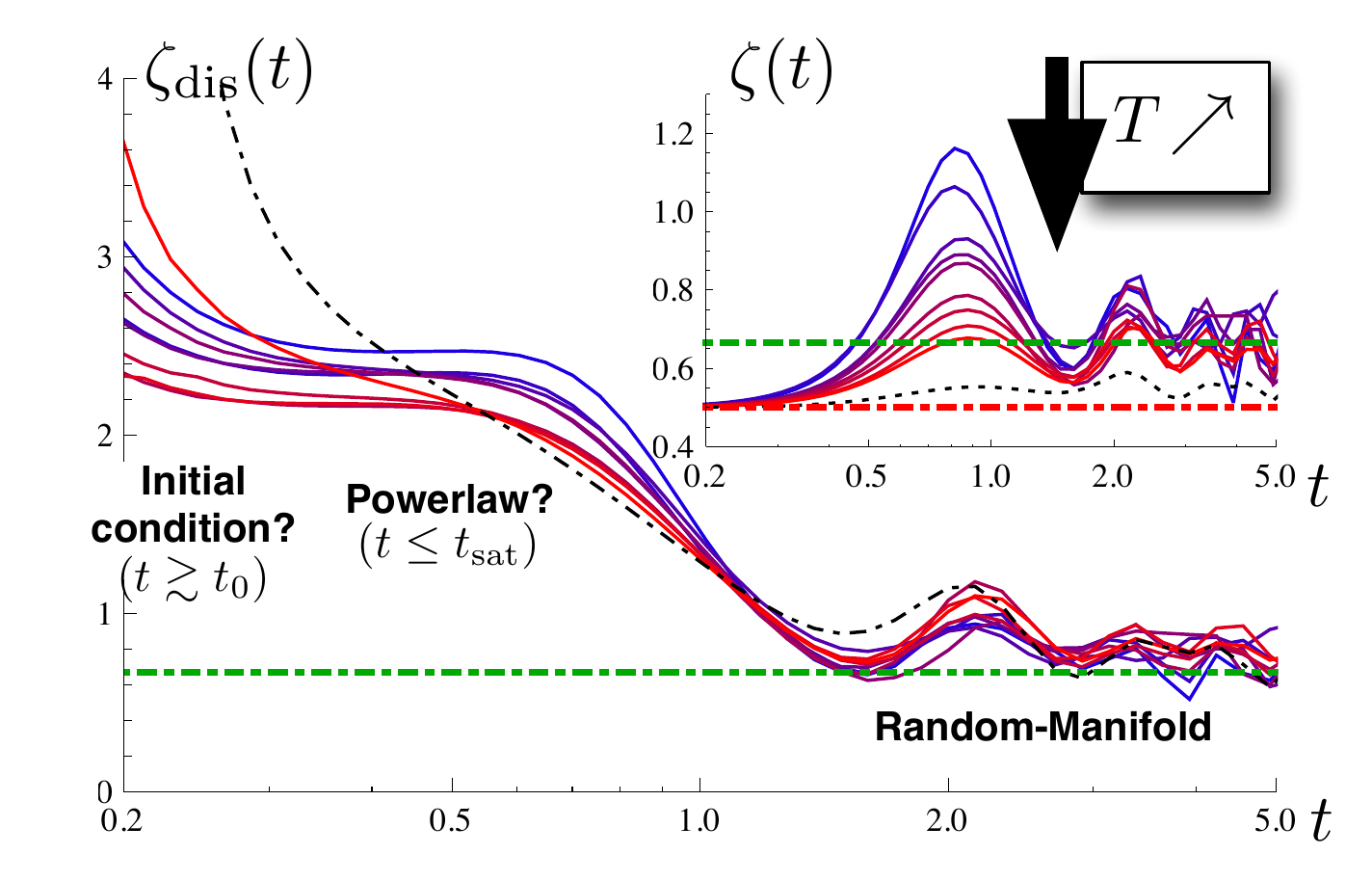}
 \caption{
 (Color online) Logarithmic slope of the roughness, ${\zeta (t)}$ and ${\zeta_{\text{dis}}(t)}$ obtained respectively from ${B(t)}$ and ${B_{\text{dis}}(t)}$ of~Fig.~\ref{fig:roughness-Bdis-allT}, for ${T \in \argc{0.35,1}}$ (blue to red for increasing temperatures) and ${T=1.8}$ (in dashed dotted black).
 The thermal exponent ${\zeta_{\text{th}}=\zeta_{\text{EW}}=\frac12}$ and the RM exponent ${\zeta_{\text{RM}}=\zeta_{\text{KPZ}}=\frac23}$ are indicated by the dashed dotted lines, respectively red and green.
 At larger `times' those logarithmic slopes are very noisy but still oscillate around the expected value ${\zeta_{\text{RM}}=\frac23}$.
  \label{fig:zeta-logderiv}
 }
\end{figure}


\section{Conclusion}
\label{section-conclusion-numerics}

In this paper, we have studied numerically the free-energy and geometrical fluctuations of a 1+1 DP growing in a quenched disordered energy landscape, uncorrelated in the `time'-direction and with a finite disorder correlation length~$\xi$ in the transverse direction~$y$ as given in~\eqref{eq-def-moydis}.
{This} model actually yields an effective description of the static 1D interface fluctuations at fixed lengthscale, which translates into the DP growing `time'.
We have specifically characterized the fluctuations of the disorder free-energy~${\bar{F}(t,y)}$ defined  
by~\eqref{eq-FzeroV}-\eqref{eq-genericdecompositionFvFbarV}, focusing on one hand
on its two-point correlator~${\bar{R}(t,y)=\overline{\partial_y \bar{F}_V(t,y) \partial_y \bar{F}_V(t,0)}}$ 
and its mean value {${\overline{\bar{F}_V(t,y)}=-\frac 1{2c}\int_0^t dt' \cdot \bar{R}(t',0)}$},
and on the other hand on its consequent geometrical fluctuations, via the roughness function~${B(t)}$.
Our numerical procedure amounts to the integration of the KPZ equation for the total free-energy~\eqref{eq-FeynmanKac-FV} with a `sharp-wedge' initial condition and a colored noise of correlator~${R^{\text{CubicS}}_{\xi}(y)}$, resulting from a cubic-splined interpolation of the random potential (as presented in~Appendix~\ref{A-effectiveCorrelatorSpline}).

We have first successfully tested the main assumption of our DP toymodel, at the core of the analytical study~\cite{agoritsas_2012_FHHtri} and the GVM roughness computations~\cite{agoritsas_2010_PhysRevB_82_184207,agoritsas_2012_ECRYS2011}:
at small transverse displacements~$y$ the correlator behaves as ${\bar{R}(t,y) \approx \widetilde{D}_t \cdot \mathcal{R}_{\tilde{\xi}_t}(y)}$,
consistently with three different fitting functions for~${\mathcal{R}(y)}$ in~Sec.~\ref{section-timeevol-Rty-fits} --~respectively {${\mathcal{R}^{\text{Gauss}}}$~\eqref{eq-def-RGauss}, ${\mathcal{R}^{\text{SincG}}}$~\eqref{eq-def-RSincG} and ${\mathcal{R}^{\text{CubicS}}}$~\-\eqref{eq-def-RCubicS-1}-\eqref{eq-def-RCubicS-2}}.
We have clearly identified two {`time'} regimes in the evolution of the free-energy fluctuations: a short-`time' regime where the correlator starts from its initial flat condition and grows while integrating the interplay between the disorder correlation and thermal fluctuations that intervenes precisely at small lengthscale~$y$; a large-`time' regime above a saturation `time'~$t_{\text{sat}}$ where the correlator is well-approximated locally by its asymptotic limit ${\widetilde{D}_{\infty} \cdot \mathcal{R}_{\tilde{\xi}_\infty}(y)}$ and thus keeps track of the short-`time' evolution from~$t_{\text{sat}}$ up to the macroscopic scale.
The combination between the disorder free-energy~${\bar{F}_V(t,y)}$ and the thermal free-energy~${F_{\text{th}}(t,y)}$ leads to two or three regimes for the roughness~${B(t)}$, respectively at high- or low-temperature with respect to the characteristic temperature~${T_c(\xi)=(\xi c D)^{1/3}}$.
A simpler physical picture is however regained by focusing on the disorder roughness ${B_{\text{dis}}(t)=B(t)-B_{\text{th}}(t)}$, which displays the same two {`time'} regimes as the disorder free-energy (see~Fig.~\ref{fig:roughness-Bdis-allT}).

We have secondly followed the temperature-dependence of the large-`time' {DP} fluctuations, whose amplitudes are found in qualitative agreement with the analytical predictions of our companion paper~\cite{agoritsas_2012_FHHtri}.
On one hand, we have multi-checked that the free-energy amplitude {${\widetilde{D}_{\infty} \sim \bar{R}(t,0) \cdot \tilde{\xi}_{\infty}}$} crosses over monotonously from the high-temperature limit~${\widetilde{D}_{\infty} \sim 1/T}$ to the low-temperature saturation~${\widetilde{D}_{\infty} \sim 1/T_c(\xi)}$ {(see~Fig.~\ref{fig:MeanMeanFbar-Peak-wrt-T})},
and on the other hand this crossover matches the temperature-dependence of the asymptotic roughness ${B_{\text{RM}} (t) \sim \widetilde{D}_{\infty}^{2/3} t^{4/3}}$ (see~Fig.~\ref{fig:roughness-3T-dotted}).
Although numerical constants stemming from the analytical approximations or from the numerical procedure clearly hinder a quantitative test, these results support a universality scenario for the amplitudes.
Consequently, even though $\xi$ might lie below accessible resolution in physical systems, it can still play a role at all lengthscales below ${T_c (\xi)}$ --~including scales much larger than~$\xi$ itself~-- and hence proves experimentally relevant.
{In other words, if the temperature-dependence of the parameters~$\arga{c,D,\xi}$ is known, changing the temperature amounts to the exploration of the disorder spatial correlation: the details of the random potential are probed with the thermal fluctuations.
This conclusion is thus encouraging to study the temperature-dependence of experimental systems, such as the magnetic domains walls or interfaces in liquids crystals as discussed in~\cite{agoritsas_2012_FHHtri}.
}

Beyond the universal large-scale exponent ${\zeta_{\text{KPZ}}=\frac23}$ and the universal crossover of the amplitude~${\widetilde{D}_{\infty}(T,\xi)}$, we believe that the precise shape ${\mathcal{R}(y)}$ also displays universality, in the form of a universal kernel that combines with the specific microscopic disorder correlator~${R_{\xi}(y)}$ {in order} to yield a generalized Airy$_2$ process~\eqref{eq-scalingFbarV}.
This guess is supported analytically by the ${\xi=0}$ limit~\eqref{eq-Rbarinfzeroxi} and the linearized case~\eqref{eq-infty-time-Rxi-lin}, and numerically by an increasing discrepancy between ${\mathcal{R}(y)}$ and~${R^{\text{CubicS}}_{\xi}(y)}$ as the temperature is lowered.
The imprint of the microscopic disorder on the macroscopic fluctuations of the 1+1 DP, or the static 1D interface, is numerically such that at high-temperature~${\mathcal{R}(y) \approx R^{\text{CubicS}}_{\xi}(y)}$, whereas at low-temperature it displays a better agreement with~${R^{\text{Gauss}}_{\xi}(y)}$.
The determination of the zero-temperature limit of ${\mathcal{R}(y)}$, let alone its complete temperature-dependence, remains an open question analytically, and will require a complete understanding of the non-Gaussian features of the free-energy fluctuations and the role of the KPZ non-linearity in~\eqref{eq-FeynmanKac-FbarV}.
{To tackle this issue numerically will require a better characterization of the short-`time' free-energy fluctuations and in particular of the initial-condition signature in~${\arga{\widetilde{D}_t,\tilde{\xi}_t}}$, ${\partial_t \overline{\bar{F}_V(t,y)}}$ and~${B_{\text{dis}}(t)}.$}

To summarize, systems belonging to the KPZ universality class are not only characterized by the large-scale exponent $\zeta_{\text{KPZ}}=\frac23$, but also by universal \textit{distributions} and universal \textit{amplitudes}.
The high-temperature limit of the amplitude $\tilde D_\infty=\frac{cD}T$ is known from the generic KPZ scaling theory (see~\cite{sasamoto_spohn_2010_JStatMech2010_P11013,kriecherbauer_krug_2010_JPhysA43_403001} for reviews), while the crossover to lower temperatures seems to be new, even at the numerical level. We suspect that the anomaly observed numerically in~\cite{lam_shin_1998_PhysRevE57_6506} for the free-energy amplitude $\widetilde D_\infty$ with respect to the expected scaling is an effect due to a small correlation length of the disordered potential, that was not taken into account since $\widetilde D_\infty$ was compared to its zero~$\xi$ (\emph{i.e.} high-temperature) expression.
Another situation where our results might prove relevant is that of the last-passage percolation~\cite{johansson_2000_CommMathPhys209_437}, which is a ``zero-temperature'' discrete DP on a lattice: the problem amounts to finding a path with minimal energy, which, in our framework, corresponds to a situation where both $\xi$ and $T$ are equal to $0$ in the macroscopic limit.
{In} the continuum formulation, those two limits do not commute, {so} a correct scaling limit remains to be found, and our findings indicate that the lattice spacing might induce observable evidences at large scale.

From a broader perspective, our numerical procedure provides a new computation frame for the continuous 1+1 KPZ equation~\eqref{eq-FeynmanKac-FV} {with a `sharp-wedge' initial condition}, which allows to study  the interplay of a colored noise with the KPZ non-linearity, whose resulting feedbacks generate new phases compared to the white-noise case.
Last, note that if the  two temperature regimes can be hinted by scaling arguments~\cite{agoritsas_2012_ECRYS2011}, the question of relating those to possible two opposite Functional-Renormalization-Group regimes of high-temperature~\cite{bustingorry_2010_PhysRevB82_140201} versus zero-temperature fixed-point~\cite{balents-fisher_1993_PhysRevB48_5949,chauve_2000_ThesePC_PhysRevB62_6241} remains open.
In the case of DP in higher dimensionalities, as recently studied  exhaustively in Ref.~\cite{halpin-healy_2012_PhysRevLett109_170602},
one might also expect manifestations of the disorder correlation length~$\xi$ at low enough temperatures.


\begin{acknowledgments}
  We would like to thank Sebastian Bustingorry, Grégory Schehr, Francis Comets and Jeremy Quastel for fruitful discussions, and Christophe Berthod for his help on the Mafalda cluster at the University of Geneva where the simulations were run. This work was supported in part by the Swiss NSF under MaNEP and Division II, by ANR 2010 BLAN 0108 and by the SCHePS interdisciplinary project of University Paris 7.
\end{acknowledgments}


\appendix


\section{Effective correlator of a 2D cubic-splined microscopic disorder}
\label{A-effectiveCorrelatorSpline}


As described in Sec.~\ref{section-numerical-recipe},
a given configuration of the microscopic disorder $V(t,y)$ is generated
first by picking up a set of random numbers $\arga{V_{j}}$ on a grid of spacing $\arga{\xi^{\text{grid}}_t,\xi^{\text{grid}}_y}$ with a normal distribution of variance $D^{\text{grid}}$,
and then by interpolating between the grid points with a 2D cubic spline (cf. Fig.~\ref{fig:RCubicS-graph-compil} \textit{top}).

In the reduced case of a 1D cubic spline between a set of points at a fixed position $t$ on the grid, it is possible to determine analytically the effective correlator of the 1D cubic-splined random potential
with a normalized function $R^{\text{CubicS}}_{\xi_y} (y)$ (cf. \eqref{eq-def-moydis}).
For the 2D spline we then assume that the effective two-point disorder correlator is given by the translation-invariant:
\begin{equation}
 \overline{V(t,y)V(0,0)}
 = D \cdot  R^{\text{CubicS}}_{\xi_t} (t) \cdot R^{\text{CubicS}}_{\xi_y} (y)
 \label{equa-def-VV-CubicS}
\end{equation} 
with $\xi_t^{\phantom{g}}=\xi^{\text{grid}}_t$ and $\xi_y^{\phantom{g}}=\xi^{\text{grid}}_y$ by construction,
and because of the passage from the discretized $y_j$ to the continuous variable $y$ an amplitude $D=D^{\text{grid}} \xi^{\text{grid}}_t \xi^{\text{grid}}_y$.

In practice, for $2n+1$ points indexed by $j= -n, \dots, n$ on a grid of spacing $\xi$, a random value $V_j$ is attached on each site $y_j=j \xi$ of the grid following the statistical distribution $\overline{V_j}=0$ and $\overline{V_j V_{j'}}=\delta_{j j'}$.
A cubic spline of $\arga{V_j}_{-n\leq j \leq n}$ is a function $V(y)$ which is
a cubic polynomial on each lattice segment $y \in \argc{y_j,y_{j+1}}$,
continuous on each lattice site $y_j$
and with its first and second derivatives also continuous at $y=y_j$.
Combining the equations of the cubic-splined parameters for a given set $\arga{V_j}$ and the disorder average over those possible sets, we obtain the following \textit{symmetric} effective correlator in the limit of $n \to \infty$:
\begin{eqnarray}
&& \begin{split}
 R^{\text{CubicS}}_{\xi} & (0\leq y \leq \xi) \\
 & = - \frac{1}{\xi^4} \argp{y-\xi}\argc{(4- 3\sqrt{3}) y^2+\xi y+\xi^2}
 \end{split}
 \label{eq-def-RCubicS-1}\\
&& \begin{split}
 R^{\text{CubicS}}_{\xi} & (y_j \leq y \leq y_{j+1}) \\
 & = -\frac{3}{\xi^4} \argp{y-y_j} \argp{y-y_{j+1}} \argp{\sqrt{3}-2}^j \\
 & \quad \quad \times \argc{y-y_{j-1}- \sqrt{3} \argp{y-y_j}}
 \end{split}
 \label{eq-def-RCubicS-2}
\end{eqnarray}
where there is a distinction between
the central segment $0\leq y \leq \xi$ which contains the auto-correlation
$$ R^{\text{CubicS}}_{\xi} (y=y_{j=0}) = \overline{V_0^2} = \overline{V_j^2}  = 1 $$
and the other segments $y_j \leq y \leq y_{j+1}$ ($j=1 \dots n$) with oscillations constrained by the cancellation at $y=y_{j \neq j}$
$$  R^{\text{CubicS}}_{\xi} (y=y_{j \neq 0}) = \overline{V_0 V_{j \neq 0}} = 0 $$
Note that this correlator has been obtained exactly by averaging over disorder the cubic splines, whose coefficients depend \textit{linearly} in the random potential and thus allow an analytical computation of this average.

\begin{figure}
 \subfigure{\includegraphics[width=\columnwidth]{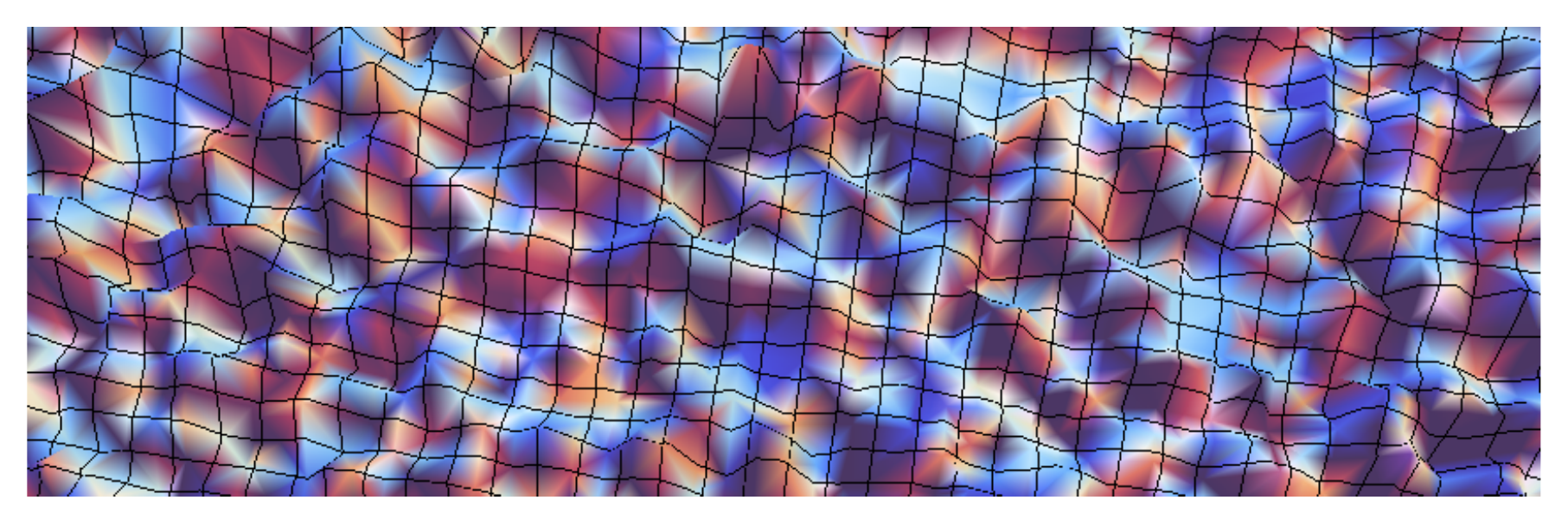}  \label{fig:random-landscape}}
 \subfigure{\includegraphics[width=\columnwidth]{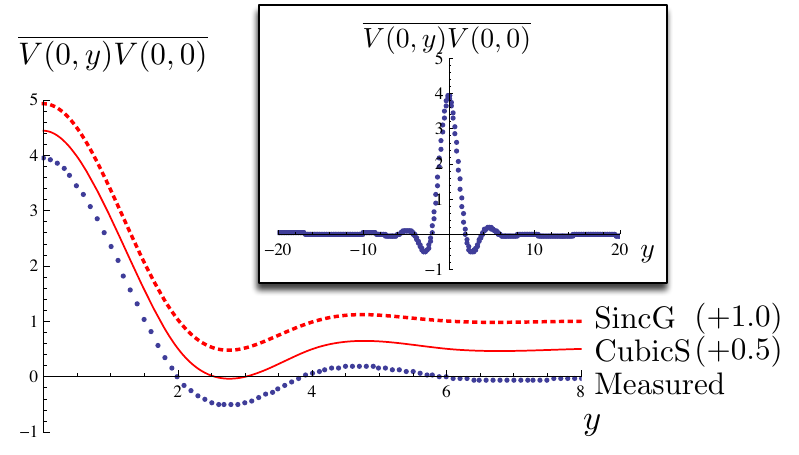}  \label{fig:RCubicS-graph}}
 \caption{
 \label{fig:RCubicS-graph-compil}
 (Color online)
 \textit{Top:} Visualization of a single disorder configuration $V(t,y)$; the mesh accounts for the grid $\arga{y_j}$, and fluctuates smoothly.
 \textit{Bottom:}
 Graph of the numerically-measured effective-1D cubic-splined disorder correlator (blue), with the `SincG' and `CubicS' (red, respectively dashed and continuous) superimposed with a slight translation for more visibility.
 \textit{Inset:}
 Full effective disorder correlator at fixed $t=0$, computed over $10'000$ disorder configurations with ${D^{\text{grid}}=4}$, ${\xi_y^{\text{grid}}=2}$ and ${\xi_t^{\text{grid}}=1}$.
 }
\end{figure}

For the microscopic disorder correlator \eqref{equa-def-VV-CubicS} the function $R_{\xi}$ is thus exactly known, but it is not the case for the effective disorder {in our DP toymodel~\eqref{eq-toymodel-def-Rbar-functional}-\eqref{eq-toymodel-def-Rbar-functional-saturation} approximating ${\bar{R}(t,y)= \overline{\partial_y \bar{F}_V(t,y) \cdot \partial_y \bar{F}_V(t,0)}}$}.
As explained in Sec.~\ref{section-numerical-recipe} we try nevertheless to extract the effective parameters at fixed `time' $\arga{\widetilde{D}_t,\tilde{\xi}_t}$ with three different fitting functions normalized to $1$:
{\textit{(i)}}~a Gaussian \eqref{eq-def-RGauss} (`Gauss'),
{\textit{(ii)}}~a sinus cardinal with a Gaussian envelope \eqref{eq-def-RSincG} (`SincG')
and {\textit{(iii)}} the exact cubic-spline correlator \eqref{eq-def-RCubicS-1}-\eqref{eq-def-RCubicS-2} (`CubicS').
The comparison between those three options accounts for a consistency check for the determination of the effective parameters via the stability of the function.
{We first test} this procedure on $\overline{V(t,y)V(0,0)}$ \eqref{equa-def-VV-CubicS},
by fitting {this correlator} separately at fixed `time' $t=0$ \textit{versus} at fixed position $y=0$:
\begin{equation}
\begin{split}
 \overline{V(t,0)V(0,0)}
  =& \argc{D \cdot  R_{\xi_y} (0)} \cdot R_{\xi_t} (t)
  \propto R_{\xi_t} (t) \\
 \overline{V(0,y)V(0,0)}
  =& \argc{D \cdot  R_{\xi_t} (0)} \cdot R_{\xi_y} (y)
  \propto R_{\xi_y} (y)
\end{split}
\end{equation}
we summarize thereafter the values obtained for the amplitude $D$ and the correlation lengths $\xi_{t,y}$ with respect to the grid parameters ({which} are kept at fixed value throughout all our computations). The errors are provided by the nonlinear-regression procedure on a correlator obtained over $10'000$ disorder configurations:
\begin{center}
\begin{tabular}{c||c|c|c|c}
\hline
 & $D$ & $\xi_y$ & $\xi_t$ & $R_{1}(0)$ \\
\hline \hline
\textbf{Grid} & $4$ & $2$ & $1$ & - \\
\hline \hline
\textbf{Gauss} & ${10.08 \pm 0.08}$ & ${0.625 \pm 0.005}$ & ${0.312 \pm 0.003}$ & $\frac{1}{\sqrt{4\pi}}$ \\
\hline
\textbf{SincG} & ${8.04 \pm 0.02}$ & ${2.020 \pm 0.004}$ & ${1.012 \pm 0.002}$ & $\frac{1}{\text{Erf}(\pi)}$ \\
\hline
\textbf{CubicS} & ${7.95 \pm 0.02}$ & ${2.006 \pm 0.004}$ & ${1.005 \pm 0.002}$ & $1$ \\
\hline
\end{tabular}
\end{center}
As we can check both in this table and visually in Fig.~\ref{fig:RCubicS-graph-compil} (\textit{bottom}),
the sinus cardinal {$\mathcal{R}^{\text{SincG}}$} catches the main phenomenological features of the exact {$\mathcal{R}^{\text{CubicS}}$} , \textit{\textit{i.e.}} the amplitude of the central peak and the position of the first oscillation,
and keeps satisfied the relations $\xi_{t,y}^{\phantom{g}}=\xi^{\text{grid}}_{t,y}$ and $D=D^{\text{grid}} \xi^{\text{grid}}_t \xi^{\text{grid}}_y$.

As for the Gaussian fit {$\mathcal{R}^{\text{Gauss}}$} ,
it overestimates the amplitude of the peak and underestimates its typical variance.
For the latter {statement,} it is simply due to the geometrical definitions of the parameter $\xi$ proper to each of these functions, and {whose discrepancy} can be merged up to a numerical constant ($\xi^{\text{CubicS}}/\xi^{\text{Gauss}}\approx 3.6$).
For the first {statement}, it stems from the fact that a normalized Gaussian function is used to fit a normalized function which has negative contributions: the integral of the central peak has thus an unnormalized integral larger than $1$, and fitting it by a Gaussian naturally yields a larger corresponding amplitude ($D^{\text{CubicS}}/D^{\text{Gauss}}\approx 0.80$).
This discrepancy will actually be present in all the similar fitting procedure{s} in Sec.~\ref{section-timeevol-Rty-fits} and~\ref{section-temp-dep-asympt-Rbar-sat}.


\section{Set of parameters for the numerical simulations}
\label{A-numericalrecipe}

We have listed thereafter the parameters of the numerical study presented throughout this paper. {These parameters are} defined in Sec.~\ref{section-numerical-recipe}.

The discretization grids for the generation of the random potential and for the recording of the data are fixed once and for all to the following values for all the computations
(the last three parameters are respectively the number of points for the linear grid in $t$, the logarithmic grid in $t$ and the linear grid in $y$ as defined in Fig.~\ref{fig:graph-numrecipe-total}):
\begin{center}
\begin{tabular}{|c|c|c||c|c|c|}
\hline
$D^{\text{grid}}$	&	$\xi_t^{\text{grid}}$	&	$\xi_y^{\text{grid}}$	&	nbptlint	&	nbptlogt	&	nbptliny \\
\hline
$4$	&	$1$	&	$2$	& $80$	&	$80$	&	$100$ \\
\hline
\end{tabular}
\end{center}

$\phantom{bla}$ \\

The following sets of data have been generated, with the elastic constant being fixed at ${c=1}$ and at different temperatures $T$.
The number of configurations `NconfV' per data set have fluctuated as a compromise between the convergence of the disorder average and a reasonable computation time, as already mentioned in Sec.~\ref{section-numerical-recipe}.
The first column gathers the sets used for the study of ${\bar{R}(t,y)}$, $\widetilde{D}_t$ and $\tilde{\xi}_t$, whereas the second column corresponds to the additional sets which explore larger `times' ${t<t_m}$ and have thus been used for the study of the roughness $B(t)$: 

\begin{center}
\begin{tabular}{|c||c||c|c|c||c|c|c|}
\hline
no	&	$T$		&	$t_m$	&	$y_m$ & NconfV	&	$t_m$	&	$y_m$	& NconfV \\
\hline \hline
1	&	$0.35$	&	$40$	&	$50$ &	$572$ 	& 	&	&	\\ 
\hline
2	&	$0.4$	&	$40$	&	$50$ &	$1092$ 	& 	&	&	\\ 
3	&	$0.5$	&	$40$	&	$50$ &	$728$ 	& 	&	&	\\ 
4	&	$0.55$	&	$40$	&	$50$ &	$540$ 	& 	&	&	\\ 
5	&	$0.6$	&	$40$	&	$50$ &	$654$ 	& 	&	&	\\ 
6	&	$0.7$	&	$40$	&	$50$ &	$1281$ 	& 	&	&	\\ 
7	&	$0.8$	&	$40$	&	$50$ &	$2665$ 	& 	&	&	\\ 
8	&	$0.9$	&	$40$	&	$50$ &	$900$ 	& 	&	&	\\ 
\hline
9	&	$1$		&	$40$ 	&	$50$ &	$1350$	& 	&	&	\\ 
\hline
10	&	$1.1$	&	$40$	&	$50$ &	$1040$ 	&	$100$	&	$70$	&	$518$	\\ 
11	&	$1.35$	&	$40$	&	$50$ &	$1050$ 	&	$300$	&	$60$	&	$390$	\\ 
12	&	$1.5$	&	$40$	&	$50$ &	$1435$ 	&	$200$ 	&	$80$	&	$1600$	\\ 
13	&	$1.6$	&	$40$	&	$50$ &	$1600$ 	& 	&	&	\\ 
14	&	$1.8$	&	$40$	&	$50$ &	$1000$ 	&	$800$ 	&	$160$	&	$159$	\\ 
15	&	$2$		&	$40$	&	$50$ &	$1750$ 	& 	&	&	\\ 
16	&	$2.25$	&	$40$	&	$50$ &	$1300$ 	& 	&	&	\\ 
17	&	$2.5$	&	$40$	&	$50$ &	$1900$ 	& 	&	&	\\ 
18	&	$2.75$	&	$40$	&	$50$ &	$1000$ 	& 	&	&	\\ 
19	&	$3$		&	$40$	&	$50$ &	$2300$ 	& 	&	&	\\ 
20	&	$3.2$	&	$40$	&	$50$ &	$1000$ 	& 	&	&	\\ 
21	&	$3.5$	&	$40$	&	$50$ &	$1825$ 	& 	&	&	\\ 
22	&	$4$		&	$40$	&	$50$ &	$1100$ 	& 	&	&	\\ 
\hline
23	&	$6$		&	$40$ 	&	$50$ &	$900$	&	$80$	&	$50$	&	$280$ \\ 
\hline
\end{tabular}
\end{center}


\section{Quantitative test of the DP toymodel: comparison between ${\mathcal{R}^{\text{Gauss}}}$, ${\mathcal{R}^{\text{SincG}}}$ and ${\mathcal{R}^{\text{CubicS}}}$}
\label{A-time-evolution-Rbar-et-al}

{
We have discussed in~Sec.~\ref{section-timeevol-Rty-fits} and~\ref{section-temp-dep-asympt-Rbar-sat}
the `time'- and temperature-dependence of the correlator~${\bar{R}(t,y)}$, from the point of view of our DP toymodel~\eqref{eq-toymodel-def-Rbar-functional}-\eqref{eq-toymodel-def-Rbar-functional-saturation}.
The determination of
the fitting parameters~$\arga{\widetilde{D}_t,\tilde{\xi}_t}$,
their average over large `times'~$\arga{\widetilde{D}_{\infty},\tilde{\xi}_{\infty}}$,
and the fitting parameters~$\arga{\widetilde{D}_{\text{sat}},\tilde{\xi}_{\text{sat}}}$ of the averaged correlator~${\bar{R}_{\text{sat}}(y)}$
depends however on the choice of the fitting function~${\mathcal{R}(y)}$.
We have thus gathered in this appendix the detailed quantitative comparison of the three fitting procedures with respectively $\mathcal{R}^{\text{Gauss}}$, $\mathcal{R}^{\text{SincG}}$ and $\mathcal{R}^{\text{CubicS}}$.
}

\subsection{Effective parameters $\argab{\widetilde{D}_t, \tilde{\xi}_t}$ and their saturation values $\argab{\widetilde{D}_{\infty}, \tilde{\xi}_{\infty}}$ \textit{versus} $\argab{\widetilde{D}_{\text{sat}}, \tilde{\xi}_{\text{sat}}}$}
\label{A-time-evolution-Rbar-et-al-part1}

{In Fig.~\ref{fig:Dtilde-3T}-\ref{fig:xitilde-3T}, we follow the evolution of the two fitting parameters $\lbrace {\tilde{\xi}_t,\widetilde{D}_t \rbrace}$ for the same fixed low- \textit{versus} high-temperatures $T\in \arga{0.35,6}$ as in Fig.~\ref{fig:Rbar-compil-compil},} assuming that the validity of the decomposition ${\bar{R}(t,y) \approx \widetilde{D}_t \cdot \mathcal{R}_{\tilde{\xi}_t}}(y)$ could be extended to smaller `times'.
{Even though the DP toymodel assumption} \textit{a priori} breaks down for ${t<t_{\text{sat}}}$,
the three fits yield consistent values both for $\widetilde{D}_t$ and $\tilde{\xi}_t$, with reasonable uncertainties even at short-`times' (see the error-bars in~\ref{fig:Dtilde-3T}-\ref{fig:xitilde-3T}).
On one hand we recover for $\widetilde{D}_t$ the two regimes observed for the correlator $\bar{R}(t,y)$, \textit{i.e.} an increase of this amplitude at short `times' and a saturation beyond $t_{\text{sat}}$ as indicated by the straight lines $\widetilde{D}_{\infty}$ in Fig.~\ref{fig:Dtilde-3T} (obtained by averaging $\widetilde{D}_t$ over ${t>t_{\text{min}}=25}$).
On the other hand $\tilde{\xi}_t$ slightly decreases at short `times' and its saturation seems to appear much sooner, especially at high-$T$.

As a self-consistency check, we can notice that the average of the fitting parameters $\arga{\tilde{\xi}_{\infty},\widetilde{D}_{\infty}}$ have the same values as the fitting parameters of the averaged correlator $\arga{\tilde{\xi}_{\text{sat}},\widetilde{D}_{\text{sat}}}$ {listed in~Tab.~\ref{tab:Dtilde-xitilde-sat-3T}.
Their} estimated errors are actually surprisingly close even though they have distinct origins, stemming respectively from the variance of $\arga{\tilde{\xi}_t,\widetilde{D}_t}$ at ${t>t_{\text{min}}}$ and from the uncertainty over the fit of $\bar{R}_{\text{sat}}(y)$.

\begin{table}
\begin{center}
\begin{tabular}{|c||c|c|c|c}
\hline
$(t > t_{\text{min}}=25)$ & $T=0.35$ & $T=1$ & $T=6$ \\
\hline \hline
$\widetilde{D}_{\text{sat}}^{\text{Gauss}}$ & ${4.6 \pm 0.2}$ & ${4.3 \pm 0.2}$ & ${1.13 \pm 0.04}$ \\
\hline
$\widetilde{D}_{\text{sat}}^{\text{SincG}}$ & ${4.4 \pm 0.2}$ & ${3.9 \pm 0.2}$ & ${0.99 \pm 0.03}$ \\
\hline
$\widetilde{D}_{\text{sat}}^{\text{CubicS}}$ & ${4.37 \pm 0.07}$ & ${3.89 \pm 0.05}$ & ${0.98 \pm 0.06}$\\
\hline
$\widetilde{D}_{\text{sat}}^{\text{SincG}}/\widetilde{D}_{\text{sat}}^{\text{Gauss}}$ & $0.95$ & $0.91$ & $0.87$ \\
\hline \hline
$\tilde{\xi}_{\text{sat}}^{\text{Gauss}}$ & ${0.61 \pm 0.04}$ & ${0.70 \pm 0.04}$ & ${0.92 \pm 0.04}$ \\
\hline
$\tilde{\xi}_{\text{sat}}^{\text{SincG}}$ & ${2.25 \pm 0.07}$ & ${2.42 \pm 0.05}$ & ${2.99 \pm 0.06}$ \\
\hline
$\tilde{\xi}_{\text{sat}}^{\text{CubicS}}$ & ${2.25 \pm 0.07}$ & ${2.40 \pm 0.05}$ & ${2.97 \pm 0.06}$\\
\hline
$\tilde{\xi}_{\text{sat}}^{\text{SincG}}/\tilde{\xi}_{\text{sat}}^{\text{Gauss}}$ & $3.72$ & $3.45$ & $3.25$ \\
\hline
\end{tabular}
\end{center}
\caption{
 Fitting parameters $\arga{\tilde{\xi}_{\text{sat}},\widetilde{D}_{\text{sat}}}$ at ${T \in \arga{0.35,1,6}}$, computed from the saturation correlators $\bar{R}_{\text{sat}}(y)$ plotted in Fig.~\ref{fig:Rbar-compil-compil} for ${t_{\text{min}}=25}$.
 {These values are almost identical to the average of the fitting parameters~$\arga{\tilde{\xi}_{\infty},\widetilde{D}_{\infty}}$ over the same `time' range~${t>t_{\text{min}}}$.}
 \label{tab:Dtilde-xitilde-sat-3T}}
\end{table}

\begin{figure}
 \subfigure{\includegraphics[width=\columnwidth]{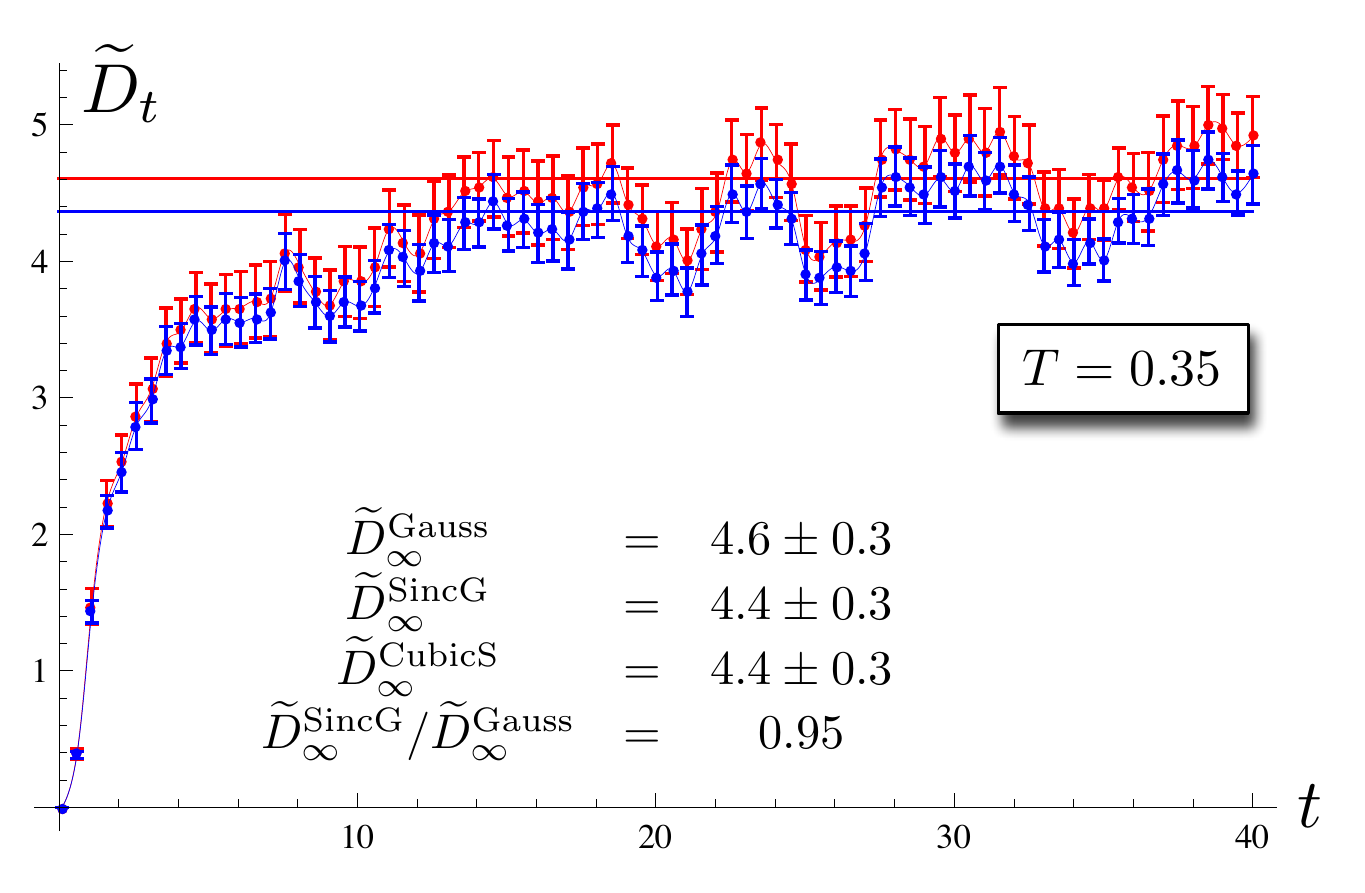} \label{fig:Dtilde-lowT}}
 \subfigure{\includegraphics[width=\columnwidth]{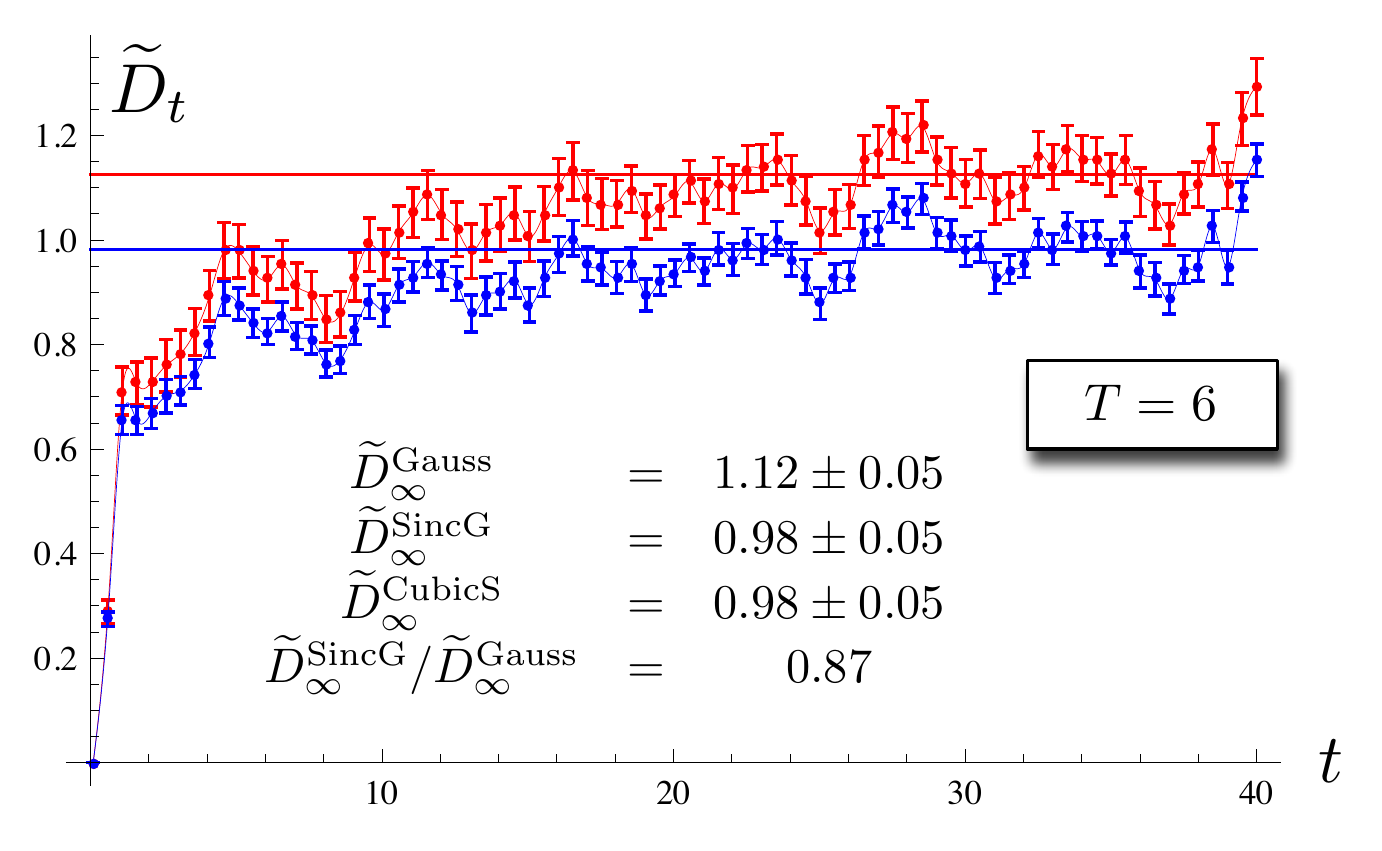} \label{fig:Dtilde-highT}}
 \caption{
 (Color online)
 Fitting parameter $\widetilde{D}_t$  at fixed {low- \textit{versus} high-}temperature ${T \in \arga{0.35,6}}$ with respect to `Gauss' (red) and `SincG' (blue), `CubicS' collapsing exactly on `SincG'.
 The straight lines indicate the respective average value $\widetilde{D}_{\infty}$ for ${t>t_{\text{min}}=25}$, whose standard deviation is given explicitly for each fit.
 \label{fig:Dtilde-3T}
 }
\end{figure}
\begin{figure}
 \subfigure{\includegraphics[width=\columnwidth]{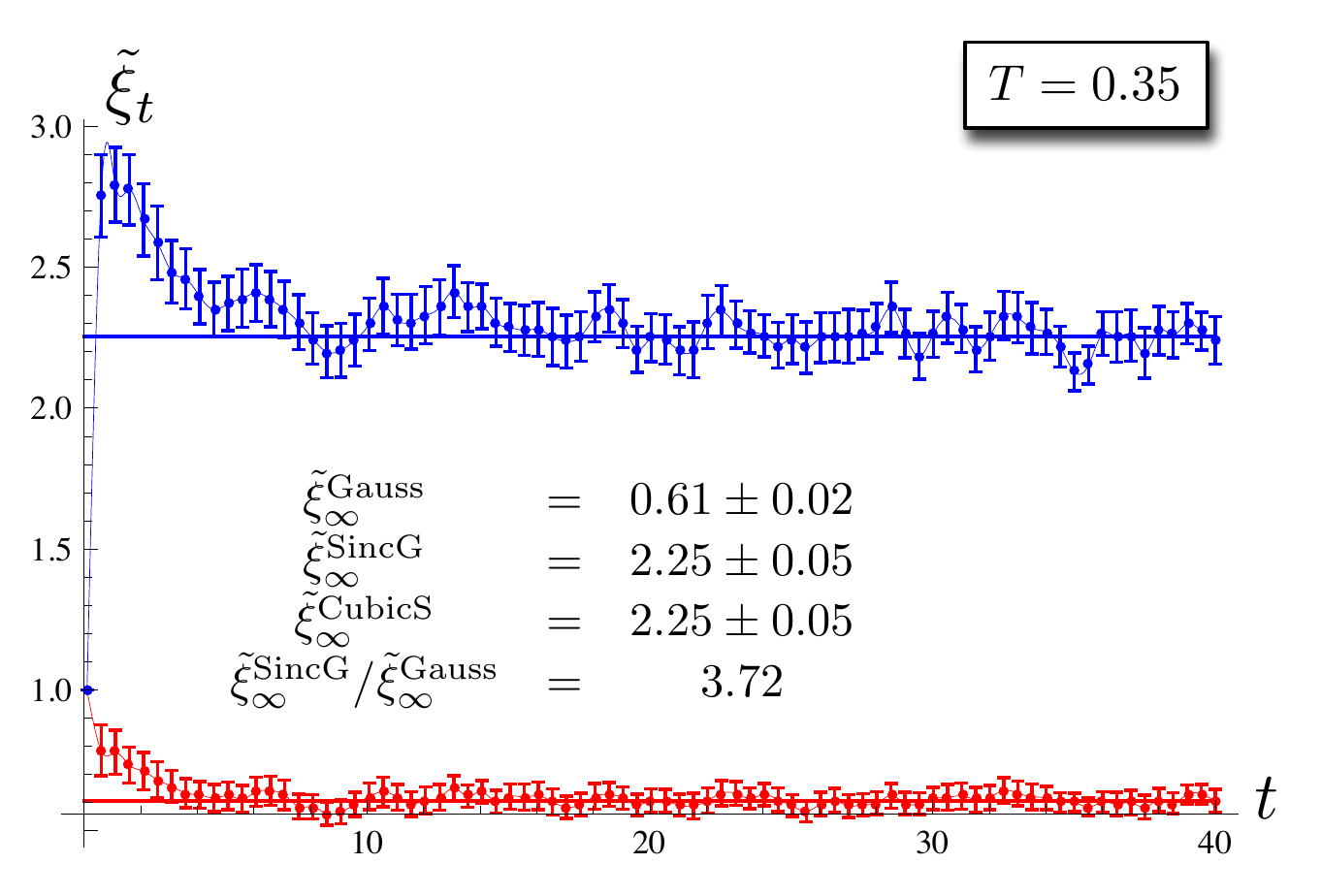} \label{fig:xitilde-lowT}}
 \subfigure{\includegraphics[width=\columnwidth]{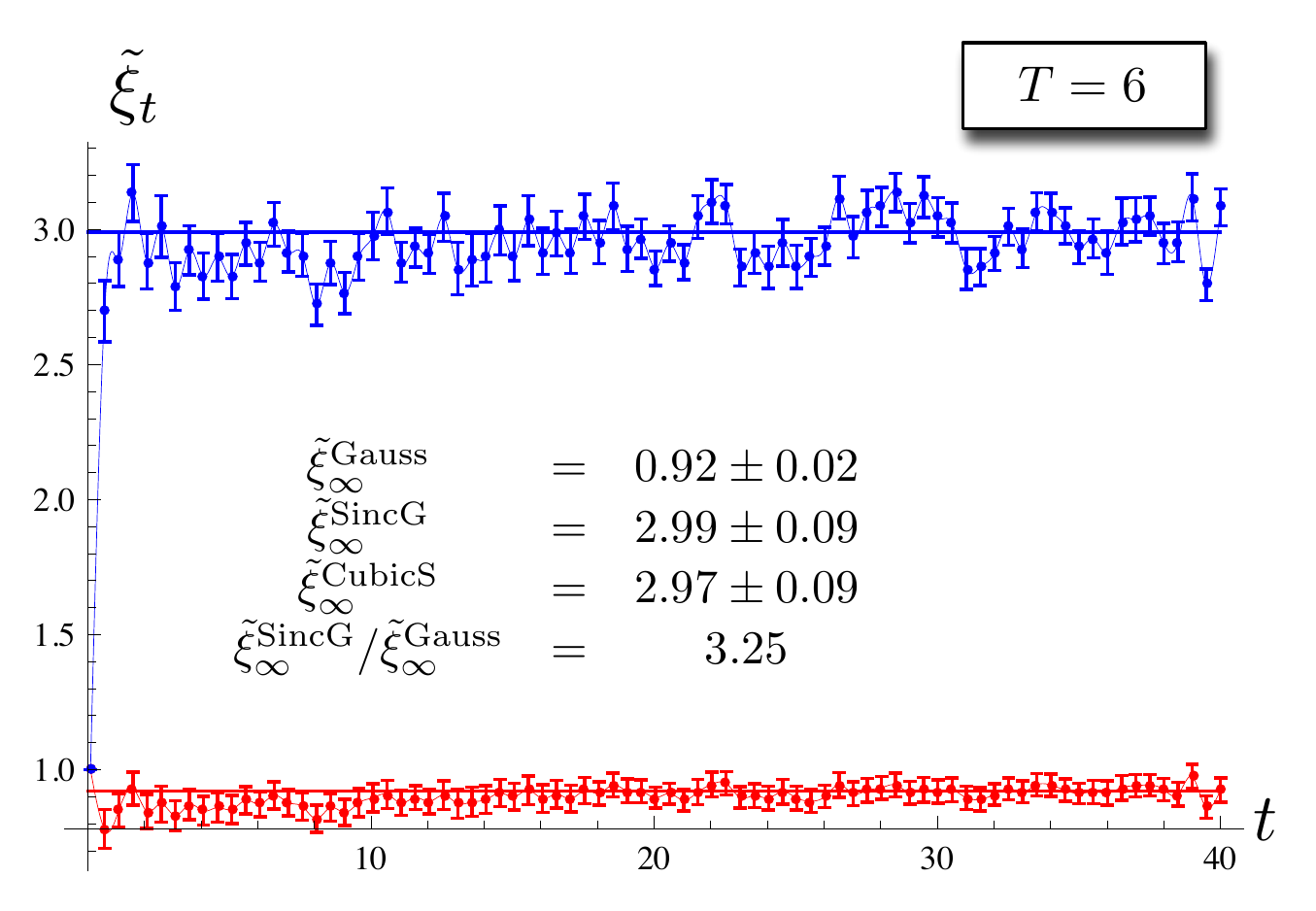} \label{fig:xitilde-highT}}
 \caption{
 (Color online)
 Fitting parameter $\tilde{\xi}_t$  at fixed {low- \textit{versus} high-}temperature ${T \in \arga{0.35,6}}$, computed conjointly to $\widetilde{D}_t$ of Fig.~\ref{fig:Dtilde-3T} and following the same convention regarding the legend.
 \label{fig:xitilde-3T}
 }
\end{figure}


If the function $\mathcal{R}_{\tilde{\xi}_t}(y)$ did coincide exactly with the microscopic disorder correlator ${R^{\text{CubicS}}_{\xi}(y)}$ at small $y$ and large `times', we would expect {to have the following geometrical ratios}
${\frac{\widetilde{D}_{\infty}^{\text{SincG}}}{\widetilde{D}_{\infty}^{\text{Gauss}}} = \frac{D^{\text{SincG}}}{D^{\text{Gauss}}} \approx 0.80}$
and
${\frac{\tilde{\xi}_{\infty}^{\text{SincG}}}{\tilde{\xi}_{\infty}^{\text{Gauss}}} = \frac{\xi^{\text{SincG}}}{\xi^{\text{Gauss}}} \approx 3.6}$
as discussed in Appendix~\ref{A-effectiveCorrelatorSpline}.
{Despite the temperature-dependence of $\arga{\tilde{\xi}_{\infty},\widetilde{D}_{\infty}}$ discussed in~Sec.~\ref{section-temp-dep-asympt-Rbar-sat}},
these ratios are numerically found to be of the expected order of magnitude and reasonably constant for {different values of~$t_{\text{min}}$} at fixed $T$ (cf.~Tab.~\ref{tab:Dtilde-xitilde-sat-3T}).
{However, as discussed in~Sec.~\ref{section-timeevol-Rty-fits}, the average procedure} over ${t>t_{\text{min}}}$ introduces an overall negative shift of the {large-$y$ asymptote of~${\bar{R}_{\text{sat}}(y)}$ and this artifact will inevitably alter the geometrical ratios.
}

{
A quantitative benchmark is nevertheless available for the comparison between~$\mathcal{R}^{\text{Gauss}}$, $\mathcal{R}^{\text{SincG}}$ and~$\mathcal{R}^{\text{CubicS}}$.
It consists in applying our whole fitting procedure on the \emph{linearized} correlator ${\bar{R}^{\text{lin}}(t,y)}$ \eqref{eq-decomposition-Rbarlin-1} corresponding to our microscopic disorder correlator~${R_\xi(y)=R_{\xi}^{\text{CubicS}} (y)}$ [depicted in Fig.~\ref{fig:finitetimescaling-Cbarlin-Rbarlin_cubicS}].
We refer the reader to our analytical paper~\cite{agoritsas_2012_FHHtri} for the exact computation of ${{\bar R}^{\text{lin}}(t,y)}$.
We thus determine the correlator~${{\bar R}^{\text{lin}}(t,y)}$ at $\arga{\xi=2, c=D=T=1}$, and we determine its fitting parameters $\arga{\widetilde{D}_t,\tilde{\xi}_t}$ using the `Gauss' and the `SincG' fitting functions.
Averaging it over a set of `times'~${t>t_{\text{min}}}$, we then extract on one hand the average of its parameters~$\arga{\widetilde{D}_\infty,\tilde{\xi}_\infty}$, and on the hand its averaged correlator~${\bar{R}^{\text{lin}}_{\text{sat}}(y)}$ of parameters~$\arga{\widetilde{D}_{\text{sat}},\tilde{\xi}_{\text{sat}}}$.
We obtain for~${T=1}$ the following geometrical ratios:
${\frac{\widetilde{D}_{\infty}^{\text{SincG}}}{\widetilde{D}_{\infty}^{\text{Gauss}}} = \frac{\widetilde{D}_{\text{sat}}^{\text{SincG}}}{\widetilde{D}_{\text{sat}}^{\text{Gauss}}} \approx 0.89}$
and
${\frac{\tilde{\xi}_{\infty}^{\text{SincG}}}{\tilde{\xi}_{\infty}^{\text{Gauss}}} = \frac{\tilde{\xi}_{\text{sat}}^{\text{SincG}}}{\tilde{\xi}_{\text{sat}}^{\text{Gauss}}} \approx 3.27}$.
These ratios are in a surprisingly good \emph{quantitative} agreement with the high-$T$ case given in~Tab.~\ref{tab:Dtilde-xitilde-sat-3T}.
This agreement supports the idea that at high-$T$ we recover the microscopic disorder correlator ${\bar{R}(\infty,y) \approx \frac{cD}{T} \cdot \mathcal{R}_{\tilde{\xi}_{\infty}}^{\text{CubicS}}(y)}$, taking in account the artifact of the large-$y$ asymptote in the determination of the fitting parameters.
Finally we have checked that these ratios depend only slightly on the value of the temperature, decreasing for both~$\widetilde{D}$ and~$\tilde{\xi}$ when $T$~increases (same trend as in~Tab.~\ref{tab:Dtilde-xitilde-sat-3T}).
}


\subsection{Evolution of the peak~${\bar{R}(t,0) \sim \widetilde{D}_t/\tilde{\xi}_t}$}
\label{A-time-evolution-Rbar-et-al-part2}

{The short-`times' behavior of the fitting parameters~$\arga{\widetilde{D}_t,\tilde{\xi}_t}$} can have either a deep physical meaning or be an artifact due to the inadequacy of the fitting function $\mathcal{R}(y)$ for the measured correlator $\bar{R}(t,y)$.
{Note for instance that} the short-`times' increase of $\widetilde{D}_t$ is compatible with the analytical predictions discussed in~\cite{agoritsas_2012_FHHtri}.

{An additional characterization of the `time'-evolution of the correlator's shape~${\mathcal{R}(y)}$ consists in comparing the evolution of the peak $\bar{R}(t,0)$ deduced from the three fits Gauss-SincG-CubicS (combining ${\mathcal{R}_{\tilde{\xi}=1}^{\text{fit}}(0)}$, ${\widetilde{D}^{\text{fit}}_t}$ and ${\tilde{\xi}^{\text{fit}}_t}$) to the value measured numerically.
This provides a consistency check, on the whole `time'-range available, that there is a temperature-crossover in the shape of the correlator.
}

In~Fig.\ref{fig:peakMeas-3T} (\textit{top}) we see that at low-$T$ the peak is correctly captured by `Gauss',
and increasing the temperature the relative difference between the fits decreases.
However, zooming on the high-$T$ (\textit{bottom left}) we see that `CubicS' and `SincG' catches more precisely the peak.
Comparing the relative differences between the fits as a function of $T$ (\textit{bottom right}),
two temperature regimes can be postulated regarding the normalized function $\mathcal{R}(y)$:
at high-$T$ we have ${\mathcal{R}(y) \approx \mathcal{R}^{\text{CubicS}}(y)}$ with the relative behavior between the fits as expected from Appendix~\ref{A-effectiveCorrelatorSpline} (geometrical collapse of ${\tilde{\xi}^{\text{SincG}}_{\infty}/\tilde{\xi}^{\text{Gauss}}_{\infty} \approx \xi^{\text{SincG}} /\xi^{\text{Gauss}}}$, overestimation of the amplitude and thus of the peak by `Gauss');
at low-$T$ the modification of the function ${\mathcal{R}(y)}$ is essentially pushed into the increasing discrepancy  ${\Delta \tilde{\xi}_{\infty}/\tilde{\xi}_{\infty}}$.
{Let's emphasize that the zero-temperature limit of~${\bar{R}(\infty,y)}$, and hence its normalized shape~${\mathcal{R}(y)}$, are not known analytically.}

\begin{figure}
 \subfigure{\includegraphics[width=\columnwidth]{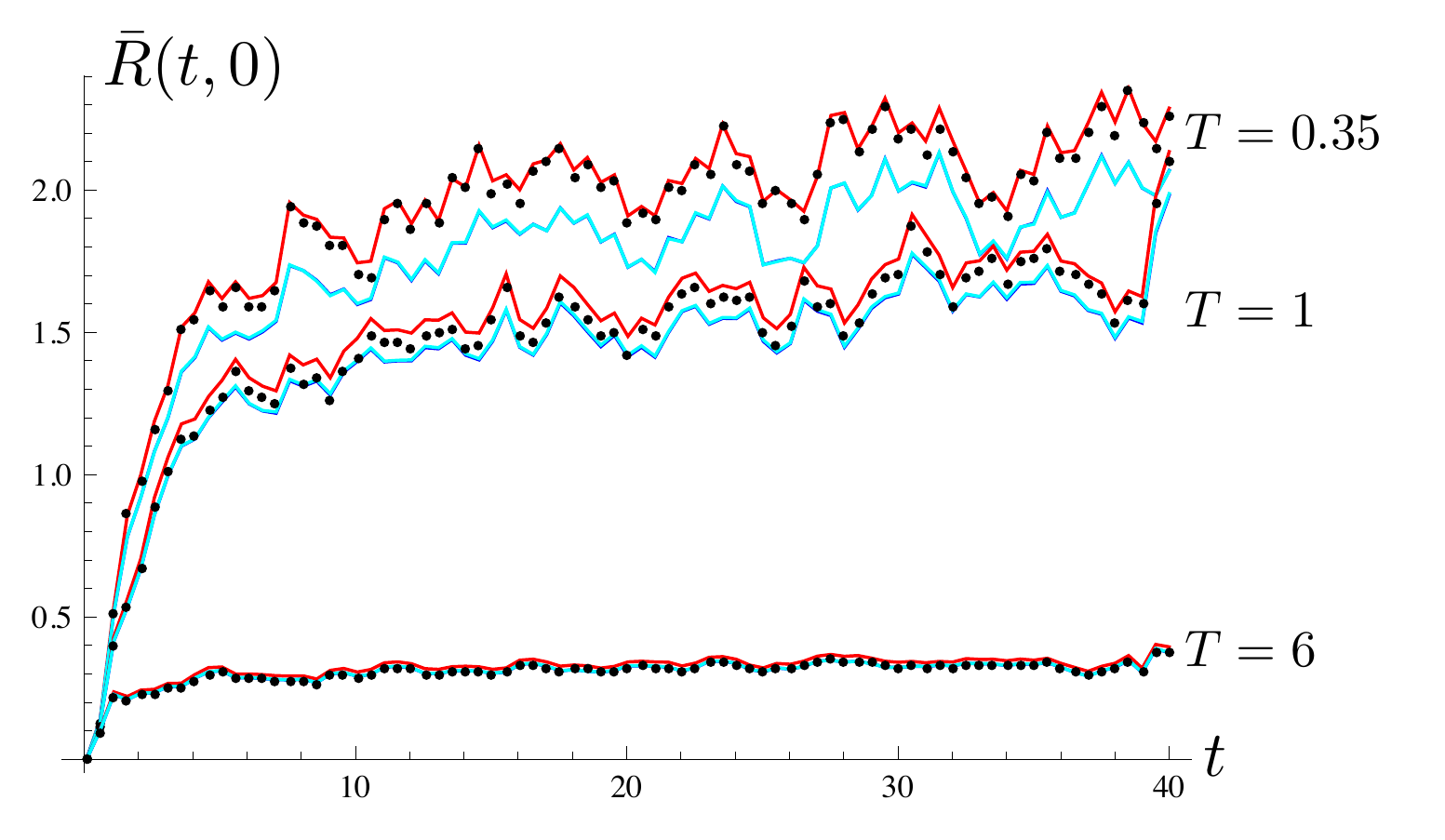} \label{fig:peak-mesGaussSincGCubicS-3T}}
  \subfigure{\includegraphics[width=\columnwidth]{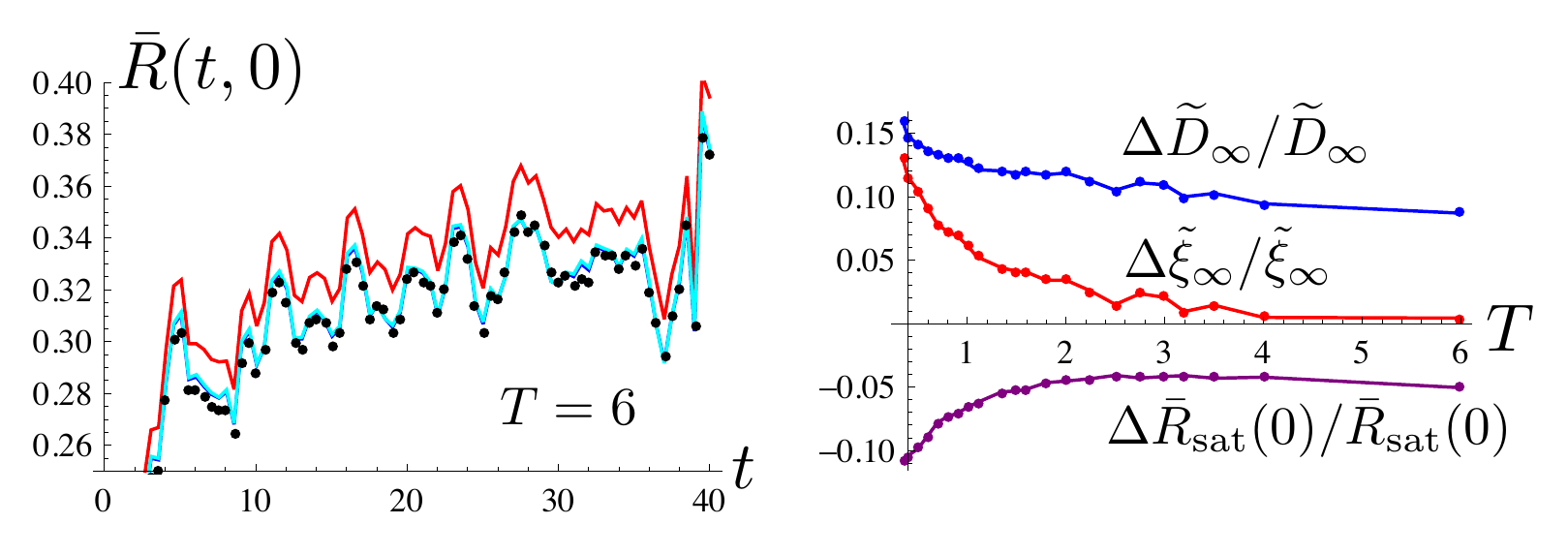} \label{fig:peak-mesGaussSincGCubicS-3T-bis}}
 \caption{
 (Color online)
 \textit{Top}:~`Time'-evolution of the maximum of the correlator peak ${\bar{R}(t,0)}$ at fixed temperature ${T \in \arga{0.35,1,6}}$, measured numerically (black dots) or predicted by the fitting functions via
 ${\bar{R}^{\text{fit}}(t,0)=\mathcal{R}_{\tilde{\xi}=1}^{\text{fit}}(0) \cdot \widetilde{D}_t^{\text{fit}}  / \tilde{\xi}_t^{\text{fit}}}$
 (red for `Gauss' \eqref{eq-def-RGauss}, blue for `SincG' \eqref{eq-def-RSincG}, cyan for `CubicS' \eqref{eq-def-RCubicS-1}-\eqref{eq-def-RCubicS-2}). The curves for `SincG' and `CubicS' systematically coincide.
 \textit{Bottom left}:~Zoom on the high-$T$ case, where the `SincG' and `CubicS' predict more accurately the peak at high-$T$, whereas at low-$T$ `Gauss' seems more suited.
 \textit{Bottom right}:~Decrease in the relative differences between the `Gauss' and `SincG' parameters with increasing temperature, averaged on the large-`time' window ${t \in \argc{25,40}}$:
${\frac{\Delta \widetilde{D}_{\infty}}{\widetilde{D}_{\infty}} = \frac{\widetilde{D}_{\infty}^{\text{SincG}} - \widetilde{D}_{\infty}^{\text{Gauss}}}{\widetilde{D}_{\infty}^{\text{SincG}}} }$,
${\frac{\Delta \tilde{\xi}_{\infty}}{\tilde{\xi}_{\infty}} = \frac{\tilde{\xi}_{\infty}^{\text{SincG}} - \tilde{\xi}_{\infty}^{\text{Gauss}} \xi^{\text{SincG}}/\xi^{\text{Gauss}}}{\tilde{\xi}_{\infty}^{\text{SincG}}}}$
 (see end of~Appendix~\ref{A-effectiveCorrelatorSpline})
 and ${\frac{\Delta \bar{R}_{\text{sat}}(0)}{\bar{R}_{\text{sat}}} = \frac{\bar{R}_{\text{sat}}^{\text{SincG}}(0) - \bar{R}^{\text{Gauss}}_{\text{sat}}(0)}{\bar{R}^{\text{SincG}}_{\text{sat}}(0)} }$.
 \label{fig:peakMeas-3T} 
}
\end{figure}


\subsection{Temperature-dependence of $\widetilde{D}_{\infty}$ and $\tilde{\xi}_{\infty}$}
\label{A-time-evolution-Rbar-et-al-part3}

{We conclude the quantitative comparison of the three fitting procedures ${\mathcal{R}^{\text{Gauss}}}$, ${\mathcal{R}^{\text{SincG}}}$ and ${\mathcal{R}^{\text{CubicS}}}$ by discussing the temperature-dependence of the asymptotic amplitude~$\widetilde{D}_{\infty}$ and typical spread~$\tilde{\xi}_{\infty}$.}

The measured amplitude ${\widetilde{D}_{\infty}}$ as a function of $T$ is reported in~Fig.~\ref{fig:Dtilde-inf-allT-compil} for both `Gauss' and `SincG' (which collapses with `CubicS') fits.
Qualitatively it decreases in $1/T$ at high-$T$ and saturates at low-$T$, {as detailed in~Ref.~\cite{agoritsas_2012_FHHtri} according to the scaling arguments and GVM predictions of the full model and the DP toymodel.}
A parametrization of this temperature crossover has been defined in the relation ${\widetilde{D}_{\infty}(T,\xi)=f(T,\xi) \frac{cD}{T}}$ \eqref{eq-Dtilde-infty-finterp} with the interpolating parameter $f(T,\xi)$.
Quantitatively its $\xi=0$ limit requires at high-$T$ {to have} ${f \lesssim 1}$ without any additional numerical prefactor,
but extracting the strength of disorder $D$ from the $1/T$ behavior of $\widetilde{D}_{\infty}$ actually yields a systematic underestimation with respect to the microscopic disorder:
${D^{\text{Gauss}}=6.72 \pm 0.08}$,
${D^{\text{SincG}}=5.92 \pm 0.05}$ and
${D^{\text{CubicS}}=5.89 \pm 0.05}$ (obtained on the three larger available temperatures),
whereas ${D=8}$ \eqref{eq-fixed-parameters-microscopic disorder}.
{As in~Sec.~\ref{A-time-evolution-Rbar-et-al-part1}, we} attribute again this discrepancy to the negative excursions of ${\bar{R}(t,y)}$ at large $y$ {(the contribution~${b(t,y)}$ in~\eqref{eq-infty-time-Rxi-bumps-1}}) which bias all the fits and preclude a quantitative test of ${f(T,\xi)}$ with respect to the GVM prediction {(recalled after~\eqref{eq-Dtilde-infty-finterp} and discussed in~\cite{agoritsas_2012_FHHtri})}.

Finally the typical spread ${\tilde{\xi}_{\infty}}$ obtained in parallel to ${\widetilde{D}_{\infty}}$ is reported in~Fig.~\ref{fig:xitilde-inf-allT-compil}.
The collapse of the different fits for ${\tilde{\xi}_{\infty}}$ has already been discussed {in~Sec.~\ref{A-time-evolution-Rbar-et-al-part1}};
it is compatible with the scenario of a high-$T$ correlator ${\mathcal{R}_{\text{sat}}(y)\approx \mathcal{R}^{\text{CubicS}}(y)}$.
However it displays a temperature-dependence that \textit{a priori} corrects to first-order the minimal assumption that ${\tilde{\xi}_{\infty} \approx \xi}$ in our DP toymodel, since the thermal fluctuations seem to increase the effective ${\tilde{\xi}_{\infty}}$ compared to the microscopic disorder correlation length ${\xi^{\text{grid}}=2}$ (which consistently remains a lower bound in our measured $\tilde{\xi}_{\infty}^{\text{CubicS}}$).
We know from the linearized solution \eqref{eq-decomposition-Rbarlin-1} that neglecting the KPZ nonlinearity we recover asymptotically ${\frac{cD}{T} \mathcal{R}^{\text{CubicS}}_{\xi}(y)}$ for the correlator,
so any modification of ${\tilde{\xi}_{\infty}}$ in this two-point correlator can only stem from the KPZ nonlinearity at high-$T$.
A numerical artifact similar to the underestimation of $\widetilde{D}_{\infty}$ is not to be excluded, but no more conclusions can be drawn from our numerical results.

\begin{figure}
 \includegraphics[width=\columnwidth]{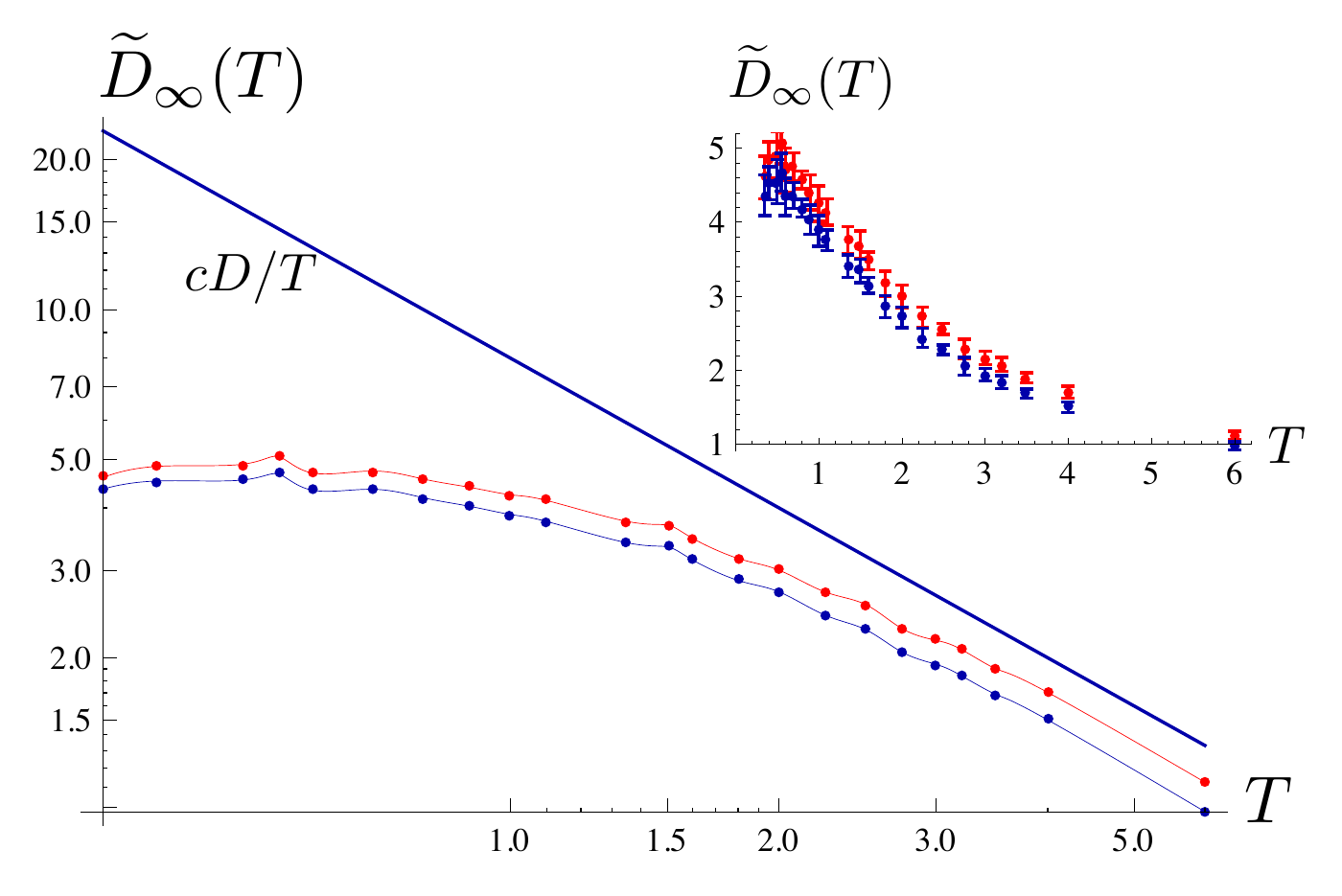}
 \caption{
 (Color online)
 Temperature-dependence of $\widetilde{D}_{\infty}$ with respect to `Gauss' (red, \textit{top}) and `SincG' (blue, \textit{bottom}).
 The straight line $cD/T$ indicates the expected behavior in the high-$T$ regime (${cD=8}$ for our data).
 \textit{Inset}:~Corresponding mean and standard deviation for the average of $\widetilde{D}_t$ over ${t>t_{\text{min}}=25}$.
 \label{fig:Dtilde-inf-allT-compil}
 }
\end{figure}

\begin{figure}
 \begin{center}
 \subfigure{\includegraphics[width=\columnwidth]{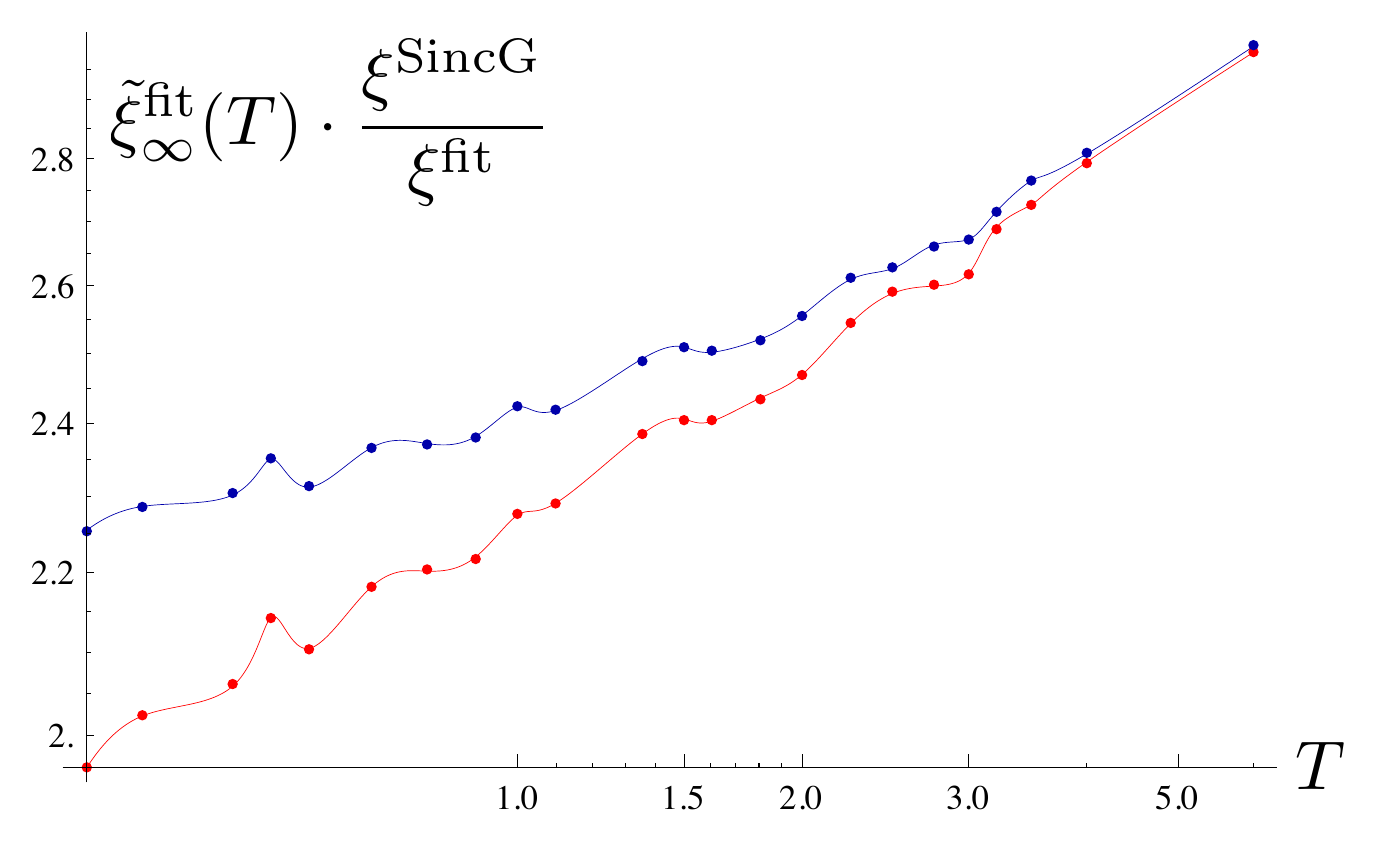}  \label{fig:xitilde-inf-allT-compil-A}}
 \subfigure{\includegraphics[width=\columnwidth]{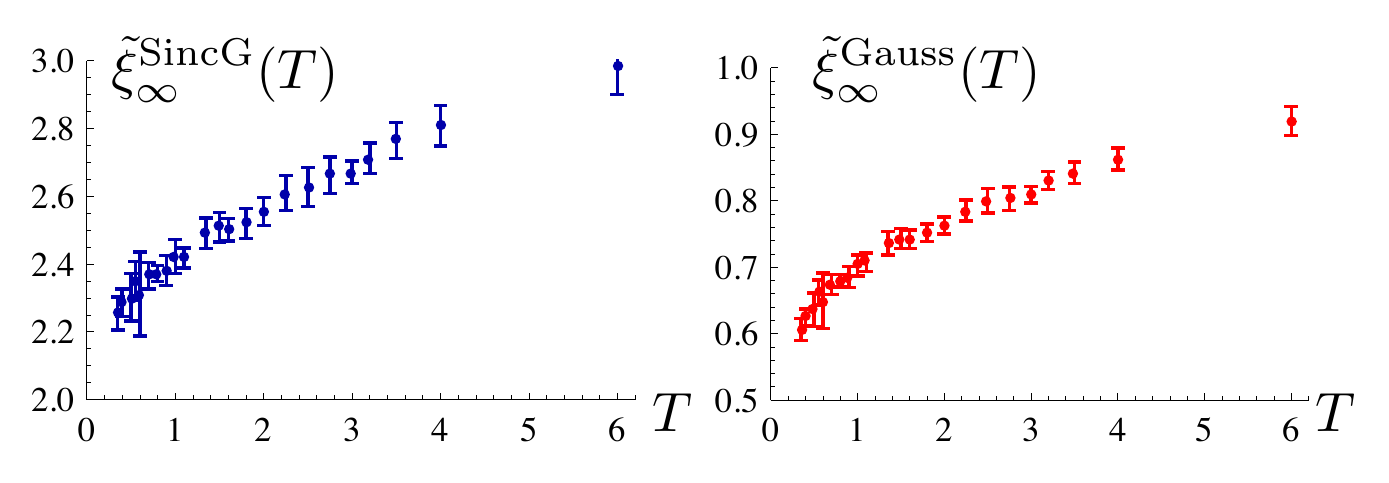}  \label{fig:xitilde-inf-allT-compil-B}}
 \end{center}
 \caption{
 (Color online)
 \textit{Top}:~Temperature-dependence of $\tilde{\xi}_{\infty}$ with respect to `Gauss' (red, \textit{bottom}) and `SincG' (blue, \textit{top}), normalized for `Gauss' with ${\xi^{\text{SincG}}/\xi^{\text{Gauss}} \approx 3.6}$ (cf.~Appendix~\ref{A-effectiveCorrelatorSpline}).
 \textit{Bottom}:~Corresponding mean and standard deviation for the average of $\widetilde{\xi}_t$ over ${t>t_{\text{min}}=25}$.
 \label{fig:xitilde-inf-allT-compil}
 }
\end{figure}


\section{Fluctuations of $\overline{\bar{F}_V(t,y)}$}
\label{A-section-fluctuation-MeanMeanFbar}

{We present in the appendix a study of the \emph{mean value} of the disorder free-energy. Its `time'-dependence yields an alternative measurement of the \emph{two-point correlator maximum}~${\bar{R}(t,0)}$. It thus provides an independent measurement of ${\bar{R}_{\text{sat}}(0) \sim \widetilde{D}_{\infty}/\tilde{\xi}_{\infty}}$, incorporated in~Fig.~\ref{fig:MeanMeanFbar-Peak-wrt-T} and discussed in~Sec.~\ref{section-temp-dep-asympt-Rbar-sat}.}

In its initial definition
the disorder free-energy ${\bar{F}_V(t,y)}$ is defined up to a constant $\text{cte}_V(t)$ depending of the chosen path-integral normalization of ${W_V(t,y)}$ in \eqref{eq-def-unnorm-Boltzmann-Wv}.
As far as statistical averages with quenched disorder are concerned, such as the roughness ${B(t)}$, this constant is irrelevant and thus usually completely skipped.
The Feynman-Kac equation \eqref{eq-FeynmanKac-FbarV} actually yields a univocal definition of $\bar{F}_V(t,y)$ which satisfies the STS \eqref{eq-STS-PFbarV}.
In our numerical approach ${\overline{\bar{F}_V(t,y)} = \text{cte}(t)}$ is in fact a tractable quantity which provides an independent way to measure ${\bar{R}_{\text{sat}}(y=0)}$ from its large-`times' linear evolution.
This is remarkable and a direct consequence of the KPZ nonlinearity, since linearizing \eqref{eq-FeynmanKac-FbarV} trivially predicts ${\overline{\bar{F}_V^{\text{lin}}(t,y)} = 0}$, while averaging {over disorder the tilted KPZ equation \eqref{eq-FeynmanKac-FbarV} yields on the contrary}
\begin{eqnarray}
 & \partial_t \overline{\bar{F}_V (t,y)}
 = -\frac{1}{2c} \overline{\argc{\partial_y \bar{F}(t,y)}^2}
 = -\frac{1}{2c} \cdot \bar{R}(t,0) &
 \label{eq-mean-FbarV}
\end{eqnarray}
{So the non-linearity induces a large-`time' behavior of $\overline{\bar{F}_V (t,y)}$ which must be affine in~$t$ if we assume that the free-energy fluctuations saturate for `times'~${t>t_{\text{sat}}}$ as in~\eqref{eq-toymodel-def-Rbar-functional-saturation}, we expect at ${t>t_{\text{sat}}}$:
\begin{equation}
 -2c\overline{\bar{F}_V (t,y)} = \int_0^{t_{\text{sat}}} dt' \cdot \bar{R}(t',0) + (t-t_{\text{sat}})\cdot \bar{R}_{\text{sat}}(0)
  \label{eq-toymodel-def-Rbar-functional-saturation-forMeanFbarV}
\end{equation}
This prediction provides an additional graphical estimation of~${t_{\text{sat}}}$ from the breakdown of the affine behavior of~$\overline{\bar{F}_V (t,y)}$.
}

In~Fig.~\ref{fig:MeanMeanFbar-3T-compil} we have plotted the disorder average of ${\bar{F}_V(t,y)}$ at {the same fixed temperature ${T \in \arga{0.35,1,6}}$ as in~Fig.~\ref{fig:Rbar-compil-compil}},
keeping first the spatial resolution in ${(t,y)}$ (\textit{top}) and then averaging over the $y$-direction (\textit{bottom}).
${\overline{\bar{F}_V(t,y)}}$ should be $y$-independent, but at short-`time' and low-$T$ it displays a slight curvature that we attribute to the artificial thermal condition at ${t_0=0.1}$.
The resulting standard-deviation is strongly reduced at higher $T$ where this initial condition is more accurate with respect to thermal fluctuations.

At large `times' ${\overline{\bar{F}_V(t,y)}}$ follows a robust linear behavior which extends down to  {${t_{\text{sat}} \lesssim 10}$}, as emphasized in~Fig.~\ref{fig:MeanMeanFbar-3T-compil}. 
Below $t_{\text{sat}}$ no clear powerlaw could be identified {although} the logarithmic scale in~Fig.~\ref{fig:MeanMeanFbar-3T-compil} makes explicit a superlinear short-`time' behavior, \textit{a priori} conditioned by the initial thermal condition and not incompatible with the analytical prediction~$\sim t^2$ in~Ref.~\cite{agoritsas_2012_FHHtri}.
{The temperature-dependence of the averaged slope ${-2c \, \partial_t \overline{\bar{F}_V(t,y)}}$ yields a independent measurement of~${\bar{R}_{\text{sat}}(0)}$ and as such has been included in~Fig.~\ref{fig:MeanMeanFbar-Peak-wrt-T} in~Sec.~\ref{section-temp-dep-asympt-Rbar-sat}.}
We find an excellent agreement between these quantities, with a slight crossover of the direct measurement from the `Gauss' to the `SincG' fits, and a systematic overestimation of $-2c \, \partial_t \overline{\bar{F}_V(t,y)}$; this last point is an artifact of the short-`times' curvature in~$y$ of $\overline{\bar{F}_V(t,y)}$.

\begin{figure}
 \subfigure{\includegraphics[width=\columnwidth]{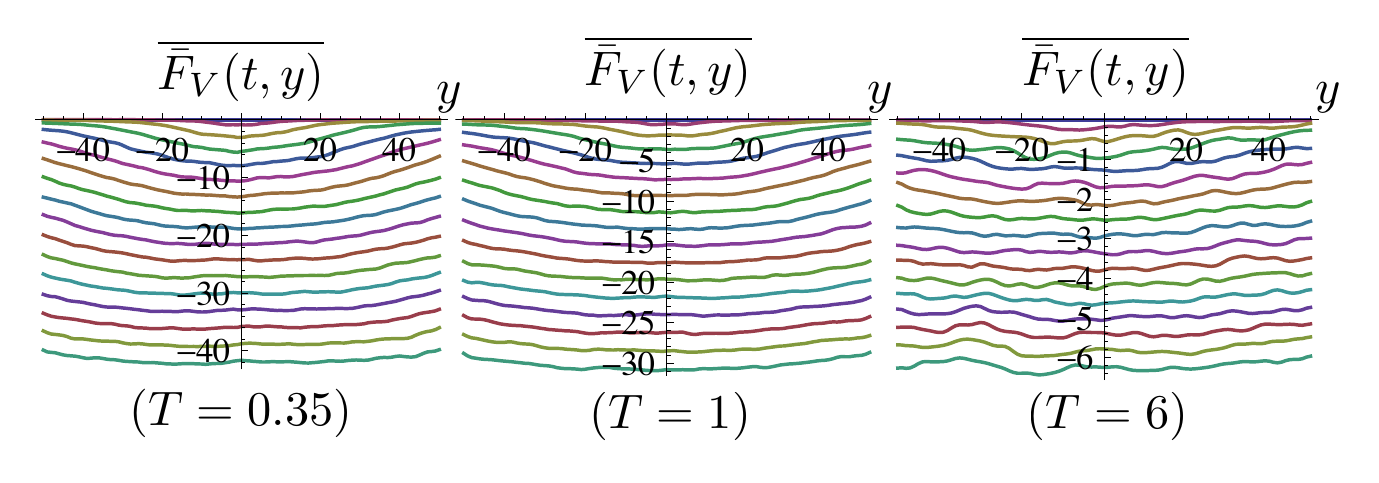}  \label{fig:MeanFbar-3T}}
 \subfigure{\includegraphics[width=\columnwidth]{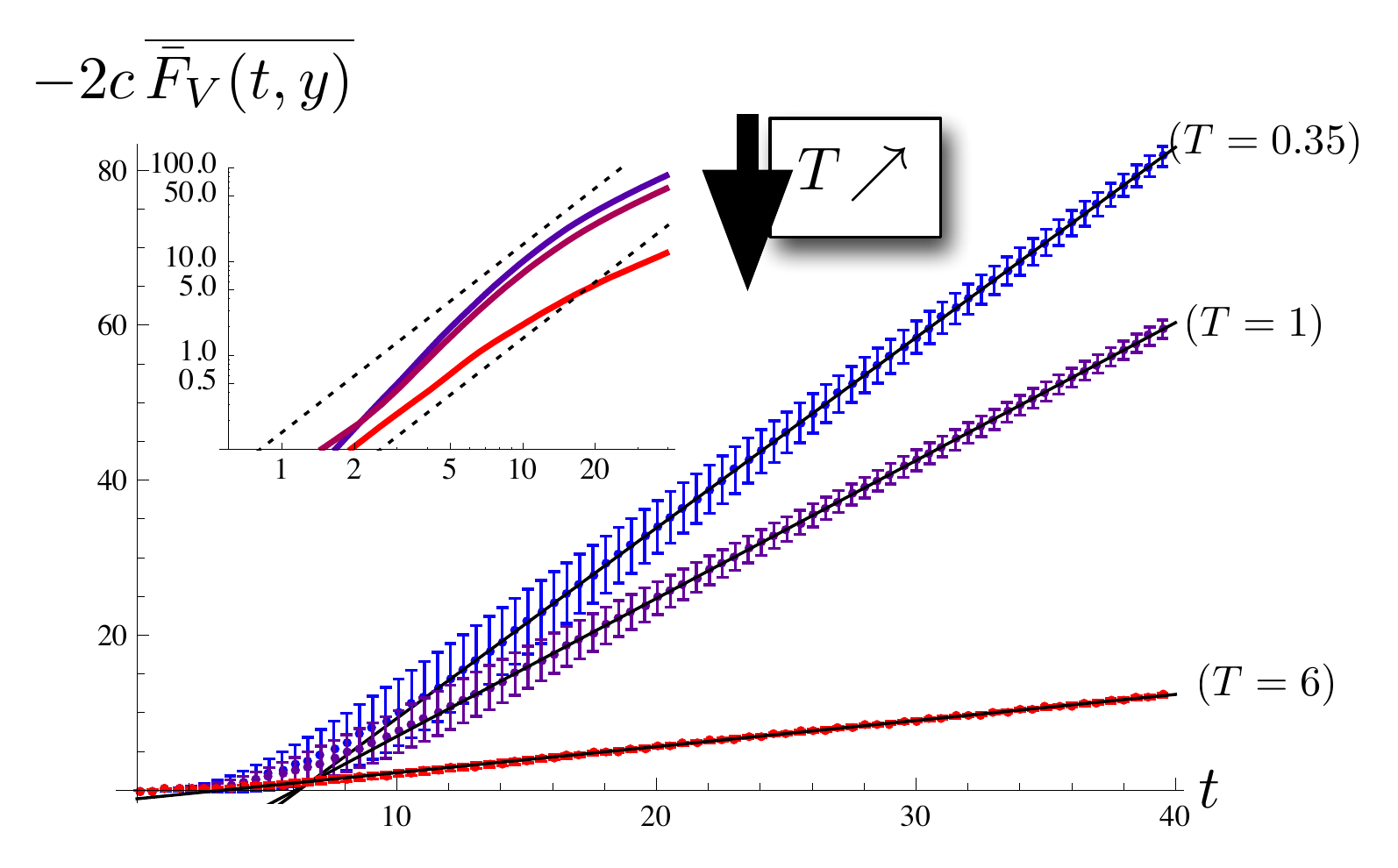}  \label{fig:MeanMeanFbar-Errorbar-fit-3T}}
 \caption{
 (Color online)
 Disorder average of ${\bar{F}_V(t,y)}$ at fixed temperature ${T \in \arga{0.35,1,6}}$.
 \textit{Top}:~${\overline{\bar{F}_V(t,y)}}$ as a function of $y$, for increasing `times' ${t \in \argc{0.1,40}}$ with `time'-steps ${\Delta t=2.5}$ (top to bottom).
 \textit{Bottom:}~${-{2c} \, \overline{\bar{F}_V(t,y)}}$ averaged over $y$ as a function `time', the error-bars indicating the corresponding standard-deviation;
 the result of the linear fit for ${t>t_{\text{min}}=25}$ is indicated by the straight lines in black.
  \textit{Inset}:~Zoom on the short-`times' behavior in a logarithmic-scale, which shows qualitatively the existence of a saturation `time' {${t_{\text{sat}} \lesssim 10}$} marking the beginning of the linear regime at large `times' (the dotted lines guide the eye for a quadratic behavior ${\sim t^2}$ as predicted analytically in~Ref.~\cite{agoritsas_2012_FHHtri}).
 \label{fig:MeanMeanFbar-3T-compil}
 }
\end{figure}

\newpage



%

\end{document}